\documentclass[numberedappendix,natbib209]{emulateapj}

\usepackage{epsfig}
\usepackage{longtable}
\usepackage{natbib}

\begin{document}

\title{The Gemini Cluster Astrophysics Spectroscopic Survey (GCLASS): The Role of Environment and Self-regulation in Galaxy Evolution at \lowercase{z} $\sim$ 1\altaffilmark{1}}

\author{Adam Muzzin\altaffilmark{2,3}, Gillian Wilson\altaffilmark{4}, H. K. C. Yee\altaffilmark{5}, David Gilbank\altaffilmark{6}, Henk Hoekstra\altaffilmark{3}, Ricardo Demarco\altaffilmark{7}, Michael Balogh\altaffilmark{8}, Pieter van Dokkum\altaffilmark{2}, Marijn Franx\altaffilmark{3}, Erica Ellingson\altaffilmark{9}, Amalia Hicks\altaffilmark{10}, Julie Nantais\altaffilmark{7}, Allison Noble\altaffilmark{11}, Mark Lacy\altaffilmark{12}, Chris Lidman\altaffilmark{13}, Alessandro Rettura\altaffilmark{4}, Jason Surace\altaffilmark{14}, Tracy Webb\altaffilmark{11}}

\altaffiltext{1}{Based on observations obtained at the Gemini Observatory, which is operated by the Association of Universities for Research in Astronomy, Inc., under a cooperative agreement with the NSF on behalf of the Gemini partnership: the National Science Foundation (United States), the Science and Technology Facilities Council (United Kingdom), the National Research Council (Canada), CONICYT (Chile), the Australian Research Council (Australia), Minist\'{e}rio da Ci\^{e}ncia e Tecnologia (Brazil) and Ministerio de Ciencia, Tecnolog\'{i}a e Innovaci\'{o}n Productiva (Argentina).}

\altaffiltext{2}{Department of Astronomy, Yale
  University, New Haven, CT, 06520-8101; adam.muzzin@yale.edu} 
\altaffiltext{3}{Leiden Observatory, Leiden University, PO Box 9513,
  2300 RA Leiden, The Netherlands}
\altaffiltext{4}{Department of Physics and Astronomy,
University of California, Riverside, CA 92521}
\altaffiltext{5}{Department. of Astronomy \& Astrophysics, University
  of Toronto, 50 St. George St., Toronto, Ontario, Canada, M5S 3H4}
\altaffiltext{6}{South African Astronomical Observatory, PO Box 9, Observatory, 7935, South Africa}
\altaffiltext{7}{Department of Astronomy, Universidad de Concepcion, Casilla 160-C, Concepcion, Chile}
\altaffiltext{8}{Department of Physics and Astronomy, University of Waterloo, Waterloo, Ontario N2L 3G1, Canada}
\altaffiltext{9}{Center for Astrophysics and Space Astronomy, 389UCB, University of Colorado, Boulder, CO 80309, USA}
\altaffiltext{10}{Department of Physics and Astronomy, Michigan State University, East Lansing, MI 48824-2320, USA}
\altaffiltext{11}{Department of Physics, McGill University, MontrŽal, QC, Canada}
\altaffiltext{12}{North American ALMA Science Center, NRAO Headquarters, 520 Edgemont Road, Charlottesville, VA 22903}
\altaffiltext{13}{Australian Astronomical Observatory, P.O. Box 296, Epping NSW 1710, Australia}
\altaffiltext{14}{Spitzer Science Center, California Institute
  of Technology, 220-6, Pasadena, CA, 91125}

\begin{abstract}
We evaluate the effects of environment and stellar mass on galaxy properties at 0.85 $< z <$ 1.20 using a 3.6$\micron$-selected spectroscopic sample of 797 cluster and field galaxies drawn from the GCLASS survey.   We confirm that for galaxies with LogM$_{*}$/M$_{\odot}$ $>$ 9.3 the well-known correlations between environment and properties such as star-forming fraction (f$_{SF}$), SFR, SSFR, D$_{n}$(4000), and color are already in place at $z \sim$ 1.  
We separate the effects of environment and stellar mass on galaxies by comparing the properties of star-forming and quiescent galaxies at fixed environment, and fixed stellar mass.  The SSFR of star-forming galaxies at fixed environment is correlated with stellar mass; however, at fixed stellar mass it is independent of environment.  The same trend exists for the D$_{n}$(4000) measures of both the star-forming and quiescent galaxies and shows that their properties are determined primarily by their stellar mass, not by their environment.  Instead, it appears that environment's primary role is to control the fraction of star forming galaxies.  
Using the spectra we identify candidate poststarburst galaxies and find that those with 9.3 $<$ LogM$_{*}$/M$_{\odot}$ $<$ 10.7 are 3.1 $\pm$ 1.1 times more common in high-density regions compared to low-density regions.   The clear association of poststarbursts with high-density regions as well as the lack of a correlation between the SSFRs and D$_{n}$(4000)s of star-forming galaxies with their environment strongly suggests that at $z \sim$ 1 the environmental-quenching timescale must be rapid.  
Lastly, we construct a simple quenching model which demonstrates that the lack of a correlation between the D$_{n}$(4000) of quiescent galaxies and their environment results naturally if self quenching dominates over environmental quenching at $z >$ 1, or if the evolution of the self-quenching rate mirrors the evolution of the environmental-quenching rate at $z >$ 1, regardless of which dominates.  

\end{abstract}

\section{Introduction}
It has been known for a long time that the properties of galaxies depend strongly on both their stellar mass and their larger-scale environment.  In the last decade, the SDSS and other spectroscopic surveys have revolutionized this field, providing high-quality data that has quantified these correlations with high precision \citep[see e.g., the review by][]{blanton09}.  It is now clear that in the local universe practically any property that can be used to characterize a galaxy shows some correlation with both its environment and stellar mass.  
\newline\indent
Galaxies in denser environments are redder \citep[e.g.,][]{Kauffmann2004, Balogh2004a, Baldry2006, vandenBosch2008, Skibba2009, vonderlinden2010}, less star forming \citep[e.g.,][]{Gomez2003, Kauffmann2004, Balogh2004b, Weinmann2006, vonderlinden2010, Peng2010, Weinmann2010}, older \citep[e.g.,][]{Smith2006, Cooper2010b}, more metal rich \citep[e.g.,][]{Smith2006,Cooper2010a}, and more frequently have elliptical/S0 morphologies \citep[e.g.,][]{Dressler1980, Postman1984, Bamford2009}.
\newline\indent
The same trends hold true for stellar mass, with more massive galaxies being redder \citep[e.g.,][]{Bower1992, Baldry2006, Thomas2010, Peng2010}, less star forming \citep[e.g.,][]{Weinmann2006, Baldry2006, Peng2010}, older \citep[e.g.,][]{Kauffmann2003, Nelan2005, Gallazzi2005, Gallazzi2006, Thomas2005, Thomas2010, Graves2009}, more metal rich \citep[e.g.,][]{Tremonti2004, Gallazzi2005, Gallazzi2006, Nelan2005, Thomas2005, Thomas2010, Graves2009} and more frequently having elliptical/S0 morphologies \citep[e.g.,][]{Bamford2009, Nair2010}.  
\newline\indent
There is also evidence for a covariance between stellar mass and environment, with more massive galaxies residing in denser environments \citep[e.g.,][]{Kauffmann2004, Baldry2006}, although this issue has not yet reached complete consensus, see e.g., \cite{vonderlinden2010} for an alternative result.  Either way, it is clear that any potential covariance between stellar mass and environment needs to be isolated in order to determine their relative influence on the galaxy population.  
\newline\indent
Recently, \cite{Peng2010} performed such an analysis and demonstrated that the star formation quenching effects of stellar mass and environment are independent of each other in the SDSS and zCOSMOS datasets up to $z \sim$ 0.6.  They used this separability to construct an empirically-based model which implies that "mass quenching" (which we will refer to in this paper as "self quenching") is the primary mechanism that ends star formation in massive galaxies (LogM$_{*}$/M$_{\odot}$ $>$ 10.6) in all environments at all redshifts; whereas "environmental quenching" is increasingly important for lower-mass galaxies at decreasing redshift.  
\newline\indent
Although we continue to acquire higher-quality measurements of the correlations between galaxy properties and their stellar mass and environment -- and even have an empirical framework for interpreting how they quench the overall galaxy population \citep{Peng2010}, we still have only circumstantial evidence about the physical processes that are responsible for quenching.  An obvious next step is to examine the evolution of galaxy properties as a function of stellar mass and environment over a wide range of epochs to determine the frequency of quenching and constrain the relevant timescale, an important clue for identifying the physical processes involved.  
\newline\indent
In the last five years extensive spectroscopic surveys of the $z \sim$ 1 universe such as GOODS \citep{Elbaz2007, Popesso2011}, DEEP2 \citep{Cooper2006, Cooper2007, Cooper2008a, Cooper2010b, Gruztbauch2011a}, zCOSMOS \citep{Iovino2010, Peng2010, Cucciati2010}, ROLES \citep{Li2011}, VVDS \citep{Cucciati2006, Scodeggio2009}, as well as the HiZELS narrow-band survey \citep{Sobral2011} have begun to quantify the effects of stellar mass and environment on galaxy properties.   Photometric surveys are now doing the same at redshifts up to $z \sim$ 2-3 \citep[e.g.,][]{Quadri2007, Grutzbauch2011b, Quadri2011}.  So far, this body of work has been somewhat controversial, with studies disagreeing on whether environment is an important factor in the quenching of star formation, and whether the star-formation-rate- (hereafter, SFR) density relation seen at $z \sim 0$ still exists at $z \sim$ 1.  
\newline\indent
Although the reason for the disagreements has not been dissected in detail, much of it may be semantic in nature, and not due to errors in data analysis or sample bias.  
This is because some authors have studied the color-density relation, others have looked at the specific-star-formation-rate- (hereafter, SSFR) density relation, while others yet have studied the actual SFR-density relation.  Some studies correlate the mean color, mean SFRs, and mean SSFRs of particular subsamples of galaxies with environment.  Others compare the change in the {\it fraction} of star-forming galaxies (hereafter f$_{SF}$) with environment, where star forming has been defined using any of those three criteria.  At some level all of these correlations have been analyzed under the moniker of the "SFR-density relation", even though they are in fact quite different quantities.  As we will show in this paper, the mean properties of galaxies and the f$_{SF}$ behave quite differently as a function of environment and stellar mass, and these different behaviors, as well as some selection effects, may be at the root of many of the past discrepancies. 
\newline\indent
Another major challenge for field galaxy surveys that has been largely ignored is that environment itself is difficult to define, and that the choice of parameterization almost certainly affects the interpretation of any analysis.  Typically 3rd or 5th nearest neighbor distance is used as a proxy for galaxy environment; however, because it counts only a few galaxies this metric is noisy and only accounts for the density of galaxies on small scales.  Using a larger number of nearest neighbor galaxies provides a better measurement of the larger scale environment; however, this comes at the cost of losing small scale information.  Furthermore, increasing the number of nearest neighbors substantially leads to problems with the edges of survey fields and requires throwing out a large fraction of the data.  A comprehensive discussion of the issues associated with defining galaxy environment is presented in \cite{Cooper2005}.  
\newline\indent
Rich galaxy clusters offer an interesting alternative for measuring the effect of environment on galaxy evolution as they cover a wide range of environmental densities.  In particular, their cores are the most extreme high-density environments at any epoch, and therefore leave little ambiguity about the measurement of local environment.  Of course, the tradeoff with using clusters is that only a small percentage of galaxies in the local or high-redshift universe live in clusters, and therefore the environmental effects seen there may not be generalizable to the galaxy population at large.  Still, they offer the possibility to define the upper limits of how strong environmental effects can be at any epoch.  
\newline\indent
A handful of clusters at $z \sim$ 1 have been studied with regard to environmental and mass evolution effects.  \cite{Patel2009, Patel2011, Koyama2008, Koyama2010, Rosati2009, Strazzullo2010, Vulcani2010, Demarco2010b} have all found that the correlation between SFR and SSFR with environment is still preserved in clusters around $z \sim$ 1.  In addition, \cite{Gobat2008}, and \cite{Rettura2010, Rettura2011} have also found that the population of quiescent galaxies in clusters at $z \sim$ 1 is more evolved than quiescent galaxies in the field.  Interestingly, there is some evidence that at $z >$ 1.4 even the cores of clusters are becoming increasingly active \citep[e.g.,][]{Tran2010, Hilton2010, Hayashi2010, Kuiper2010, Hatch2011}, although this effect is not seen in all clusters \citep[e.g.,][]{Tanaka2010, Bauer2011}.  Whether this is truly the epoch where the expected change in the star formation properties of cluster galaxies actually occurs, or is simply a selection effect is still difficult to tell as our current knowledge is based on just a handful of clusters.
\newline\indent
Here we present the most comprehensive study of the relative effects of stellar mass and environment on both cluster and field galaxies at $z \sim$ 1 to date using data from the Gemini Cluster Astrophysics Spectroscopic Survey (GCLASS, PIs: Wilson + Yee).  GCLASS is a spectroscopic survey of 10 rich clusters at 0.85 $< z < $ 1.34 using the GMOS instruments on Gemini-North and Gemini-South.  The GCLASS cluster sample has been drawn from the Spitzer Adaptation of the Red-sequence Cluster Survey \citep[SpARCS, see][]{Muzzin2009a, Wilson2009, Demarco2010}.  In this paper we use 9 of the clusters, and examine the properties of cluster galaxies as a function of stellar mass and environment for the first time using a large sample of clusters.
\newline\indent
One of the major improvements the GCLASS data offer over other spectroscopic studies is that the spectroscopic targets have been selected based on their observed IRAC 3.6$\micron$ flux.  This is different from other $z \sim$ 1 field galaxy surveys such as DEEP2, which is an R-band selected survey (R $<$ 24.1), and zCOSMOS, which is an I-band selected survey (I $<$ 22.5 for zCOSMOS-bright).  At $z \sim$ 1 these produce rest-frame U- and B-band selected samples, whereas the 3.6$\micron$ selection from GCLASS produces a rest-frame H-band selected sample and is therefore much closer to a stellar-mass-selected sample.  
\newline\indent
This paper is laid out in the following manner.  In $\S$ 2 we define the cluster sample, $\S$ 3 discusses the reduction of the spectroscopic data, and in $\S$ 4 we explain how physical quantities such as stellar masses and SFRs are extracted from the data.  Readers interested in the main results can skip those sections and refer to $\S$ 5 where we look at the correlation of galaxy properties for the sample of galaxies with stellar masses (LogM$_{*}$/M$_{\odot}$) $>$ 9.3 as a function of environment and stellar mass, and $\S$ 6 where we look at the same properties but holding galaxy type, environment, and stellar mass fixed in order to untangle their interdependence.  In $\S$ 7 we discuss the population of poststarburst galaxies as a function of environment and stellar mass, and in $\S$ 8 we present a simple model for the evolution of the quiescent galaxy population.  We conclude with a summary in $\S$ 9.   Throughout this paper we assume a $\Lambda$CDM cosmology with $\Omega_{m}$ = 0.3, $\Omega_{\Lambda}$ = 0.7, and H$_{0}$ = 70 km s$^{-1}$ Mpc$^{-1}$.

\section{GCLASS Cluster Sample}
\indent
The GCLASS cluster sample is selected from the 42 deg$^2$ SpARCS optical/IR cluster survey.  SpARCS detects clusters using the cluster red-sequence method developed by \cite{Gladders2000, Gladders2005} for the Red-sequence Cluster Surveys \citep[RCS1 \& RCS2, see][]{Yee2007,Gilbank2011}.  Those surveys select clusters as overdensities of galaxies in an R - z$^{\prime}$ color space because the R - z$^{\prime}$ color spans the 4000\AA$ $ break feature in early-type galaxies to $z \sim$ 1.  SpARCS selects clusters using a z$^{\prime}$ - 3.6$\micron$ color selection, which spans the break at $z >$ 1.  The z$^{\prime}$ data for the SpARCS survey were obtained using the CFHT and CTIO 4m telescopes and the 3.6$\micron$ data used in SpARCS is taken from the 50 deg$^2$ Spitzer Wide-area Extragalactic Survey \citep[SWIRE, ][]{Lonsdale2003}.   An overview of the SpARCS observations, field layouts, and photometry is presented in the companion papers by \cite{Muzzin2009a} and \cite{Wilson2009}.  The SpARCS cluster detection algorithm is a slightly modified version of the \cite{Gladders2000, Gladders2005} algorithm and is discussed in \cite{Muzzin2008}.  The SpARCS cluster catalog will be presented in a future paper (A. Muzzin et al., in preparation).  
\newline\indent
The 10 clusters selected for followup as part of GCLASS were chosen in the following manner.  Clusters with red-sequence photometric redshifts between 0.85 $< z < $ 1.25 were ranked in order of increasing detection significance.  From this list we hand-selected a sample that attempted to include the richest clusters, but also to have an even distribution both in photometric redshift and right ascension (R.A.).  The resulting sample of nine clusters are evenly divided into three redshift bins, 0.85 $< z <$ 1.0, 1.0 $< z <$ 1.1, 1.1 $< z <$ 1.25, and are generally the three richest clusters in SpARCS in their respective redshift range.  Lastly, we included SpARCS J003550-431224 at $z = 1.34$, a cluster for which we had already some observations, \citep{Wilson2009} as the 10th cluster to extend the redshift range of the sample.  The high-redshift of SpARCS J003550-431224 means the spectra available for it are limited to $\lambda_{rest} <$ 4050\AA.  This wavelength coverage is too short for the spectral properties to be analyzed in the same manner as the other clusters, so in the current paper we exclude SpARCS J003550-431224 and focus on the remaining 9 clusters.  All statistics of the spectroscopy quoted throughout the remainder of the paper exclude the spectra from SpARCS J003550-431224.
\newline\indent
We note that there are richer clusters in SpARCS that were not included as part of GCLASS because of the redshift and R.A. distribution requirements and therefore, formally GCLASS is not a richness-limited sample of clusters at 0.85 $< z_{phot} <$ 1.25.  The GCLASS clusters comprise 10/24 of the richest clusters at $z_{phot} > 0.85$ in SpARCS and therefore in general, should be considered a fair sampling of IR-selected rich clusters within that redshift range.  Color images of the GCLASS clusters in the gz$^{\prime}$[3.6$\micron$] bands are shown in Figure 1.
\newline\indent
As Figure 1 shows, even rich clusters at $z \sim$ 1 have a diverse range of morphologies, from roughly spherically-symmetric and centrally-concentrated with a clear central galaxy, to asymmetric with filamentary-like structure and no clear central galaxy.  Figure 1 highlights the importance of using a sample of clusters when studying the effect of the cluster environment on galaxy evolution, so as to average over the peculiarities of the assembly history of individual systems.  
\newline\indent
Given the complexity of the galaxy distribution in the clusters, the centroid for 8 of the 9 clusters has been defined as the position of the brightest cluster galaxy (BCG), where the BCG is defined as the brightest 3.6$\micron$ source that has a spectroscopic redshift consistent with being a cluster member.  As Figure 1 illustrates, the location of the BCG for those 8 clusters is close to the center of the luminosity-weighted red-sequence galaxy distribution in the observed z$^{\prime}$ - 3.6$\micron$ vs. 3.6$\micron$ color-magnitude space, and hence makes a logical choice of centroid.  In the remaining cluster, SpARCS J105111+581803 at $z = 1.035$, the BCG is substantially offset from the center of the luminosity-weighted red-sequence galaxy distribution.  For this cluster, the cluster center has been defined as the centroid of the luminosity-weighted red-sequence galaxy distribution \citep[a discussion of how this is computed is presented in][]{Muzzin2008}
\begin{deluxetable*}{lcrrccccr}
\tabletypesize{\scriptsize}
\tablecolumns{9}
\tablecaption{GCLASS Cluster Sample}
\tablewidth{7.3in}
\tablehead{\colhead{Name} & \colhead{ $z_{spec}$ } &
\colhead{R.A.} & \colhead{Decl.} & \colhead{Masks} & \colhead{Total} &
\colhead{Total} & \colhead{Cluster} & \colhead{Reference} \\
\colhead{} & \colhead{} & \colhead{J2000} &
\colhead{J2000} & \colhead{} & \colhead{Slits} & \colhead{Redshifts} &  \colhead{Members\tablenotemark[1]}  & \colhead{} \\
\colhead{(1)}& \colhead{(2)}& \colhead{(3)}& \colhead{(4)}&
\colhead{(5)}& \colhead{(6)} &
\colhead{(7)} & \colhead{(8)} & \colhead{(9)}
}
\startdata
SpARCS J003442-430752 & 0.867 & 00:34:42.086 & -43:07:53.360 & 4 & 159 & 137 & 45 (40) &  \cr
SpARCS J003645-441050 & 0.869 & 00:36:45.039 & -44:10:49.911 & 4 & 170 & 119 & 47 (47) & \cr
SpARCS J161314+564930 & 0.871 & 16:13:14.641 & 56:49:29.504 & 5 & 211 & 161 & 92 (85) & Demarco et al. (2010a) \cr
SpARCS J104737+574137 & 0.956 & 10:47:37.463 & 57:41:37.960 & 4 & 200 & 147 & 31 (31) & \cr
SpARCS J021524-034331 & 1.004 & 02:15:23.200 & -03:43:34.482 & 4& 167 & 125 & 48 (48) & \cr
SpARCS J105111+581803 & 1.035 & 10:51:11.232 & 58:18:03.128 & 4 & 194 & 145 & 34 (32) & \cr
SpARCS J161641+554513 & 1.156 & 16:16:41.232 & 55:45:25.708 & 5 & 242 & 162 & 46 (46) & Demarco et al. (2010a) \cr
SpARCS J163435+402151 & 1.177 & 16:34:35.402 & 40:21:51.588 & 6 & 205 & 125 & 50 (44) & Muzzin et al. (2009) \cr
SpARCS J163852+403843 & 1.196 & 16:38:51.625 & 40:38:42.893 & 6 & 189 & 112 & 44 (39) & Muzzin et al. (2009) \cr
SpARCS J003550-431224\tablenotemark[2,3] & 1.335 & 00:35:49.700 & -43:12:24.160 & 3 & 91 & 49 & 20 & Wilson et al. (2009)
\enddata

\tablenotetext{1}{The number of confirmed cluster members with LogM$_{*}$/M$_{\odot}$ $>$ 9.3 is shown in brackets.}
\tablenotetext{2}{This cluster is omitted in the current analysis because the usable wavelength coverage of the spectra is limited due to the high redshift.} 
\tablenotetext{3}{Additional masks for this cluster have been observed and will be presented in the GCLASS catalog paper.}  
\end{deluxetable*}

\begin{figure*}
\plotone{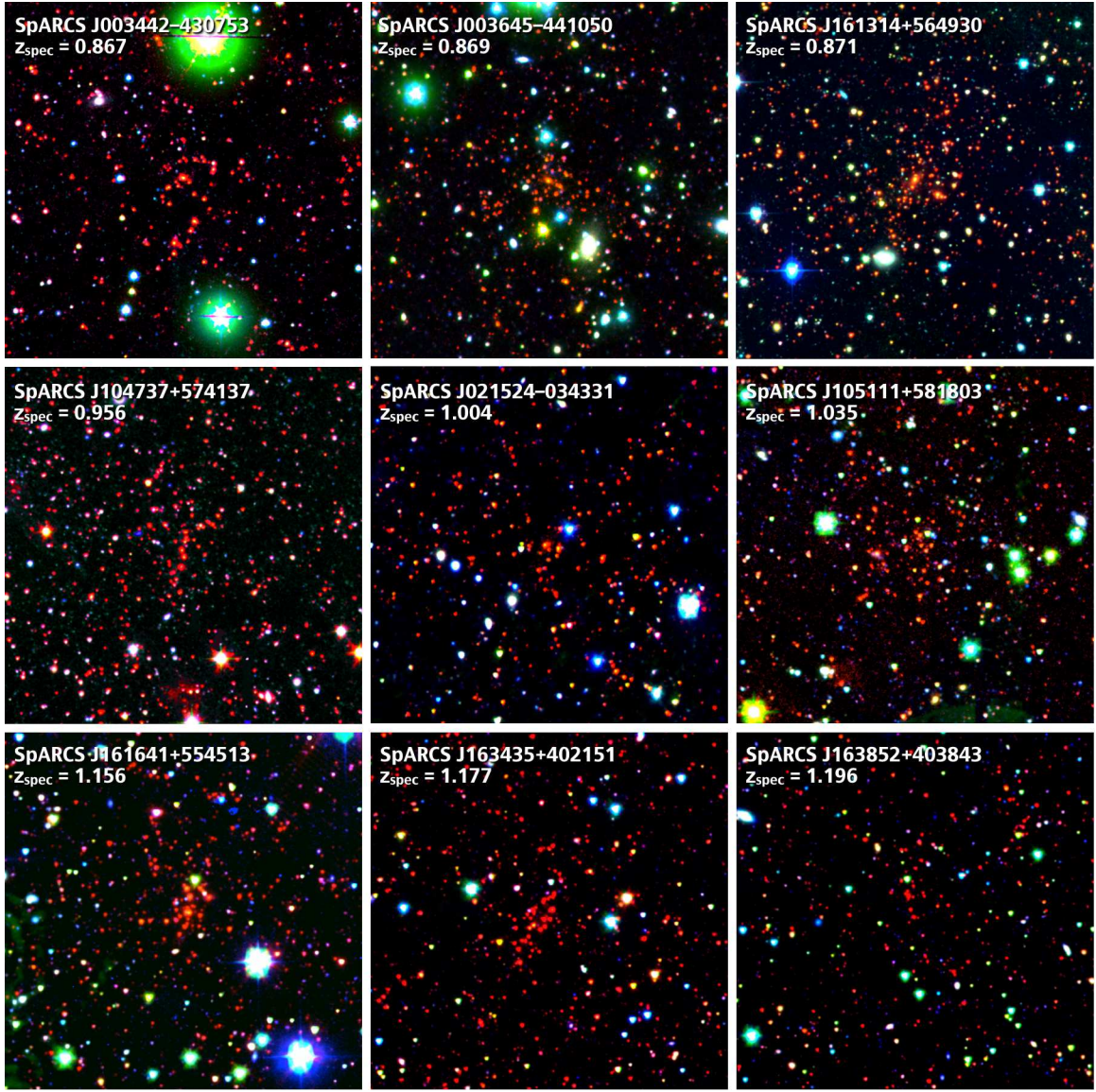}
\caption{\footnotesize gz[3.6$\micron$] color images of the nine GCLASS clusters used in this analysis.  The field of view of each image is 7$'$ $\times$ 7$'$, which is approximately the area covered by the spectroscopic observations and corresponds to 3.5 Mpc on a side at $z = 1$.  Rich clusters at $z \sim$ 1 show a range of morphologies, from roughly spherical and centrally-concentrated with a clear central galaxy (e.g., SpARCS J003645-441050, $z = 0.869$), to asymmetric with filamentary-like structure and no clear central galaxy (e.g., SpARCS 104737+574137, $z = 0.956$).}
\end{figure*}
\section{Spectroscopic Data}
The goal of the GCLASS spectroscopic observations was to obtain spectra for $\sim$ 50 members in each of the ten clusters.  Based on the cluster color-magnitude diagrams and richnesses, we estimated this would provide a spectroscopic completeness of $>$ 50\% for galaxies with M$_{3.6\micron}$ $<$ M*$_{3.6\micron}$ + 1.0 (see $\S$ 4.6 for a complete discussion of the spectroscopic completeness).  Geometric slit placement restrictions make obtaining 100\% complete spectroscopic samples unrealistic, so the sampling in GCLASS was chosen as a trade off between completeness and efficiency.  The completeness is sufficient to determine high-quality velocity dispersions for each cluster (see G. Wilson in preparation) and also allows us to compute reliable completeness corrections as a function of both stellar mass and clustercentric radius that can be used for defining complete samples of galaxies for galaxy evolution studies.  Although the completeness level per cluster is not as high as some previous surveys of individual clusters at similar redshift \citep[e.g.,][]{Tran2007, Demarco2007}, GCLASS uses substantially fewer masks per cluster which in turn permits the study of a larger sample of clusters and cluster galaxies.  
\newline\indent
Full details of the GCLASS mask design process, and data reduction will be presented in a future catalog release paper, here we outline the most relevant steps of the observational setup and data reduction.
\subsection{Target Selection and Mask Design}
Spectroscopic targets were prioritized using three criteria, which in order of importance were: 1) clustercentric radius 2) observed z$^{\prime}$ - 3.6$\micron$ color, 3) 3.6$\micron$ flux.  Highest priority was given to bright 3.6$\micron$ sources in the cluster core with colors close to the observed red-sequence, lowest priority was given to faint 3.6$\micron$ sources on the cluster outskirts with z$^{\prime}$ - 3.6$\micron$ colors 1.0 magnitude redder or 1.5 magnitudes bluer than the cluster red-sequence\footnote{The precise color prioritization varies from cluster-to-cluster by a few tenths of a magnitude around these numbers.  The color selection was broader for the higher-redshift clusters due to the bluer rest-frame probed by z$^{\prime}$ - 3.6$\micron$}.  As an illustration of how the slit priorities were assigned, in Figure 2 we plot the observed z$^{\prime}$ - 3.6$\micron$ vs. 3.6$\micron$ color-magnitude diagram for SpARCS J105111+581803.  
\newline\indent
In total there were 12 priority levels for targets (P1 - P12).  P1 - P10 were the primary selection criteria optimized for obtaining the maximum number of cluster redshifts, and P11 and P12 were included as "filler" slits to maximize the number of slits per mask as well as provide a sample of galaxies without a specific color selection to test for possible selection biases.    P1-P4 were the highest priority slits, targeting galaxies within 550 kpc of the cluster center and z$^{\prime}$ - 3.6$\micron$ colors close to, or on the observed cluster red sequence (see Figure 2).  P5 were sources within R $<$ 550 kpc that met no specific color criteria but with detections in the MIPS 24$\micron$ band.  These sources were given higher priority than filler slits because of the possibility that they may be cluster galaxies with colors substantially different than the primary cluster color priorities.  P6 - P10 were identical to P1 - P5, but for galaxies at R $>$ 550 kpc.  The filler slits, P11 and P12, were galaxies in any location with z$^{\prime} <$ 22 mag, and z$^{\prime} >$ 22 mag, respectively.  
\newline\indent
Slitmasks were designed using custom software created by us to produce output files which could also be read by the {\it gmmps} software provided by the Gemini observatory.  Masks contained between 15 and 59 slits (excluding alignment stars), with a median of 42 slits per mask.   Table 1 lists the clusters with the total number of masks and slits per cluster.
\newline\indent
The greatest concern with the selection of the spectroscopic targets is that the use of a z$^{\prime}$ - 3.6$\micron$ color cut could introduce some bias into the galaxy sample.  A pure 3.6$\micron$ flux-limited sample would be preferable; however, due to the high-redshift of the clusters such a selection would be dominated by low-redshift foreground galaxies and is not an efficient use of telescope resources.  
\newline\indent
As Figure 2 shows, the color priority selection was chosen to be quite broad so that it would span the properties of even the most extreme red and blue cluster members, and only exclude the most obvious very low/high redshift interlopers.  It is also important to note that the color criteria only provided priority levels for the mask design algorithm, not hard color cuts.  Many filler slits (P11 \& P12) that did not meet any color criteria were placed and we used these to test the if the color prioritization provides an optimum sampling of cluster members.  Filler slits resulted in a total of 274 redshifts across all clusters, 15 of which were cluster members.  This suggests that the color selection prioritized cluster members well (422/437 of the cluster members met the color criteria), but also provided an efficient use of overall telescope time (only 15/274 galaxies that did not meet the color criteria were members).
\subsection{Observations and Data Reduction}
The spectroscopic masks were observed with GMOS-N and GMOS-S using the R150 grating with a central wavelength of 7500\AA~with 1$"$ wide slits which gives a resolution of 17\AA, equivalent to an R = 440.  For all masks we used 3\arcsec~long microslits with the nod-and-shuffle option \citep[N\&S, see][]{Glazebrook2001, Abraham2004} available with GMOS.  For each cluster, approximately half the masks were observed in the "band shuffle" mode, and the other half were observed with the "micro shuffle" mode.   The main advantage of the band shuffle mode is that the shuffled charge is stored on the top and bottom third of the chip while observing, not next to the slit and therefore the 3\arcsec~slits can be packed directly beside each other in the cluster core where the density of galaxies is highest.  The disadvantage is that because of the need to store the shuffled charge, only a strip covering the central 1.7$'$ of y-axis the chip can be used.  In the micro shuffle mode, charge is stored beside the slit, which means slits must be at least 3\arcsec~away from each other; however, the full 5.5$'$ FOV of GMOS can be used.  We found that the combination of half band shuffle and half micro shuffle masks produced a distribution of slits that provided both a good sampling of the cluster core, but also fair coverage of cluster galaxies out to a radius of 3$'$, which is $\sim$ 1.5 Mpc at $z \sim$ 1. 
\newline\indent
For all observations we used the OG610 filter which blocks light blueward of $\sim$ 6100\AA$ $ observed frame.  This made the spectra shorter so that two tiers of slits could be placed per mask.  
Each mask was observed with six exposures, each with 15 N\&S cycles with 60s exposure time at each of the two nod positions.  We nodded along the slit, which resulted in 30 min on-sky, per exposure, for a total of 3 hrs on-sky per mask.  We used three positions for the central wavelength of the grating, 7380\AA, 7500\AA, and 7620\AA~which dithers the spectra in the dispersion direction and effectively fills in the GMOS detector gaps.  We also used the DTX offsetting in the spatial direction, which provides additional dithering, mitigating some of the noise from the N\&S charge traps.
\newline\indent
Data reduction was performed using the IRAF GMOS package provided by the Gemini observatory.  Each frame was subtracted using N\&S darks.  These frames follow the same shuffling pattern as the science data, but are taken with the shutter closed.  Subtracting them removes the bias pattern, dark current, as well as charge that builds up and becomes trapped on certain pixels that do not shuffle the charge efficiently.  Unlike typical spectroscopic observations, we did not flat field the exposures before final combination because the GMOS chips, particularly GMOS-S, suffer from significant fringing.  For GMOS-S the fringing is $\sim$ 10\% peak-to-peak starting at $\lambda$ $>$ 8000\AA~but gets to as high as 55\% peak-to-peak at $\lambda$ $>$ 9000\AA.  The N\&S observation mode does a superior job of removing the fringing in the spectra because the sky subtraction is determined using the same pixels that the object spectra is on (due to the on/off nodding pattern); however, the flat fields taken with an internal lamp also have significant fringing.  Attempting to correct for pixel-to-pixel sensitivity using the lamp exposures re-introduces pixel-to-pixel variations from the fringing that was removed with the N\&S sky subtraction.  The fringing is stronger than the pixel-to-pixel variations which are further mitigated by the dithering in both the spatial and dispersion directions, therefore the quality of data that is not flat fielded is substantially higher than that which is flat fielded.
\newline\indent
The six individual frames per mask were coadded into a final image of 2d spectra, and then 1d spectra were extracted using the iGDDS software developed for the Gemini Deep Deep Survey \citep[GDDS,][]{Abraham2004}.  Wavelength calibration is performed by identifying sky lines, and redshifts are obtained by comparison with templates available in the iGDDS software.  Relative flux calibration was determined using long slit observations of standard stars in each program.  Comparison of the standard stars shows the relative flux calibration remained stable over the several years of observations.  We apply the same flux calibration curve to all spectra, regardless of position, which assumes that the chromatic response of all detector pixels is the same.  This may introduce errors in the continuum shape of the spectra if there is variance in the GMOS chromatic response that are position-dependent, although at present there is no evidence to suggest this is a significant issue.
\newline\indent
Based on the completed observations up to March 2011 we created a v1.0 spectroscopic catalog for the clusters.  The v1.0 catalog contains observations for 45/46 total masks in GCLASS.  Table 1 shows the number of slits and successful redshifts in the field of each cluster in the sample, based on the v1.0 catalogs.
\begin{figure*}
\plotone{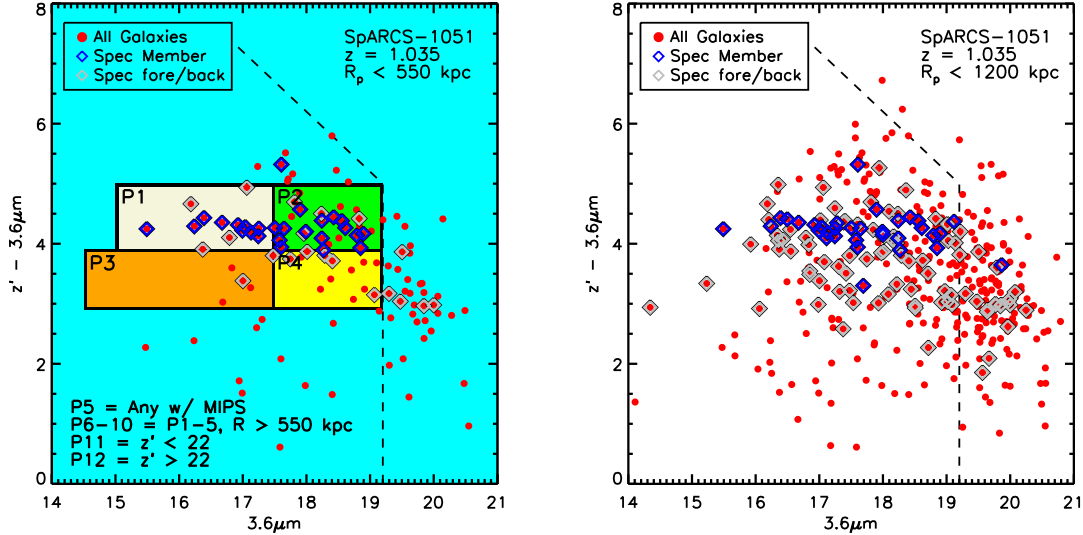}
\caption{\footnotesize Left Panel: Observed z$^{\prime}$ - 3.6$\micron$ color vs. 3.6$\micron$ magnitude of galaxies in the field of the cluster SpARCS J105111+581803 (red points).  Only galaxies with R $<$ 550 kpc are plotted.  The dashed lines denote the 5$\sigma$ completeness limits on the photometry.  Galaxies identified as spectroscopic members are plotted with blue diamonds, and those that are confirmed foreground/background galaxies are plotted with grey diamonds.  The shaded boxes represent the priority levels for targets in the mask design process.  Right Panel: Same as left panel but for galaxies at R $<$ 1200 kpc.  Overall, the spectroscopic completeness at the various priority levels is high, and the priority levels are well-tuned to obtaining the maximum number of cluster redshifts.}
\end{figure*}  
\section{Determination of Galaxy Properties}
\subsection{Cluster Membership}
Galaxies are considered to be cluster members if their velocity relative to the cluster velocity in the rest-frame is $\Delta$v $\leq$ 1500 km s$^{-1}$.  We experimented with more sophisticated methods of defining membership based on fractional values of the cluster velocity dispersion, but found this made little difference to the final sample of cluster members.  This is primarily because the clusters are well-defined in velocity space (see G. Wilson, in preparation) with the vast majority of objects within 2$\sigma$ of line-of-sight velocity dispersion, values that are typically less than $\Delta$v $\leq$ 1500 km s$^{-1}$.   Therefore, in the interest of simplicity, we adopt this method for defining membership.  This definition is also conveniently similar to the line-of-sight cuts used for defining environment in field spectroscopic surveys \citep[e.g.,][]{Kauffmann2004, Cooper2005, Peng2010}.  Based on this criteria, the total number of cluster members across the sample of 9 clusters is 437.  The number of members per cluster is listed in Table 1.
\subsection{Field Sample}
Throughout this paper we study the properties of galaxies as a function of environment, where environment is defined by the clustercentric radius of the galaxy.  We have chosen to use physical clustercentric radius in kpc as our measure of the cluster environment, rather than rescale to the virial radius of each cluster.  The clusters have a fairly modest range of velocity dispersions (a factor of $\sim$ 3, G. Wilson et al., in preparation), so physical clustercentric radii should scale roughly the same as virial radius for each cluster.  A physical clustercentric radius is also appealing because it does not require the assumption of virial equilibrium, something which may not yet be achieved in young galaxy clusters.
\newline\indent
Clustercentric radius is a robust measure of environment and correlates well with other environmental estimators such as local density or Nth nearest neighbor; however, even at large clustercentric radius clusters contain a  population of backsplash galaxies which have already completed an orbit that brought them into or near the cluster core.  Therefore, it is useful to define a control population of "field" galaxies that have no association whatsoever with a rich galaxy cluster.  The field population will be drawn from a wide range of environments: from small groups to voids; however, as clusters grow they accrete galaxies from all environments, hence the field sample should be representative of the population of galaxies that are infalling into the clusters at that epoch.
\newline\indent
The GCLASS cluster sample spans 0.85 $< z <$ 1.20 so data from the survey itself can be used to define a population of field galaxies that have a similar redshift range as the clusters, as well as similar color/magnitude selection criteria.  The very broad color prioritization in the mask design means that numerous field galaxies with similar colors as the cluster galaxies also have reliable spectroscopy.  We consider galaxies at 0.85 $< z <$ 1.20 that have $\Delta$v $>$ 2000 km s$^{-1}$ from the cluster velocity to not be associated with the cluster and are assigned to the field sample.  This selection results in a sample of 273 field galaxies with identical color/magnitude selection criteria as the cluster galaxies.  
\newline\indent
Given that massive clusters often form in overdense regions such as super clusters, there is some concern that the field sample will not be completely representative of a blank-field field sample, because line-of-sight groups may be more prevalent when looking towards a cluster, particularly clusters selected as overdensities in an optical/IR survey.  As a test, we drew an additional field sample of galaxies in the same redshift range from the GDDS survey \citep{Abraham2004}.  The GDDS spectroscopy is taken with the same instrument as GCLASS and has the same resolution and uses the N\&S mode, hence the spectra can be compared directly with the GCLASS spectra.  
\newline\indent
The GDDS is a K-selected sample (K $<$ 20.6) in four independent GMOS fields and has substantially deeper spectra than GCLASS (20-30 hrs on sky compared to 3 hrs).   In the 0.85 $< z <$ 1.20 redshift range the observed K - 3.6$\micron$ colors of galaxies vary weakly with spectral type, hence it is straightforward to consider the GDDS sample in terms of a 3.6$\micron$-selected sample.  Based on an average of \cite[hereafter BC03]{Bruzual2003} models with different star-formation histories and ages, we infer that at $z = 1$, the typical K-3.6$\micron$ color is K-3.6$\micron$ =  1.2 and convert the K-band fluxes of the GDDS galaxies to 3.6$\micron$ fluxes using this transformation.  
\newline\indent
Once the K-band flux is converted to a 3.6$\micron$ flux, the GDDS selection is 3.6$\micron$ $<$ 19.4 (Vega magnitudes), which is nearly identical to the depth of the SWIRE observations, 3.6$\micron$ $<$ 19.2 (Vega magnitudes).  We compared the GCLASS field sample to the GDDS field sample by performing a stacking analysis of the spectra (see $\S$ 5).  The equivalent width (EW) of the [OII] emission line, and D$_{n}$(4000) measures for the stacks are the same to within 5\%, suggesting that the GCLASS field sample is representative of a blank-field field sample.  Given the superior signal-to-noise ratio (S/N) of the GDDS spectra, we include the 87 GDDS galaxies at 0.85 $< z <$ 1.20 in our field sample, bringing the total number of galaxies in the field sample to 360 galaxies.  Colors, SFRs, and stellar masses for the GDDS galaxies are computed using the same methods as for the GCLASS galaxies.
\begin{figure}
\plotone{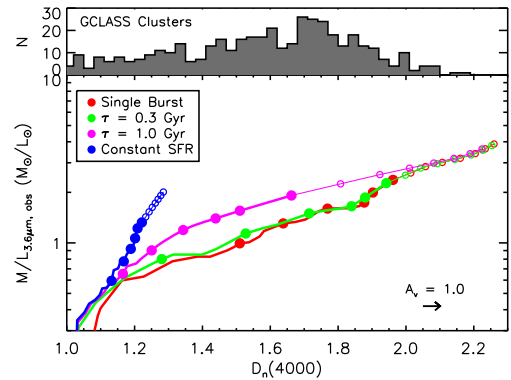}
\caption{\footnotesize Bottom Panel: Age tracks of models that show the M/L ratio in the observed 3.6$\micron$ band at $z \sim$ 1 as a function of D$_{n}$(4000).  The age in Gyr (1 - 13) starts at the left side and is denoted by the circles on each track.  Ages $>$ 6 Gyr are forbidden by the age of the universe at $z \sim$ 1 and are shown as open circles.  Top Panel: The distribution of D$_{n}$(4000) values in the GCLASS cluster galaxies.  The models show that galaxies can only obtain D$_{n}$(4000) $>$ 1.7 by $z \sim$ 1 if they have short star formation timescales $and$ ages nearly as old as the universe.  Even for the extreme star-formation histories considered here the range in M/L$_{3.6\micron}$ values at fixed D$_{n}$(4000) is at most a factor of a few.  Thefore, the M/L$_{3.6\micron}$ for all galaxies is determined based on the $\tau$ = 0.3 Gyr model. }
\end{figure}
\subsection{Stellar Masses}
Stellar masses are calculated for all galaxies by using the spectra to determine the stellar mass-to-light ratio (M$_{\odot}$/L$_{\odot}$) in a manner similar to that developed by \cite{Kauffmann2003} for the SDSS spectroscopic sample.  \cite{Kauffmann2003} used the BC03 models to make synthetic spectra of galaxies with a range of star-formation histories.  From these synthetic spectra they constructed a grid of observed z$^{\prime}$-band stellar mass-to-light ratios (M$_{\odot}$/L$_{z^{\prime}}$) as a function of D(4000) and the H$\delta$ absorption measure.   
Our method is similar, but with a few modifications given the different photometric bands in the surveys, and the fact that the typical S/N for GCLASS spectra is lower than that in the SDSS.  
\newline\indent
We also use the BC03 stellar population models with solar metallicity and a \cite{Chabrier2003} initial mass function (IMF); however, we determine the observed 3.6$\micron$ stellar mass-to-light ratios (M$_{\odot}$/L$_{3.6\micron}$).  At $z \sim$ 1 this corresponds to the rest-frame H-band, which is similar to the z$^{\prime}$-band for SDSS galaxies in that it has a weak dependence on galaxy spectral energy distribution (SED) type and suffers little from dust extinction.  
\newline\indent
\cite{Maraston2005} has released an alternative set of stellar population models which use a different treatment of the thermally-pulsating asymptotic giant branch stars (TP-AGB) than the BC03 models.  Those models predict M$_{\odot}$/L$_{\odot}$ in the rest-frame near-infrared that are approximately a factor of $\sim$ 1.6 smaller than those from the BC03 models.  At present it is unclear which of the available models provides a better description of high-redshift galaxies \citep[e.g.,][]{Wuyts2007,Muzzin2009b,Kriek2010}, so we have chosen to use the BC03 models to provide continuity with previous studies. We note that the choice of models does not affect our analysis in any significant way, because the difference in stellar masses between the BC03 and the Maraston models is a systematic offset of a factor of $\sim$ 1.6, with little dependence on the stellar population \cite[e.g.,][]{Wuyts2007,Muzzin2009b}.  
\newline\indent 
The main difference between our method for determining stellar mass and \cite{Kauffmann2003} is that we use the D(4000) measures to estimate the M$_{\odot}$/L$_{3.6\micron}$, but do not use the H$\delta$ measures.  
H$\delta$ is a relatively weak absorption feature with $\sim$ 1-5\AA~EW for most galaxies.  The S/N for the brighter galaxies in GCLASS is high and measuring H$\delta$ is trivial; however, this is not the case for the fainter spectra.  The difficulty is that at the redshift range of the galaxy sample the rest-wavelength of H$\delta$ (4102\AA) corresponds to observed wavelengths of 7590\AA~to 9025\AA, a region that has numerous bright sky lines that at the low-resolution of the spectra cause significant noise.  As we will show below, H$\delta$ is an indicator of a young stellar population, but the fact that the universe is much younger at $z = 1$ than $z =$ 0 means that H$\delta$ is less important when determining M$_{\odot}$/L$_{3.6\micron}$.
\newline\indent
Like \cite{Kauffmann2003} we adopt the \cite{Balogh1999} definition of D(4000) which is the ratio of in the flux density at 3850\AA-3950\AA~to the flux density at 4000\AA-4100\AA, and hereafter refer to this as D$_{n}$(4000).  We parameterize the star-formation history of the galaxies using the standard declining exponential $\tau$-models of the form SFR $\propto$ e$^{-t/\tau}$, where $t$ is the time since the onset of star formation, and $\tau$ sets the star formation timescale.  
As an example, Figure 3 shows the evolution of the observed M$_{\odot}$/L$_{3.6\micron}$ at $z = 1$ as a function of D$_{n}$(4000) for models with a range of $\tau$.   Each point on the tracks indicates an additional Gyr in age starting from the left side of the plot.  Ages older than the age of the universe at $z = 1$ (t = 6 Gyr) are forbidden and are indicated by open circles.  The measured D$_{n}$(4000) values from the GCLASS spectra are shown as the histogram in the top panel of Figure 3.  
\newline\indent
The arrow in Figure 3 shows the effect of 1 magnitude of V-band extinction (A$_{v}$ = 1.0) on the D$_{n}$(4000) of the models assuming a \cite{Calzetti2000} dust law.  This weak dependence on dust illustrates an advantage of using D$_{n}$(4000) as an indicator of the M$_{\odot}$/L$_{\odot}$ compared to SED-fitting.  This weak dependence occurs because of the narrow wavelength range used to measure D$_{n}$(4000), compared to the much broader dust extinction curve.  Of course, the disadvatange of using D$_{n}$(4000) as compared to SED-fitting is that it does not provide additional information on the star-formation history of the galaxy, i.e., which $\tau$-model should be used.
\newline\indent
Figure 3 shows the maximum range in M$_{\odot}$/L$_{3.6\micron}$ between the most extreme models (the youngest constant star formation vs. the oldest single-burst) is a factor of $\sim$ 13. The age of the universe does not allow for populations with t $>$ 6 Gyr, and therefore for galaxies in the GCLASS sample the maximum range in allowed M$_{\odot}$/L$_{3.6\micron}$ is only half as large, a factor of $\sim$ 7.  This is still substantial; however, an additional restriction on the M$_{\odot}$/L$_{3.6\micron}$ is the fact that not all star-formation histories are consistent with the data.  Models with $\tau$ $\geq$ 1 Gyr do not provide sufficiently strong D$_{n}$(4000)'s to match many of the cluster galaxies, even if the stellar populations are as old as the age of the universe.  Many of the D$_{n}$(4000)'s are so strong they can only be obtained with relatively short star formation timescales ($\tau$ $\leq$ 0.3 Gyr), and old ages.   For the 94\% of galaxies with D$_{n}$(4000) $>$ 1.2, the maximum difference in allowed M$_{\odot}$/L$_{3.6\micron}$ at fixed D$_{n}$(4000) is only a factor of $\sim$ 4.  This factor is the most extreme range in allowed M$_{\odot}$/L$_{3.6\micron}$ across all models and suggests that the stellar masses would be reasonably accurate even without any additional information on the star formation history from the D$_{n}$(4000).  
\newline\indent
Similar to other high-redshift stellar mass studies \citep[e.g.,][]{ForsterSchreiber2004, Marchesini2009} we adopt the $\tau$ $=$ 0.3 Gyr model as our default model and thereafter determine the M$_{\odot}$/L$_{3.6\micron}$ based on the D$_{n}$(4000).  This approach may underestimate the M$_{\odot}$/L$_{3.6\micron}$ for populations with D$_{n}$(4000) $\leq$ 1.2 by as much as a factor of $\sim$ 2, but should work well for the majority of the galaxies in the sample.  Given the other systematic uncertainties involved in stellar mass modeling \citep[e.g.,][]{Maraston2006, Muzzin2009b, Conroy2009}, and that our primary use of the stellar masses is to rank-order galaxies in increasing mass, this should not affect our analysis.  We note that it may cause systematic offsets when comparing our stellar masses of galaxies with weak D$_{n}$(4000) to those determined by other studies with the more typical SED fitting method, and advise caution if doing so. 
\newline\indent
The full spectroscopic sample of galaxies consists of 437 cluster members and 360 field galaxies.  Of the 437 cluster members, 9 of the galaxies are not detected at 3.6$\micron$, and of the 360 field galaxies, 26 are not detected at 3.6$\micron$.  Based on the SWIRE 3.6$\micron$ flux limit we estimate that these galaxies have LogM$_{*}$/M$_{\odot}$ $<$ 9.3.  For these galaxies we do not attempt to calculate stellar masses using the D$_{n}$(4000) models and their z$^{\prime}$-band magnitudes.  At $z \sim$ 1 the z$^{\prime}$-band corresponds to rest-frame B-band.  The M$_{\odot}$/L$_{z^{\prime}}$ at $z = $ 1 have a substantially larger range at fixed D$_{n}$(4000) than the M$_{\odot}$/L$_{3.6\micron}$, and also a strong dependence on dust content and hence are uncertain. For the remainder of this analysis we remove these galaxies from the sample and only consider galaxies above a limiting mass of LogM$_{*}$/M$_{\odot}$ $>$ 9.3.  This requires removing and additional 16 cluster galaxies and 40 field galaxies that have 3.6$\micron$ detections, but LogM$_{*}$/M$_{\odot}$ $<$ 9.3; however, it provides a firm 3.6$\micron$-limited sample, which can be corrected for completeness to a stellar-mass-limited sample (see $\S$ 4.6).  Once the limit of LogM$_{*}$/M$_{\odot}$ $>$ 9.3 has been applied, the final sample of cluster members contains 412 galaxies and the final field sample contains 294 galaxies.  The number of cluster members in each cluster above the mass limit is indicated in Table 1.
\subsection{Star Formation Rates}
Star formation rates are determined using the luminosity of the [OII] emission line (L$_{\mbox{\footnotesize 3727\AA,line}}$), which has been determined from the aperture-corrected line flux (F$_{\mbox{\footnotesize 3727\AA,line}}$).  The F$_{\mbox{\footnotesize 3727\AA,line}}$ is itself derived from the measurement of the EW of the [OII] line.   The first step in determining the EW([OII]) is to fit the emission line to a Gaussian in the rest-frame.  The continuum level F$_{\mbox{\footnotesize 3727\AA,cont}}$ is then determined using the region of the spectrum within 50\AA~on both the red and blue side of the emission line.  
The EW([OII]) is then converted to an aperture-corrected line flux using the total 3.6$\micron$ flux.   We determine the ratio between the continuum at rest-frame 3727\AA~and observed-frame 3.6$\micron$ (F$_{\mbox{\footnotesize 3727\AA,cont}}$/F$_{\mbox{\footnotesize 3.6\micron,obs}}$)$_{\mbox{\footnotesize model}}$ from the BC03 model implied by the D$_{n}$(4000) measurement (see $\S$ 4.3).  This is measured for each galaxy individually because the rest-frame probed by 3.6$\micron$ is slightly different depending on the redshift of the galaxy.  The [OII] line flux is then F$_{\mbox{\footnotesize 3727\AA,line}}$ = F$_{\mbox{\footnotesize 3.6\micron}}\times$(F$_{\mbox{\footnotesize 3727\AA,cont}}$/F$_{\mbox{\footnotesize 3.6\micron,obs}}$)$_{\mbox{\footnotesize model}}$$\times$EW([OII]), where F$_{\mbox{\footnotesize 3.6\micron}}$ is the observed total 3.6$\micron$ flux of the galaxy.  
\newline\indent
Once the F$_{\mbox{\footnotesize 3727\AA,line}}$ is determined we convert this to a L$_{\mbox{\footnotesize 3727\AA,line}}$ using the spectroscopic redshift and apply the mass-dependent [OII]-SFR relation determined by \cite{Gilbank2010}.  The \cite{Gilbank2010} relation was empirically calibrated using Balmer-decrement corrected H$\alpha$ SFRs in the SDSS.  This relation therefore encodes the extinction correction and metallicity dependence of the [OII] line implicitly by assuming these quantities are correlated with stellar mass.  As shown in Figure 3 of Gilbank et al., the mass-dependence of the conversion is weak for galaxies with LogM$_{*}$/M$_{\odot}$ $>$ 10.5, which comprise much of our sample, but gets increasingly larger for galaxies with LogM$_{*}$/M$_{\odot}$ $<$ 10.5.  We note that although the mass-dependence of the correction is weak at LogM$_{*}$/M$_{\odot}$ $>$ 10.5, the correction itself is significant, approximately a factor of 3 larger than a constant-luminosity [OII] SFR.
\subsection{Rest-frame U-B$_{short}$ Colors}
Synthetic rest-frame colors for the galaxies are calculated by observing the flux-calibrated spectra in the rest-frame with the appropriate filter response functions.  The wavelength coverage of the spectra in the observed frame is 6100\AA~- 9500\AA, which corresponds to rest-frame 3300\AA~- 5140\AA~for the lowest-redshift galaxies ($z = 0.85$), and 2700\AA~- 4320\AA~for the highest-redshift galaxies ($z = 1.20$).  The wavelength coverage is sufficient to compute a synthetic U-band flux for galaxies at all redshifts; however, the B-band is only completely within the spectral window for the lowest-redshift galaxies.  The peak transmission for the B-band is $\sim$ 4100\AA; however, it has a long tail of transmission redward that does not decrease to $<$ 10\% of maximum transmission until $\sim$ 5250\AA.  Therefore, in order to compute a rest-frame color, we define a modified B-band filter where we set the transmission redward of 4450\AA~equal to zero.  This synthetic filter, which hereafter we refer to as "B$_{short}$" allows us to compute a U - B color for all but the highest redshift galaxies in the sample, but still retains the general characteristics of the standard B-band.
\newline\indent
We tested the effect of shortening the B-band on galaxy colors by comparing the U - B colors of synthetic spectra from the BC03 models with a range of colors to the U - B$_{short}$ colors for the same models.  The transformation changes as a function of galaxy color, but for galaxies with U - B$_{short}$ $>$ 0.6, the majority of objects in the GCLASS sample, the transformation is U - B $\sim$ U - B$_{short}$ + 0.15.  
\begin{figure*}
\plottwo{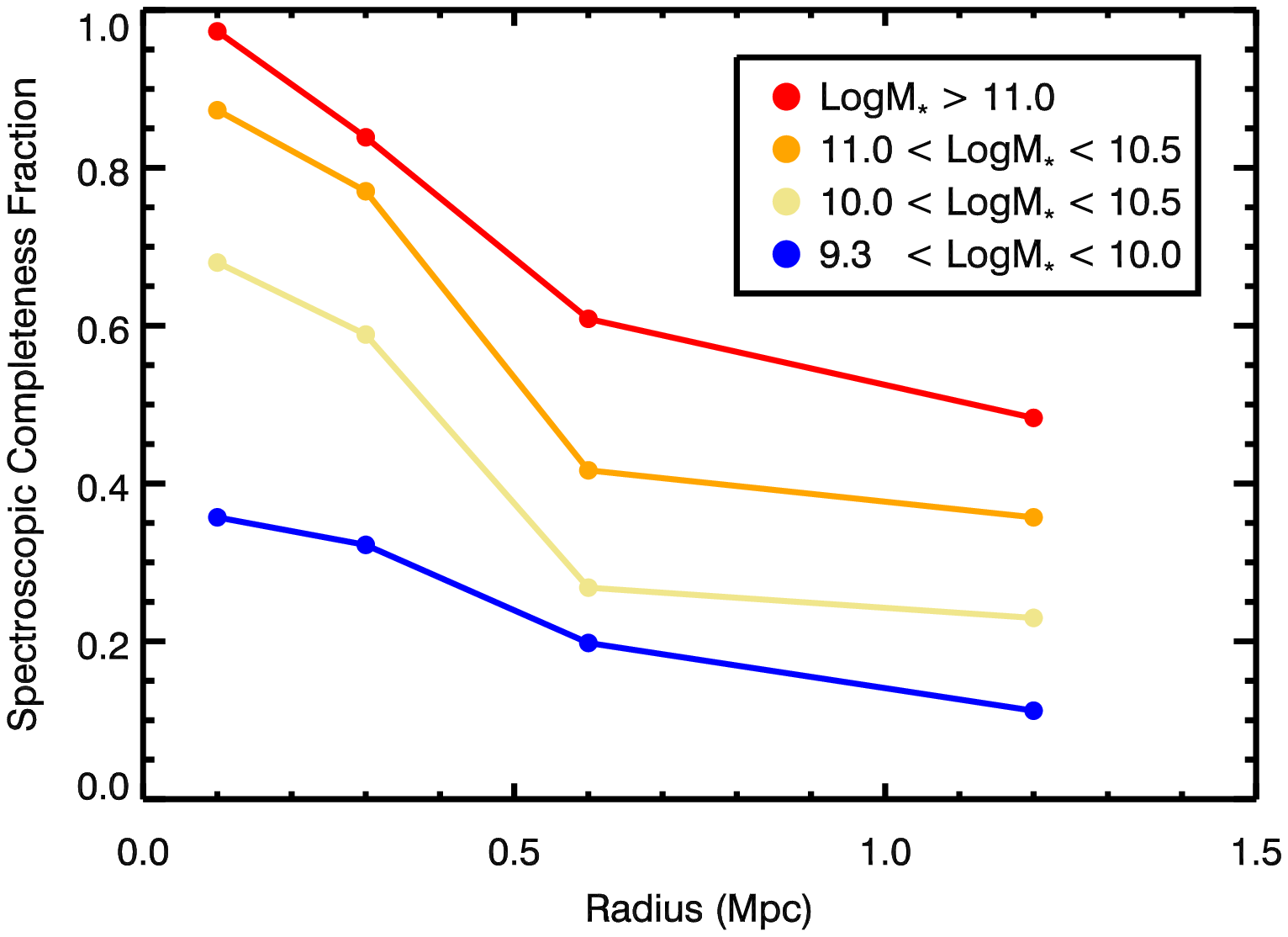}{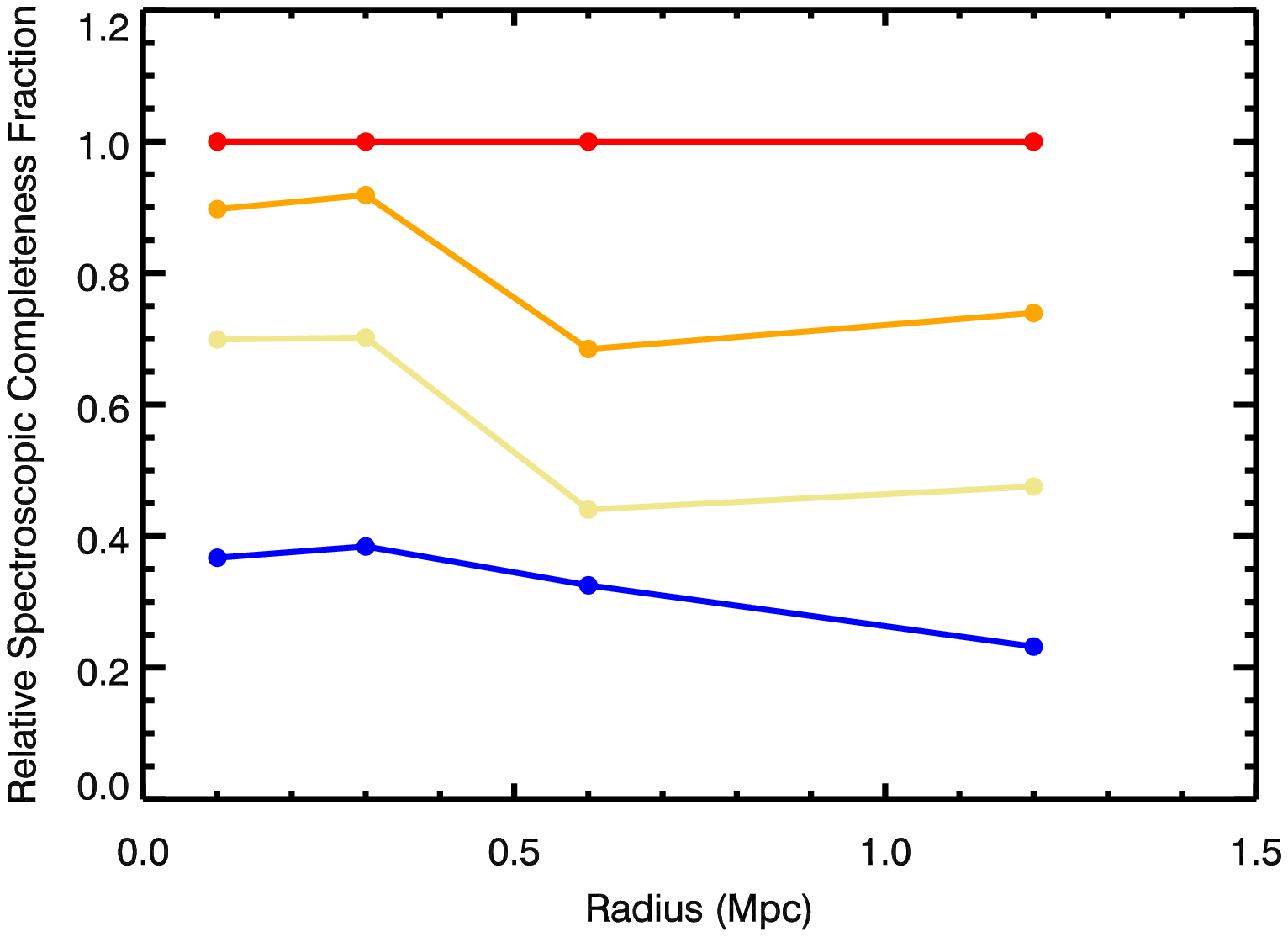}
\caption{\footnotesize Left Panel: Spectroscopic completeness fraction of cluster galaxies as a function of clustercentric radius and stellar mass.  The completeness is calculated based on the color and magnitude space occupied by cluster members in each cluster, and the relative fraction of galaxies in the color space that have redshifts (see text).  Spectroscopic completeness is highest for massive galaxies and galaxies in the cluster core.   Right panel:  Spectroscopic completeness as a function of radius and stellar mass normalized to the completeness for massive galaxies.  This shows that the primary completeness bias is stellar mass; however, this bias is similar at all radii.}
\end{figure*}
\subsection{Spectroscopic Completeness}
The completeness of a spectroscopic sample depends both on the sampling frequency of the targets, as well as redshift success rate.  In turn, both of these depend on observable quantities of the galaxies such as flux, color, position, redshift, and the presence/lack of observable emission lines.  The sampling frequency depends on observables because galaxies are prioritized in the mask design process based on their properties.  The redshift success rate depends on observables, particularly flux and redshift, because these quantities increase or decrease the confidence of the redshift determination.  Brighter targets provide higher S/N spectra, and galaxies at particular redshifts have their prominent spectral features within the observed bandpass.  If the primary source of incompleteness is the sampling frequency, not the redshift success rate, it is straightforward to compute spectroscopic completeness corrections.  This condition can usually be met by limiting the overall sample to within a flux and redshift limit where the redshift success rate is high.
\newline\indent
Due to the relatively deep exposures and good N\&S sky subtraction the redshift success rate for all slits, regardless of brightness and redshift is quite good (71\%).  In particular, the success rate for the primary priorities, P1 - P10, ranged between a minimum of 70\% (P7) to a maximum of 96\% (P1).  Even the bright filler slits (P11) had a good redshift success rate of 69\%.  The majority of slits where no redshift could be obtained were the faint filler slits (P12) which had a redshift success rate of only 21\%.  The majority of the P12 objects fall below the flux/stellar mass limit that we have adopted for the sample (LogM$_{*}$/M$_{\odot}$ $>$ 9.3, see below), and only 1 was a cluster member.  We examined the spectra of P1 - P10 galaxies in order to ascertain why the redshift success rate was less than 100\%.  For many P1 - P10 galaxies where no redshift was determined the spectrum did have clearly detected continuum emission; however, there were no identifiable spectral features.  It is likely that most of these targets are either galaxies at $z <$ 0.5 where the band-limiting filter at 6100\AA~makes it difficult to identify features, or galaxies at $z >$ 1.5 where there are few strong spectral features in the observed bandpass.  Therefore, provided we limit the sample to the redshift range where the spectral features are in the observed bandpass (0.85 $< z <$ 1.20), and to galaxies above the flux/stellar mass limit required for a successful redshift (LogM$_{*}$/M$_{\odot}$ $>$ 9.3), the completeness corrections can be calculated under the assumption that the sampling frequency is the dominant source of incompleteness.
\newline\indent
Despite the fact that the completeness depends on observable quantities, the spectroscopic completeness corrections need to be calculated as a function of the physical quantities that are used to define the galaxy sample.  For the current study, where we examine the properties of galaxies as a function of stellar mass and environment, we calculate completeness correction in terms of those two quantities.  
\newline\indent
As was shown in $\S$ 4.3, the stellar mass of galaxies is closely related to their 3.6$\micron$ flux, the primary selection criteria for spectroscopic targets.  In principle, this means that defining the completeness above a stellar mass limit is nearly analogous to defining it above a  3.6$\micron$ flux limit. 
\newline\indent
For each cluster, the quantity of interest is the spectroscopic completeness for cluster galaxies, not all galaxies in the field of view.  This means we are not interested in the completeness of a flux-limited sample, but a flux-limited sample within a particular region of color-magnitude space, i.e., the color-magnitude space occupied by $z \sim$ 1 cluster members.  This region of color-magnitude space will be different from cluster-to-cluster because they are at a range of redshifts and the k-correction affects the color of the observed red-sequence and blue cloud.  
\newline\indent
The location of the cluster in color-magnitude space is determined by examining the location of all confirmed cluster members in the z$^{\prime}$ - 3.6$\micron$ vs. 3.6$\micron$ color-magnitude diagram.  We then define the color-magnitude space inhabited by cluster galaxies using several adaptive-size boxes that are adjusted to cover the color space occupied by all cluster members.  The completeness for cluster members as a function of 3.6$\micron$ flux within that region of color space is then defined as 
\begin{equation}
f_{cluster,z} = \frac{N_{cluster,z}}{N_{gal} (N_{cluster,z}/(N_{field,z}+N_{cluster,z}))},
\end{equation}
where f$_{cluster,z}$ is the fraction of cluster members in that flux range that have a spectroscopic redshift, N$_{cluster,z}$ is the total number of cluster members in the color-magnitude space with a redshift, N$_{field,z}$ is the total number of field galaxies in the color-magnitude space with a spectroscopic redshift, and N$_{gal}$ is the total number of galaxies with or without a redshift in the color-magnitude space.
\newline\indent
Once the completeness in this color space is computed, we convert this to a completeness as a function of stellar mass using a fit to the correlation between 3.6$\micron$ and stellar mass.  The correlation between these two parameters is linear in log space and has an rms scatter of 0.15 dex.  For the final completeness calculations we combine the data from all 9 clusters into an ensemble cluster and consider the completeness in several bins of stellar mass, and clustercentric radius.  The completeness curves for galaxies of different stellar masses are plotted as a function of radius in the left panels of Figure 4.  
\newline\indent
Figure 4 shows clearly that the spectroscopic completeness is a function of both stellar mass, and clustercentric radius.  The completeness is highest for massive galaxies in the cluster cores, and lowest for lower-mass galaxies in the cluster outskirts.  Both of these incompleteness effects are straightforward to understand. 
\newline\indent
The increasing completeness with increasing stellar mass is primarily a product of the 3.6$\micron$ flux prioritization and the fact that bright galaxies have the highest S/N spectra, hence obtaining redshifts is easiest.  The decreasing completeness with increasing clustercentric radius is primarily an area effect.   Due to the microslits available with the band shuffle spectroscopy mode, the $total$ number of redshifts as a function of clustercentric radius is roughly constant; however, the area grows as $\pi$r$^2$, and hence obtaining redshifts for the same fraction of galaxies becomes increasing difficult at larger radii.  
\newline\indent
In the right panel of Figure 4 we plot the spectroscopic completeness of galaxies as function of clustercentric radius, scaled to the same completeness as a function of clustercentric radius for the most massive galaxies.  These curves are not perfectly flat but have only mild slopes, demonstrating that the main incompleteness in GCLASS is the stellar mass incompleteness, and that the stellar mass incompleteness is nearly independent of environment.  
\newline\indent
The main focus of this paper is to examine the properties of galaxies at fixed stellar mass, and fixed environment (see $\S$ 6).  As the right panel of Figure 4 shows, the sampling for galaxies of a given stellar mass relative to other stellar masses is similar in all environments.  Therefore, completeness corrections are not required in that analysis.  In $\S$ 5, where we examine the properties of galaxies for a mass-limited sample, we do employ the completeness corrections as a function of stellar mass and environment to correct for those biases.
\section{Results}
Most of the analysis presented in this paper is based on measurements made from stacking the spectra.  The main advantage of stacking is that it increases the S/N which improves the precision of the measurement of the mean properties of galaxies.   Of course, the tradeoff in better measurements of the mean is that information about the distribution in properties is lost.  Given that there is a large range in S/N for the spectra, the distribution of properties is not always easy to interpret, and therefore we have concentrated on robust measurements of the mean properties of the galaxies.  
\newline\indent
When stacking, each spectrum is normalized using the median flux in the range 4050\AA$ $-4100\AA, rest-frame.  This part of the spectrum is redward of the 4000\AA~break, but does not contain substantial absorption or emission lines.  After normalization, each galaxy is weighted by its position and stellar mass completeness weight (see $\S$ 4.6).  The rest-frame wavelength coverage of the spectra depends on the redshift of the galaxy, and therefore additional weighting is performed based on how many galaxies contribute to the stack on the red and blue ends.  Once all weighting has been performed, the individual spectra are combined into a mean stacked spectrum.  
\newline\indent
Throughout the analysis, the uncertainties in properties measured from the spectral stacks (e.g., SFR and D$_{n}$(4000)) are determined using 300 bootstrap resamplings of the data.  Many of the stacks contain $>$ 40 galaxies and hence the random errors in most properties are small, typically $<$ 1\%.  In these cases systematic errors are almost certainly the dominant source of error.  The systematics of greatest concern are the models employed in the determination of physical parameters, such as the conversion of [OII] line fluxes to SFRs, or the determination of stellar masses from synthetic BC03 spectra.  At present, these types of systematics are difficult to quantify, and indeed are probably the dominant source of error in most studies that determine physical properties of galaxies from observables.  Therefore,  throughout this paper the quoted errors are always the bootstrap random errors, but we note that systematic errors are likely to be substantially larger and advise caution if comparing quantities such as SFRs and stellar masses determined in this paper to quantities determined in other papers using different model assumptions.
\begin{figure*}
\plotone{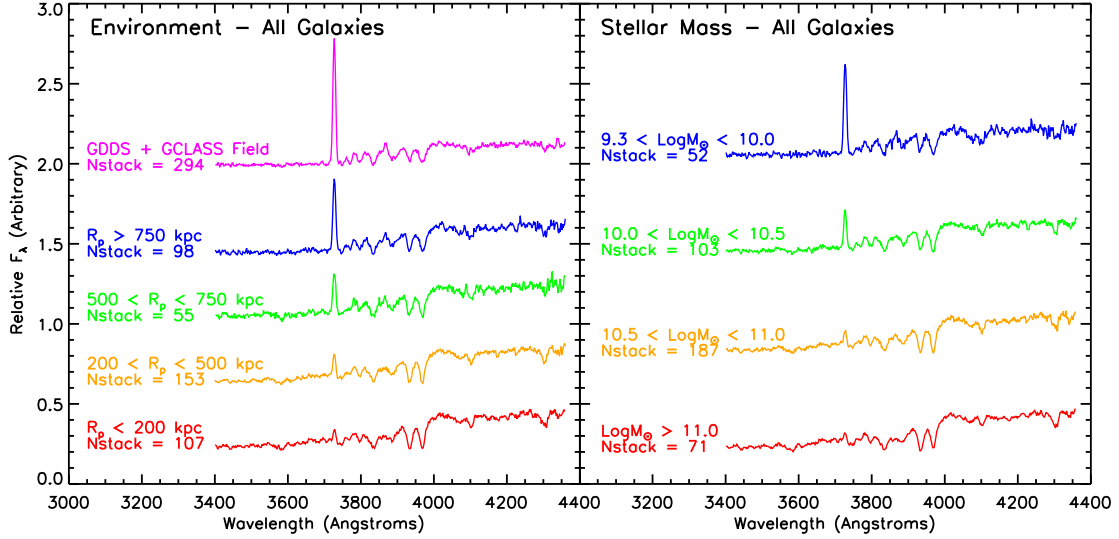}
\caption{\footnotesize Left Panel: Completeness-corrected mean stacked spectra of cluster galaxies with LogM$_{*}$/M$_{\odot}$ $>$ 9.3 as a function of clustercentric radius.  Field galaxies are taken from both GCLASS and the GDDS (see text).  There is a clear change in the spectral properties of galaxies with clustercentric radius, moving from 4000\AA~break dominated systems with a weak [OII] 3727\AA~in the core to Balmer-break dominated systems with a strong [OII] 3727\AA~in the field.  Right Panel: Completeness-corrected mean stacked spectra of cluster galaxies (R $<$ 1.5 Mpc) as a function of stellar mass.  The change in properties with increasing stellar mass appears to mirror the change in properties with increasing environmental density.}
\end{figure*}
\begin{deluxetable*}{ccccccr}
\tabletypesize{\scriptsize}
\scriptsize
\tablecolumns{7}
\tablecaption{Galaxy Parameters for the Mass-limited Sample as a Function of Environment}
\tablewidth{5.3in}
\tablehead{\colhead{Radius} & \colhead{ f$_{SF}$ } &
\colhead{Log(SSFR)} & \colhead{Log(SFR)} & \colhead{D$_{n}$(4000)} & \colhead{U - B$_{short}$} &
\colhead{Nstack} \\
\colhead{Mpc} & \colhead{} & \colhead{yr$^{-1}$} &
\colhead{M$_{\odot}$ yr$^{-1}$} & \colhead{} & \colhead{} & \colhead{} \\
\colhead{(1)}& \colhead{(2)}& \colhead{(3)}& \colhead{(4)}&
\colhead{(5)}& \colhead{(6)} & \colhead{(7)} 
}
\startdata
0.10 & 0.22$^{+0.05}_{-0.05}$ & -9.87$^{+0.08}_{-0.10}$ & 0.91$^{+0.08}_{-0.10}$ & 1.606$^{+0.019}_{-0.019}$ & 1.019$^{+0.016}_{-0.016}$ & 107  \nl
0.35 & 0.25$^{+0.04}_{-0.04}$ & -9.72$^{+0.06}_{-0.07}$ & 0.66$^{+0.06}_{-0.07}$ & 1.520$^{+0.008}_{-0.008}$ & 0.980$^{+0.010}_{-0.010}$ & 153  \nl
0.65 & 0.46$^{+0.11}_{-0.11}$ & -9.20$^{+0.09}_{-0.12}$ & 1.07$^{+0.15}_{-0.24}$ & 1.417$^{+0.019}_{-0.019}$ & 0.888$^{+0.019}_{-0.019}$ & 55 \nl
1.00 & 0.67$^{+0.10}_{-0.10}$ & -8.85$^{+0.07}_{-0.08}$ & 1.49$^{+0.07}_{-0.08}$ & 1.355$^{+0.017}_{-0.017}$ & 0.780$^{+0.018}_{-0.018}$ & 98 \nl
Field & 0.82$^{+0.07}_{-0.07}$ & -8.54$^{+0.01}_{-0.01}$ & 1.58$^{+0.01}_{-0.01}$ & 1.213$^{+0.003}_{-0.003}$ & 0.641$^{+0.007}_{-0.007}$ & 294 \nl
\enddata

\end{deluxetable*}
\begin{deluxetable*}{ccccccr}
\tabletypesize{\scriptsize}
\scriptsize
\tablecolumns{7}
\tablecaption{Galaxy Parameters for the Mass-limited Sample as a Function of Stellar Mass}
\tablewidth{5.3in}
\tablehead{\colhead{Log(Mass)} & \colhead{ f$_{SF}$ } &
\colhead{Log(SSFR)} & \colhead{Log(SFR)} & \colhead{D$_{n}$(4000)} & \colhead{U - B$_{short}$} &
\colhead{Nstack} \\
\colhead{M$_{\odot}$} & \colhead{} & \colhead{yr$^{-1}$} &
\colhead{M$_{\odot}$ yr$^{-1}$} & \colhead{} & \colhead{} & \colhead{} \\
\colhead{(1)}& \colhead{(2)}& \colhead{(3)}& \colhead{(4)}&
\colhead{(5)}& \colhead{(6)} & \colhead{(7)} 
}
\startdata
\cutinhead{Cluster}
11.2 & 0.27$^{+0.07}_{-0.07}$ & -10.0$^{+0.08}_{-0.10}$ & 1.15$^{+0.10}_{-0.13}$ & 1.767$^{+0.015}_{-0.015}$ & 1.043$^{+0.011}_{-0.011}$ & 71 \nl
10.7 & 0.25$^{+0.04}_{-0.04}$ & -9.70$^{+0.10}_{-0.14}$ & 1.04$^{+0.10}_{-0.14}$ & 1.554$^{+0.036}_{-0.036}$ & 0.980$^{+0.016}_{-0.016}$ & 187 \nl
10.3 & 0.47$^{+0.08}_{-0.08}$ & -9.20$^{+0.04}_{-0.04}$ & 1.09$^{+0.05}_{-0.05}$ & 1.419$^{+0.011}_{-0.011}$ & 0.874$^{+0.014}_{-0.014}$ & 103 \nl
9.7 & 0.56$^{+0.13}_{-0.13}$ & -8.98$^{+0.13}_{-0.18}$ & 0.66$^{+0.15}_{-0.24}$ & 1.251$^{+0.023}_{-0.023}$ & 0.670$^{+0.021}_{-0.021}$ &  52 \nl
\cutinhead{Field}
11.2 & 0.58$^{+0.19}_{-0.19}$ & -9.62$^{+0.13}_{-0.20}$ & 1.53$^{+0.13}_{-0.20}$ & 1.721$^{+0.069}_{-0.069}$ & 1.057$^{+0.040}_{-0.040}$ & 26 \nl
10.7 & 0.69$^{+0.12}_{-0.12}$ & -9.08$^{+0.10}_{-0.13}$ & 1.63$^{+0.09}_{-0.12}$ & 1.442$^{+0.021}_{-0.021}$ & 0.887$^{+0.020}_{-0.020}$ & 77 \nl
10.3 & 0.78$^{+0.09}_{-0.09}$ & -8.58$^{+0.06}_{-0.08}$ & 1.67$^{+0.06}_{-0.08}$ & 1.236$^{+0.012}_{-0.012}$ & 0.706$^{+0.013}_{-0.013}$ & 82  \nl
9.7 & 0.96$^{+0.13}_{-0.13}$ & -8.51$^{+0.02}_{-0.02}$ & 1.17$^{+0.04}_{-0.04}$ & 1.138$^{+0.005}_{-0.005}$ & 0.534$^{+0.010}_{-0.010}$ & 109 \nl
\enddata
\end{deluxetable*}

\subsection{Star Formation Rates as a Function of Environment and Stellar Mass}
Here we investigate the relationship between the SFRs and SSFRs of galaxies with their environment and stellar mass.  In the left panel of Figure 5 we plot the completeness-corrected mean stacked spectra of galaxies with LogM$_{*}$/M$_{\odot}$ $>$ 9.3 as a function of increasing clustercentric radius culminating with the field sample.  
\newline\indent
Without making any quantitative measurements it is immediately clear from the left panel of Figure 5 that the average stellar populations of galaxies are dramatically different depending on whether they are located in the field, the cluster outskirts, or the cluster core.  Remarkably, it appears that the transition in stellar populations as a function of clustercentric radius is quite smooth and that there is no preferred location where the properties of galaxies change distinctly.
\newline\indent
Galaxies in the cluster cores (R $<$ 200 kpc) have clear 4000\AA~breaks and show the MgI (3830\AA) and G-band (4304\AA) absorption features.  Cluster core galaxies also exhibit weak [OII] emission.  All of these features are indications of an evolved stellar population with little ongoing star formation.  
\newline\indent
The galaxies in the field sample are considerably different.  They show the Balmer series absorption lines and have a Balmer-break rather than a 4000\AA~break.  They also show evidence of [NeIII] (3869\AA) emission and have strong [OII] emission, all of which are indications of a relatively young stellar population with ongoing star formation and possible AGN activity.  
\newline\indent
The SSFRs and SFRs of the stacks are determined from the measured EW([OII]) using the method described in $\S$ 4.4 and are plotted in the left panels of Figure 6, as well as listed in Tables 2 and 3.  There is a 1.33 $\pm$ 0.09 dex decline in the SSFR of galaxies as their environment changes from the field into the cores of rich clusters, making it clear that the average star formation properties of galaxies are strongly correlated with their environment at $z \sim$ 1.  Figure 6 shows that not only does the mean SSFR decrease with increasing galaxy density, but that the mean SFR of galaxies also declines from the field toward the cluster core.  This confirms that both the SSFR-density relation and the SFR-density relation are already in place at $z \sim$ 1, at least in the highest-density environments.  Interestingly, the SFR-density relation only declines by 0.67 $\pm$ 0.03 dex between the cluster core and field, not nearly as much as the SSFR-density relation.  This suggests that the mean stellar masses of cluster and field galaxies are not the same, and may drive some of these correlations, an issue we will return to in $\S$5.4.
\newline\indent
The trends in the SSFRs and SFRs with environment are quite clear; however, we know that these properties also correlate with galaxy stellar mass.  In the right panels of Figure 5 we plot the mean stacked spectra of the cluster galaxies in several bins of stellar mass.  Figure 5 shows that the same trends that exist between stellar populations and increasing environmental density also exist with increasing stellar mass.  More massive galaxies appear to be a more evolved population with little ongoing star formation, whereas lower-mass galaxies are younger with clear signs of active star formation.
\newline\indent
In the right panels of Figure 6 we plot the inferred SSFR and SFR from these stacks as a function of stellar mass.  For reference we also performed a stacking analysis of the field sample as a function of stellar mass and also show those SSFRs and SFRs in Figure 6.  As with environment, there is a clear trend of decreasing SSFR with increasing galaxy stellar mass both for galaxies in clusters as well as in the field.  Interestingly, the middle panel of Figure 6 shows that the mean SFR of the galaxies is a fairly weak function of stellar mass.  This independence of SFR on galaxy mass is also seen in the HiZELS narrow-band H$\alpha$ survey at $z \sim$ 0.84 \citep{Sobral2011}, and is most likely a co-incidence of the overall process of galaxy quenching as a function of stellar mass and redshift and the particular redshift range considered here.
\newline\indent
The results shown in Figure 6 are in good agreement with most of the previous work on $z \sim$ 1 clusters, but do not agree as well with the $z \sim$ 1 field galaxy studies.  The 1.33 $\pm$ 0.09 dex decline in SSFR seen with environment is similar to the study of \cite{Patel2009, Patel2011}, who measured the SSFRs and SFRs of galaxies around a massive galaxy cluster at $z = 0.83$ using low-resolution grism spectroscopy.  \cite{Patel2011} measured a 0.85 dex decline in the SSFR for all galaxies in their mass-limited sample (LogM$_{*}$/M$_{\odot}$ $>$ 10.25) from the low-density field to the core of their rich cluster.  
\newline\indent
The trend of decreasing SSFR with increasing galaxy density is qualitatively similar to the DEEP2 study of \cite{Cooper2008a}; however, quantitatively it is quite different.  \cite{Cooper2008a} measured only a 0.14 dex decline in SSFR between the highest and lowest density environments found in DEEP2 at $z \sim$ 1, a negligible environmental effect.  As a comparison we plot the mean field value from the DEEP2 study beside our field sample in Figure 6.  Considering the differences between GCLASS and DEEP2: 3.6um-spectroscopic selection vs. R-band, and a different [OII] extinction correction, these values are in reasonably good agreement.  This suggests that the discrepancy between our studies is not from different methods of computing SSFRs.  It may be that the effects of environment on the galaxy population at $z \sim$ 1 are relatively minor in all environments except for the richest galaxy clusters, something that has been suggested by other authors \citep[e.g.,][]{Sobral2011}.
\subsection{The Fraction of Star-Forming and Non-Star-Forming Galaxies as a Function of Environment and Stellar Mass}
In the bottom panels of Figure 6 we plot the fraction of galaxies with detectable [OII] emission as a function of environment and stellar mass, and hereafter refer to these galaxies as star-forming galaxies.  The star-forming galaxies have been identified by examination of individual object spectra.  We have tested our ability to identify [OII] emission at different S/N levels by degrading the S/N of the best spectra with [OII] emission.  These tests show that [OII] emission can be identified down to $\sim$ 1\AA~equivalent width for the highest S/N noise spectra, and down to $\sim$ 3\AA~equivalent width for the lowest S/N spectra.  For the typical galaxy in our sample an EW([OII]) of 3\AA~corresponds to a SSFR of $\sim$ 5 x 10$^{-11}$ yr$^{-1}$.  
Formally this means that some galaxies we consider quiescent may not be completely quenched; however, given that the SSFRs of galaxies decline quite sharply at $z <$ 1, galaxies with SSFRs this low are unlikely to add significantly more stellar mass through star formation in the future.  We tested if we may have misclassified a substantial population of star-forming galaxies with low SSFRs as quiescent by stacking the spectra of the quiescent galaxies.  We find no evidence for a detection of [OII] in the quiescent stacks at a level of $<$ 1\AA, suggesting that $>$ 90\% of the galaxies we classify as quiescent have SSFRs $<$ 5 x 10$^{-11}$ yr$^{-1}$.  
\newline\indent
Figure 6 shows there is a strong correlation between the fraction of star-forming galaxies with both environment and stellar mass.  The trends with environment are the strongest, with 82\% $\pm$ 7\% of field galaxies being star forming, but only 22\% $\pm$ 5\% of galaxies in the cores of clusters showing star formation.  The correlation between f$_{SF}$ and environment has been observed ubiquitously in the local universe \citep[e.g.,][]{Kauffmann2004, Balogh2004a, Peng2010} and our data confirms it is still in place at $z \sim$ 1.  Our f$_{SF}$ as a function of environment also agrees well with \cite{Patel2011}, who found 79\% $\pm$ 4\% in low density environments and 32\% $\pm$ 3\% in the cores of rich clusters.  The \cite{Patel2011} galaxies are classified based on their location in the rest-frame U-V vs. V-J color-color diagram, not [OII] emission, but the consistency between the measurements suggest both methods identify similar galaxies.
\newline\indent
There is also a change in the f$_{SF}$ as a function of galaxy stellar mass both for field and cluster galaxies.  For cluster galaxies, the percentage of galaxies in the highest stellar mass range (LogM$_{*}$/M$_{\odot}$ $>$ 11.0) that are star forming is 27\% $\pm$ 7\%, whereas in the lowest stellar mass range (9.3 $<$ LogM$_{*}$/M$_{\odot}$ $<$ 10.0), 56\% $\pm$ 13\% of the galaxies are star forming.  For field galaxies the corresponding numbers are 58\% $\pm$ 18\% and 96\% $\pm$ 13\%, respectively.  This shows that a significant portion of the correlation between both SSFR and SFR with stellar mass and environment must be driven by the changing fraction of star-forming galaxies with these parameters.  This is an important distinction that is different from an overall decline in the SSFRs and SFRs of the star-forming galaxies.  In $\S$ 6 we will examine the mean properties of star-forming galaxies alone to ascertain what portion of the decline in SSFR and SFR with environment and stellar mass are caused by the change in f$_{SF}$ compared to a decline in the SSFRs and SFRs of the star-forming galaxies.
\begin{figure*}
\plottwo{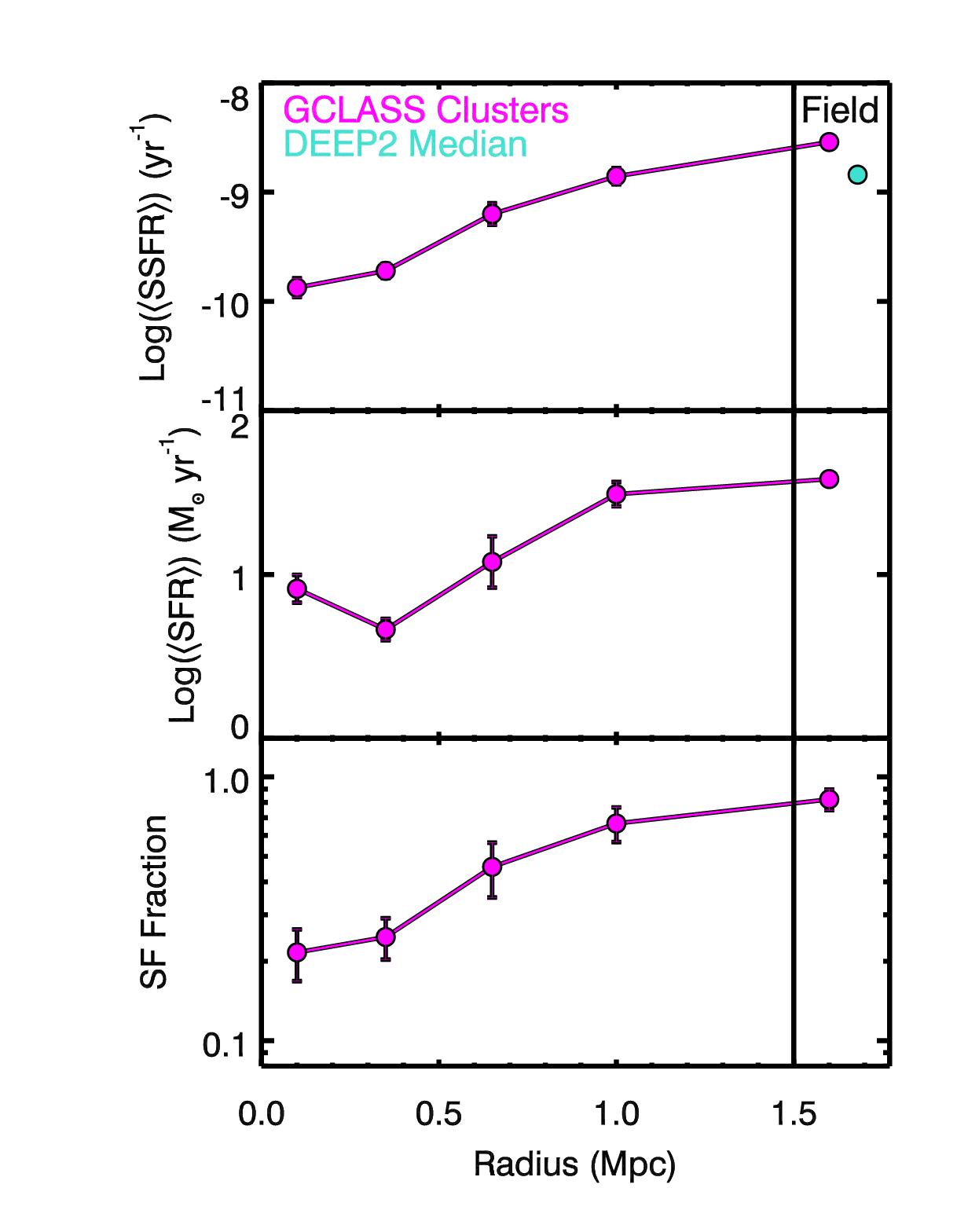}{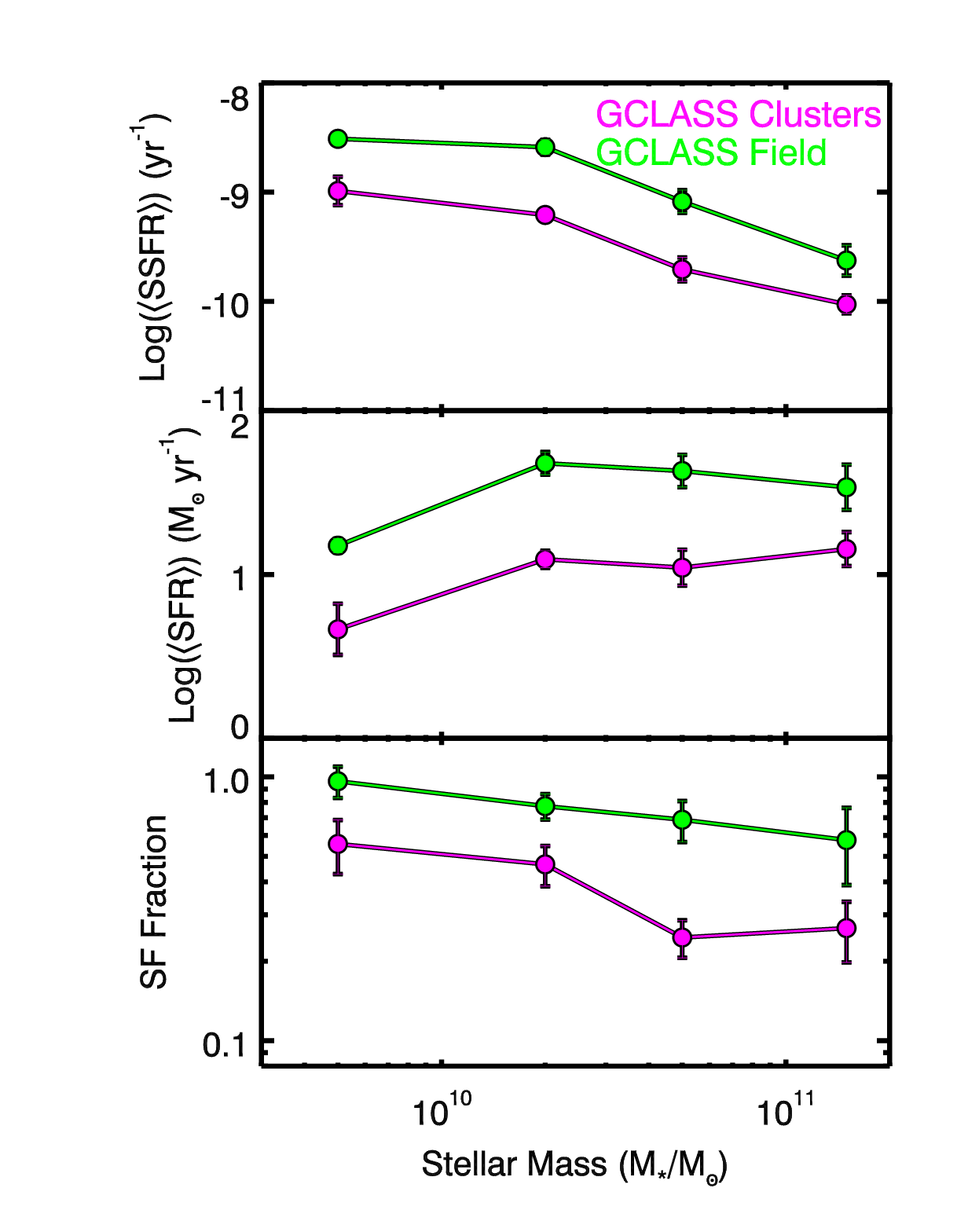}
\caption{\footnotesize Left Panels: Log(SSFR), Log(SFR), and f$_{SF}$ for galaxies with LogM$_{*}$/M$_{\odot}$ $>$ 9.3 as a function of clustercentric radius.  The SSFR and SFR are determined from the stacked spectra in Figure 5.  Error bars are calculated from 300 bootstrap resampling of the stacks but in same cases are smaller than the data points.  The median value of the DEEP2 field sample which has as a similar redshift to the GCLASS sample is shown for comparison.  At $z \sim$ 1 there is clearly a strong SSFR-density and f$_{SF}$-density relation in place.  There is also a SFR-density relation, but it is weaker than the other two correlations.  Right Panels: Same as left panels but plotted as a function of stellar mass both for cluster galaxies (magenta) and field galaxies (green).  The cluster galaxies display the typical "downsizing" trends with more massive galaxies being less frequently star forming and having lower SSFRs.}
\end{figure*}
\begin{figure*}
\plottwo{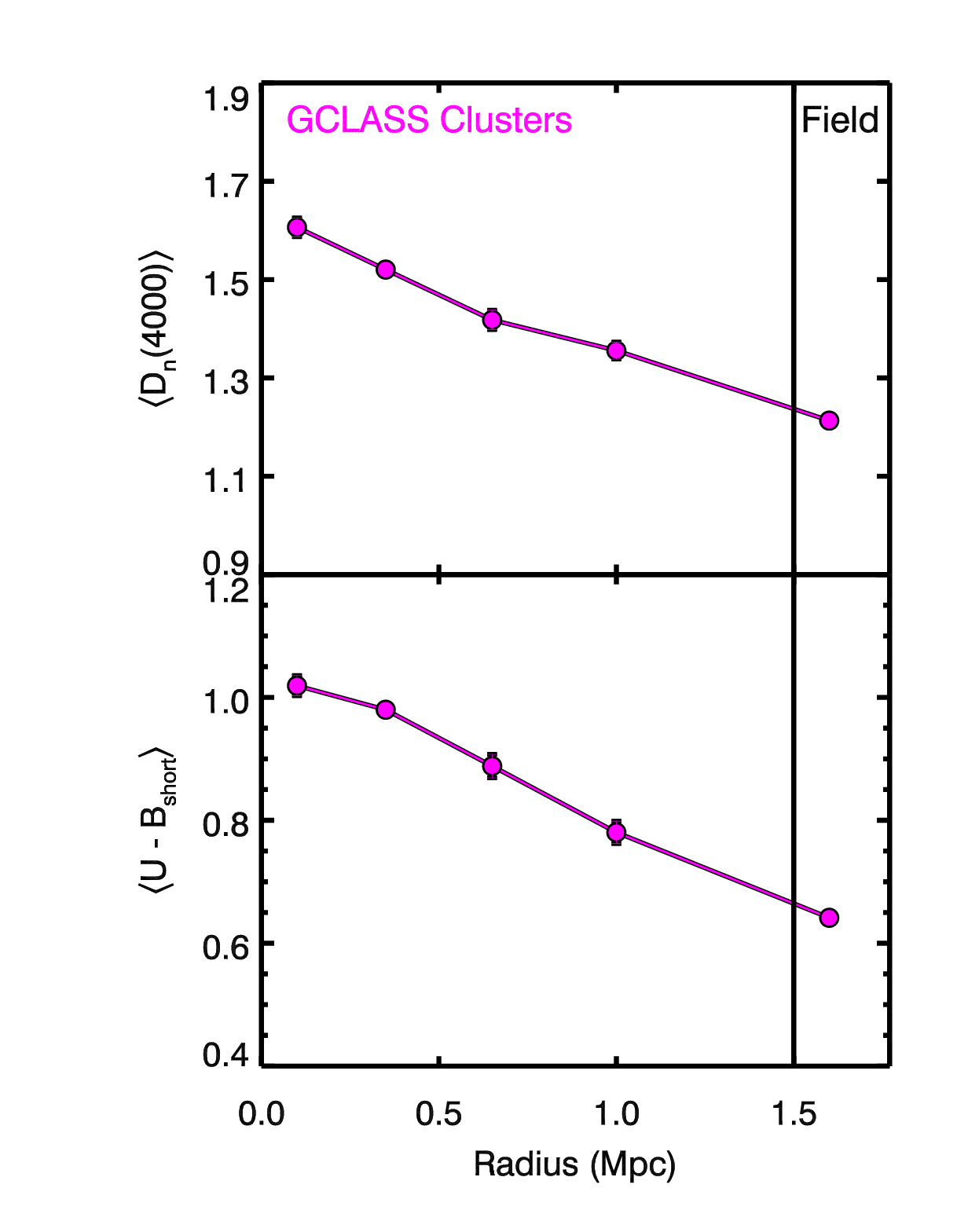}{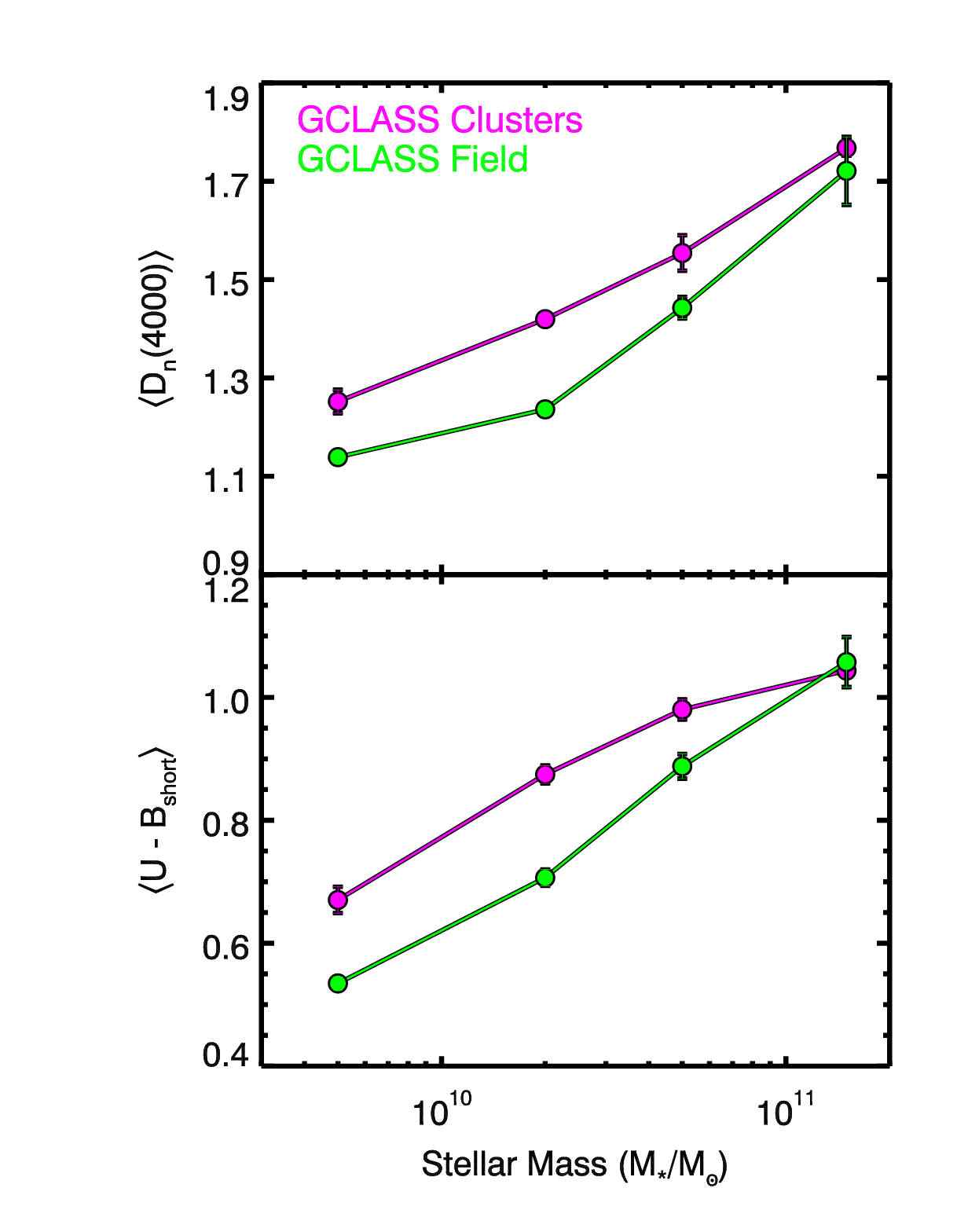}
\caption{\footnotesize Left Panels:  D$_{n}$(4000) and rest-frame U - B$_{short}$ color for galaxies with LogM$_{*}$/M$_{\odot}$ $>$ 9.3 as a function of clustercentric radius.  The D$_{n}$(4000) and U-B$_{short}$ color are determined from the stacked spectra in Figure 5.  Error bars are calculated from 300 bootstrap resampling of the stacks but in same cases are smaller than the data points.  There is a clear D$_{n}$(4000)-density and color-density relation in place at $z \sim$ 1.  Right Panels: Same as left panels but plotted as a function of stellar mass.  Similar to the local universe both parameters are a strong function of galaxy stellar mass. }
\end{figure*}
\subsection{D$_{n}$(4000) and U-B$_{short}$ colors as a function of Environment and Stellar Mass}
In the left panel of of Figure 7 we plot the measured D$_{n}$(4000)s and U-B$_{short}$ colors of the stacked spectra as a function of clustercentric radius.  There is a correlation of D$_{n}$(4000) with clustercentric radius, with D$_{n}$(4000) changing from a value of 1.213 $\pm$ 0.003 in the field to 1.606 $\pm$ 0.019 in the cluster cores.   The D$_{n}$(4000) is proportional to the luminosity-weighted age and metallicity of a galaxy, hence we can infer that the galaxies in the cluster cores have on average older and/or more metal-rich stellar populations than those in the field.  This inference is fully consistent with the spectral features observed in the stacked spectra (Figure 5), where field galaxies are dominated by Balmer lines, and cluster core galaxies have the metal lines commonly found in older stellar populations. 
\newline\indent
There is also clear evidence for a color-density relation, with galaxies in the cluster cores having U-B$_{short}$ colors 0.38 $\pm$ 0.02 mag redder than those in the field.   This is similar to the color-density relation seen at $z \sim$ 1 in DEEP2 by \cite{Cooper2006}.  They found that the typical U-B color of galaxies changes by $\sim$ 0.8 mag between the highest and lowest density environments.  It is surprising that the DEEP2 color-density relation is stronger than the color-density relation measured in the GCLASS clusters, but that the DEEP2 SSFR-density relation is substantially weaker than that measured in the GCLASS clusters ($\S$ 5.1).  Nonetheless, both studies do confirm a significant color-density relation at $z \sim$ 1. 
\newline\indent
In the right panel of Figure 7 we plot the measured D$_{n}$(4000) and U-B$_{short}$ color for galaxies in different stellar mass bins.  The trends in these parameters with stellar mass are remarkably similar to the trends with environment, and it is clear that there is also a D$_{n}$(4000)-stellar mass and color-stellar mass relation at $z \sim$ 1 as well.  
\subsection{The Dependence of Stellar Mass on Environment}
In the previous sections we have shown that the correlations between galaxy properties such as SSFR, SFR, D$_{n}$(4000), U-B$_{short}$, and f$_{SF}$ with both their environment and their stellar mass at $z \sim$ 1 are quite strong.  It also appears that the effect of increasing stellar mass on the properties of galaxies is identical to the effect from increasing environmental density.  Indeed, the correlations of all parameters with stellar mass and environment are even $quantitatively$ quite similar -- although it is important to note that this may be a coincidence of the particular range of environments and stellar masses considered in this study.  
\newline\indent
Nonetheless, from the analysis up to this point we can safely conclude that the properties of galaxies are strongly correlated with both their stellar mass and environment at $z \sim$ 1.  However, we cannot yet take the next step and conclude how $causal$ those correlations are.  If the stellar mass of galaxies and their environment are themselves correlated it may be that only one of them is causally implicated in galaxy evolution and that the correlation between the other parameter and galaxy properties is simply a result of a more fundamental stellar mass vs. environment correlation.   Indeed, a correlation between galaxy stellar mass and environment is observed by most studies of galaxies in the local \citep[e.g.,][]{Kauffmann2004, Baldry2006}, and high-redshift universe \citep[e.g.,][]{Bolzonella2010}.   
\newline\indent
In Figure 8 we plot the stellar mass of galaxies as a function of increasing clustercentric radius from the cluster cores to the field sample.  Points in grey are galaxies with spectroscopic redshifts, but that fall below the mass limit of LogM$_{*}$/M$_{\odot}$ $>$ 9.3, and are shown to illustrate the completeness.  Galaxies detected in the z$^{\prime}$-band but not at 3.6$\micron$ are plotted as arrows.  The vertical dashed line shows twice the FWHM of the IRAC PSF, and indicates where there may be incompleteness in lower-mass galaxies due to blending with the central galaxy.  The yellow line in Figure 8 shows the running mean determined by the 10 nearest galaxies, and does not include the grey points.  We emphasize that the yellow line represents the mean stellar mass of galaxies with LogM$_{*}$/M$_{\odot}$ $>$ 9.3, not the mean stellar mass of all galaxies in $z \sim$ 1 clusters.
\newline\indent
Figure 8 shows that the mean stellar mass of galaxies decreases by $\sim$ 1 dex from the cluster cores out to the field.  This decline is substantial; however, it is driven primarily by the extremely massive cluster central galaxies which have few counterparts at larger radii.  Ignoring the centrals and considering only satellites at R $>$ 0.1 Mpc shows that there is still a decrease in the mean stellar mass with environment, but it is much less, approximately a factor of 2.  This covariance between mean stellar mass and environment demonstrates that in order to understand which parameter is most important in shaping the properties of galaxies, we must look at environmental effects at fixed stellar mass, as well as stellar mass effects at fixed environment.  
\newline\indent
One concern with Figure 8 is that the stellar-mass incompleteness in the GCLASS spectroscopic sample may create a correlation between environment and stellar mass that does not exist.  As we showed in $\S$ 4.6, lower-mass galaxies are under-represented by a factor of $\sim$ 2-3 compared to the most massive galaxies.  Despite this, the large number of masks per cluster and ability to pack many slits in the cluster core with the band-shuffle mode means that for each cluster we are able to sample the same galaxies equally well in all parts of the cluster, and the stellar mass selection effects are basically independent of clustercentric radius (see Figure 4).  Indeed, if anything, lower-mass galaxies are slightly under-represented at large radius. This means that not only is the trend of decreasing galaxy stellar mass with increasing clustercentric radius seen in Figure 8 is not a selection effect, but it may be even more significant than we have inferred.  
\newline\indent
\begin{figure}
\plotone{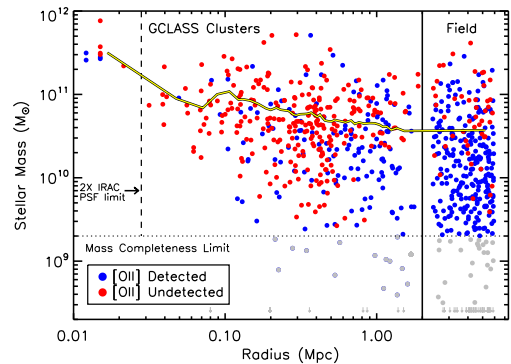}
\caption{\footnotesize  Stellar mass of galaxies as a function of clustercentric radius.  Points in grey are galaxies that fall below the LogM$_{*}$/M$_{\odot}$ $>$ 9.3 completeness limit.  The running mean in bins of 10 galaxies is shown as the yellow line.  Field galaxies are shown at the right and have been offset randomly along the x-axis for clarity.  Including the central galaxies, there is evidence for mass segregation in the clusters with central galaxies being an order of magnitude more massive than satellites on the outskirts.  Considering only the satellites, the mass segregation is weaker.  On average, satellites in the cluster core are twice as massive as satellites on the cluster outskirts.  In this plot the completeness is a strong function of both stellar mass and radius in that it is higher for massive galaxies and galaxies in the core; however, the relative completeness in stellar mass is nearly independent of radius (see $\S$ 4.6) so the mass segregation is a real and not a selection effect. }
\end{figure}
\section{Properties of Galaxies at Fixed Type, Fixed Environment, and Fixed Stellar Mass}
Given the covariance between galaxy stellar mass and environment, it is clear that in order to ascertain which has the largest causal effect on galaxy evolution, the properties of galaxies need to be measured as a function of environment at fixed stellar mass, and vice versa.  
\newline\indent
In addition to this, in $\S$ 5.2 we showed that the f$_{SF}$ was also a function of stellar mass and environment.  This means that in addition to any correlations induced from stellar-mass-environment covariance, it is possible that part (or all) of the trends of the mean {\it properties} of galaxies with environmental density (see Figures 6 and 7) may be caused only by the changing f$_{SF}$, and not by a change in the properties of the star-forming galaxies and quiescent galaxies themselves.  This is an important distinction, because if stellar mass or environment are causally linked to the quenching of star formation, we would expect to measure a change in the SSFRs of star-forming galaxies with those parameters, beyond that seen simply from the changing f$_{SF}$ with environment.  
\newline\indent
Therefore, in order to untangle the interdependency of stellar mass, environment and f$_{SF}$, in this section we first examine how the f$_{SF}$ depends on environment at fixed stellar mass, and vice versa.  Thereafter we separate star-forming and quiescent galaxies and measure how their SSFRs and D$_{n}$(4000)s vary with environment at fixed stellar mass, and vice versa, in order to see which of these leaves the largest imprint on their stellar populations.
\begin{deluxetable}{ccccr}
\tabletypesize{\scriptsize}
\scriptsize
\tablecolumns{5}
\tablecaption{Galaxy Parameters for Star Forming Galaxies as a Function of Mass and Environment}
\tablewidth{3.3in}
\tablehead{\colhead{Radius} & \colhead{ f$_{SF}$ } &
\colhead{Log(SSFR)} & \colhead{D$_{n}$(4000)} &
\colhead{Nstack} \\
\colhead{Mpc} & \colhead{} & \colhead{yr$^{-1}$} &
\colhead{} & \colhead{} \\
\colhead{(1)}& \colhead{(2)}& \colhead{(3)}& \colhead{(4)} &
\colhead{(5)} 
}
\startdata
\cutinhead{Log(M$_{*}$/M$_{\odot}$) $>$ 10.7}
0.1 & 0.20$^{+0.06}_{-0.06}$ & -9.24$^{+0.07}_{-0.09}$ & 1.624$^{+0.028}_{-0.028}$ & 14 \nl
0.5 & 0.18$^{+0.06}_{-0.06}$ & -9.12$^{+0.10}_{-0.14}$ & 1.542$^{+0.045}_{-0.045}$ & 10 \nl
1.0 & 0.37$^{+0.11}_{-0.11}$ & -9.24$^{+0.08}_{-0.10}$ & 1.624$^{+0.028}_{-0.028}$ & 15 \nl
Field & 0.63$^{+0.12}_{-0.12}$ & -9.20$^{+0.02}_{-0.02}$ & 1.519$^{+0.010}_{-0.010}$ & 43 \nl
\cutinhead{10.0 $<$ Log(M$_{*}$/M$_{\odot}$) $<$ 10.7}
0.1 & 0.19$^{+0.05}_{-0.05}$ & -8.90$^{+0.17}_{-0.31}$ & 1.256$^{+0.044}_{-0.044}$ & 6 \nl
0.5 & 0.26$^{+0.04}_{-0.04}$ & -8.72$^{+0.04}_{-0.04}$ & 1.332$^{+0.012}_{-0.012}$ & 32 \nl
1.0 & 0.54$^{+0.10}_{-0.10}$ & -8.82$^{+0.02}_{-0.02}$ & 1.327$^{+0.007}_{-0.007}$ & 34 \nl
Field & 0.85$^{+0.11}_{-0.11}$ & -8.87$^{+0.03}_{-0.03}$ & 1.219$^{+0.003}_{-0.003}$ & 99 \nl
\cutinhead{9.3 $<$ Log(M$_{*}$/M$_{\odot}$) $<$ 10.0}
0.1 & 0.60$^{+0.40}_{-0.44}$ & -8.69$^{+0.06}_{-0.07}$ & 1.117$^{+0.045}_{-0.045}$ & 3 \nl
0.5 & 0.37$^{+0.13}_{-0.13}$ & -8.46$^{+0.05}_{-0.06}$ & 1.138$^{+0.011}_{-0.011}$ & 11 \nl
1.0 & 0.88$^{+0.12}_{-0.31}$ & -8.68$^{+0.07}_{-0.08}$ & 1.209$^{+0.021}_{-0.021}$ & 15 \nl
Field & 0.96$^{+0.04}_{-0.13}$ & -8.47$^{+0.01}_{-0.01}$ & 1.137$^{+0.001}_{-0.001}$ & 105 \nl
\enddata
\end{deluxetable}
\begin{deluxetable}{ccr}
\tabletypesize{\scriptsize}
\scriptsize
\tablecolumns{3}
\tablecaption{Galaxy Parameters for Quiescent Galaxies as a Function of Mass and Environment}
\tablewidth{2.3in}
\tablehead{\colhead{Radius} &
\colhead{D$_{n}$(4000)} &
\colhead{Nstack} \\
\colhead{Mpc} &
\colhead{} & \colhead{} \\
\colhead{(1)}& \colhead{(2)}& \colhead{(3)} 
}
\startdata
\cutinhead{Log(M$_{*}$/M$_{\odot}$) $>$ 10.7}
0.1 & 1.794$^{+0.003}_{-0.003}$ & 57 \nl
0.5 & 1.779$^{+0.005}_{-0.005}$ & 50 \nl
1.0 & 1.725$^{+0.006}_{-0.006}$ & 27 \nl
Field & 1.734$^{+0.012}_{-0.012}$ & 25 \nl
\cutinhead{10.0 $<$ Log(M$_{*}$/M$_{\odot}$) $<$ 10.7}
0.1 & 1.575$^{+0.009}_{-0.009}$ & 25 \nl
0.5 & 1.598$^{+0.003}_{-0.003}$ & 74 \nl
1.0 & 1.628$^{+0.010}_{-0.010}$ & 15 \nl
Field & 1.579$^{+0.021}_{-0.021}$ & 18 \nl
\cutinhead{9.3 $<$ Log(M$_{*}$/M$_{\odot}$) $<$ 10.0}
0.1 & 1.489$^{+0.037}_{-0.037}$ & 3 \nl
0.5 & 1.443$^{+0.011}_{-0.011}$ & 19 \nl
1.0 & 1.326$^{+0.004}_{-0.004}$ & 3 \nl
Field & 1.234$^{+0.064}_{-0.064}$ & 4 \nl
\enddata
\end{deluxetable}
\begin{figure*}
\plotone{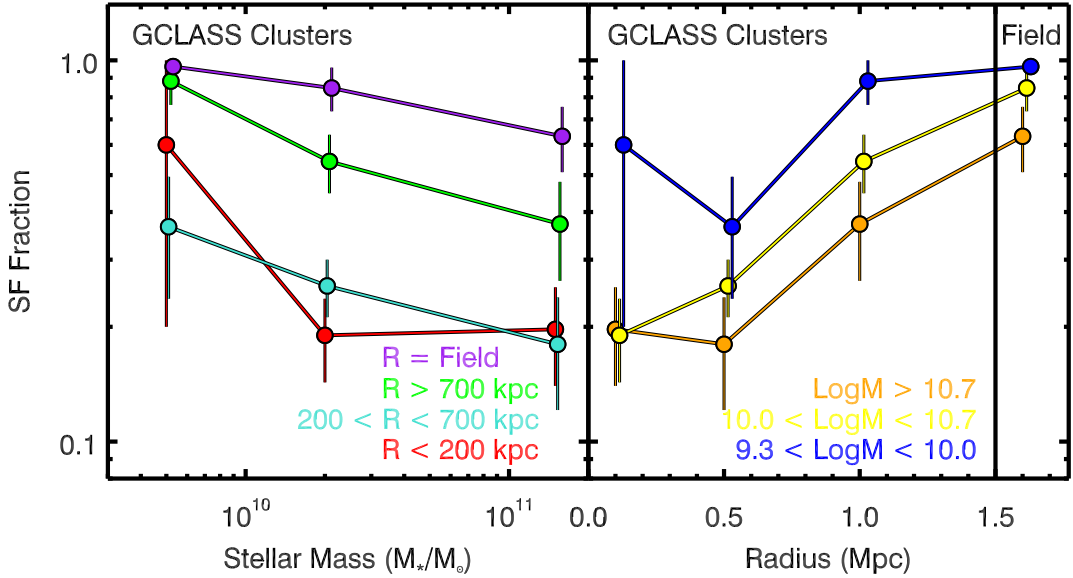}
\caption{\footnotesize Left Panel: The fraction of star-forming galaxies (f$_{SF}$) as a function of galaxy stellar mass for galaxies in different environments.  Right Panel: The f$_{SF}$ as a function of environment for galaxies with different stellar masses.  The fractional decline in f$_{SF}$ as a function of environment is a factor of $\sim$ 3 at all stellar masses.  Conversely, the fractional decline in f$_{SF}$ as a function of stellar mass is a factor of $\sim$ 2 in all environments.  This shows that the effects of stellar mass and environment on the f$_{SF}$ are independent and separable at $z \sim$ 1. }
\end{figure*}
\begin{figure*}
\plotone{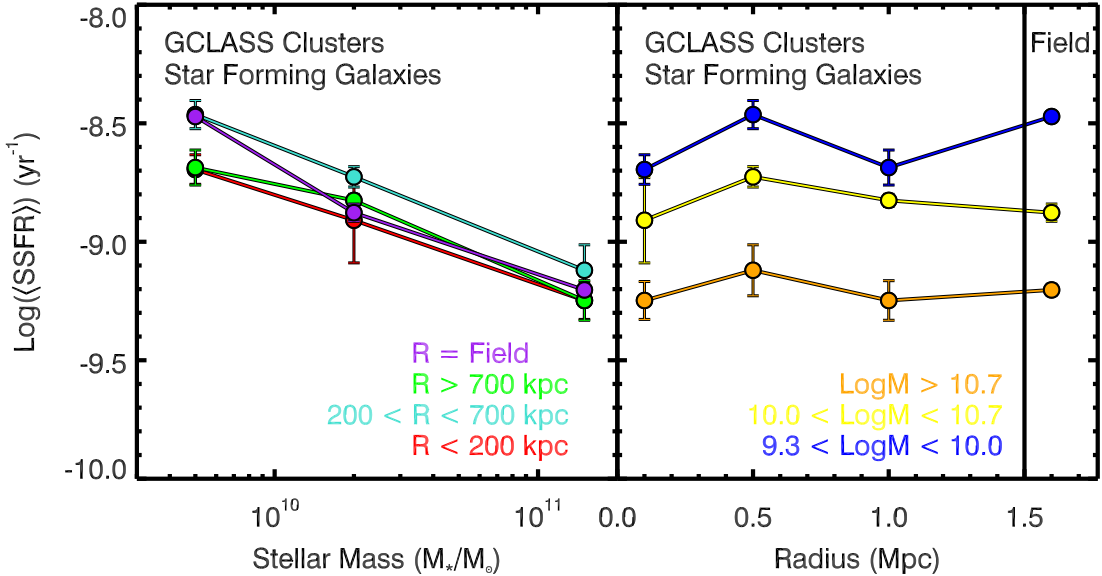}
\caption{\footnotesize Left Panel: Logarithm of the SSFR of star-forming galaxies as a function of stellar mass for galaxies in different environments.  Right Panel: Logarithm of the SSFR of star-forming galaxies as a function of environment for galaxies with different stellar masses.  The SSFR is correlated with stellar mass in all environments; however, but it is independent of environment at all stellar masses.  This suggests that the primary factor in determining the SFRs of star-forming galaxies is their stellar mass, not their environment.}
\end{figure*}
\begin{figure*}
\plotone{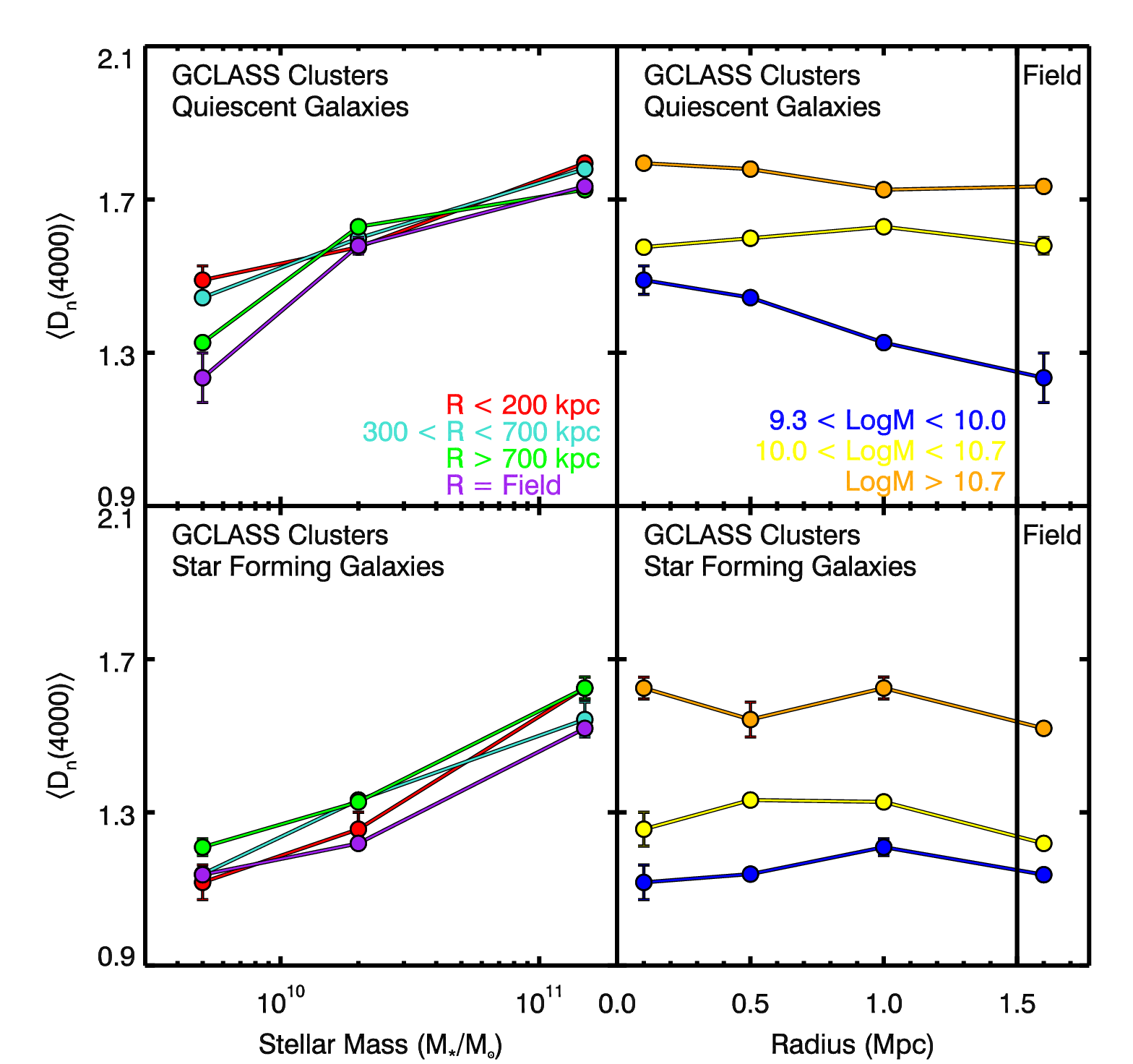}
\caption{\footnotesize Left Panels: D$_{n}$(4000) of quiescent and star-forming galaxies as a function of stellar mass for galaxies in different environments.  Right Panels: D$_{n}$(4000) of quiescent and star-forming galaxies as a function of environment for galaxies with different stellar masses.  The D$_{n}$(4000) of both quiescent and star-forming galaxies is a function of stellar mass in all environments; however, it is independent of environment at all stellar masses.  Given the correlation between D$_{n}$(4000) and the age of the stellar population, this suggests that stellar mass is the primary factor in determining the age and/or metallicity of a galaxy, not its environment.}
\end{figure*}
\subsection{The Effect of Environment and Stellar Mass on the Star-Forming Fraction}
In the left panel of Figure 9 we plot the f$_{SF}$ as a function of stellar mass in four environments: The field, R $>$ 700 kpc, 200 $<$ R $<$ 700 kpc, R $<$ 200 kpc.  In the right panel we plot the f$_{SF}$ as a function of clustercentric radius in three bins of stellar mass: LogM$_{*}$/M$_{\odot}$ $>$ 10.7, 10.0 $<$ LogM$_{*}$/M$_{\odot}$ $<$ 10.7, and 9.3 $<$ LogM$_{*}$/M$_{\odot}$ $<$ 10.0.  These are slightly larger mass and environmental bins than those used in $\S$ 5, but given the additional separation of star-forming and quiescent galaxies, these adjustments are necessary to ensure adequate numbers of both in all stellar mass and environmental bins.
\newline\indent
Figure 9 shows that even at fixed environment, the f$_{SF}$ is still correlated with stellar mass, with more massive galaxies being less likely to be star forming.  Interestingly, the $relative$ decline in the f$_{SF}$ as a function of increasing stellar mass is consistent with being the same in all environments, roughly a factor of $\sim$ 2-3 in all four environments.  
\newline\indent
Although the relative decline in f$_{SF}$ with increasing stellar mass occurs independently of environment, the absolute number of star-forming galaxies does depend on environment and this is demonstrated in the right panel of Figure 9 which shows the f$_{SF}$ as a function of environment in the three stellar mass bins.  The f$_{SF}$ is correlated with environment for galaxies over a wide range in stellar mass.  The relative decline in f$_{SF}$ from the field to the cluster core is a factor of $\sim$ 2-4, and within the uncertainties is consistent with being independent of stellar mass.  The trend in decreasing f$_{SF}$ with increasing environmental density appears to be slightly stronger than the trend with increasing stellar mass, suggesting environment may control this ratio slightly more than stellar mass; however, the uncertainties are still too large to determine if one is more dominant than the other.
\newline\indent
It is then quite clear from this analysis that {\it both environment and stellar mass independently play a causal role in determining the fraction of star-forming galaxies at z $\sim$ 1}.  It also appears that the quenching effects of stellar mass (self-quenching) and environment (environmental-quenching) on the f$_{SF}$ are separable.  Whatever process by which galaxies self-quench at a given stellar mass appears to be equally efficient regardless of which environment those galaxies are found in.  Likewise, whatever process by which environment quenches star formation in galaxies appears equally efficient regardless of the stellar mass of the system.
\newline\indent
The identical conclusion was reached in the recent study by \cite{Peng2010} who showed using both the SDSS at $z \sim$ 0, and zCOSMOS at $z \sim$ 0.6 that the relative increase in the quenched fraction of galaxies with increasing environmental density is independent of stellar mass, and vice versa.  Our data confirm that this separability of environment and stellar mass on the f$_{SF}$ continues to hold quite clearly up to $z \sim$ 1.  
\subsection{The Specific Star Formation Rates of Star-forming Galaxies at Fixed Environment and Fixed Stellar Mass}
Now that we have established that both environment and stellar mass play an important and causal role in the quenching of galaxies at $z \sim$ 1, we would like to understand the details of how those processes might work.   In this subsection we consider the properties of galaxies of each type individually -- either star-forming or quiescent -- at fixed environment, and at fixed stellar mass.  As in $\S$ 5, this analysis is based on measurements made from stacking galaxies within different stellar mass and environmental bins.
\newline\indent
In the left panel of Figure 10 we plot the SSFRs of the star-forming galaxies in different environments as a function of stellar mass and in the right panel we plot their SSFRs as a function of clustercentric radius in different stellar mass bins.  Figure 10 shows that the SSFR of star-forming galaxies is correlated with their stellar mass, but this correlation appears to be basically independent of the environment they live in.  Given the clear dependence of the f$_{SF}$ on both stellar mass and environment, this is a surprising result.  If both stellar mass and environment are responsible for quenching star formation it might be expected that the SSFRs of star-forming galaxies should decrease both as a function of environment and stellar mass, not just stellar mass alone.  
\newline\indent
A lack of dependence of the SSFR of star-forming galaxies on environment is also observed in the local universe.  \cite{Balogh2004b} found that in the 2dFGRS and SDSS the galaxy population is bi-modal in terms of the EW(H$\alpha$).  They found that galaxies with EW(H$\alpha$) $<$ 4\AA~dominate the high density regions, and galaxies with EW(H$\alpha$) $>$ 4\AA~dominate the low density regions.  Similar to our results at $z \sim$ 1, the fraction varies strongly with environment, but the mean EW(H$\alpha$), which is directly related to the SSFR, is nearly independent of environment.  Independently, both \cite{Kauffmann2004} and \cite{Peng2010} have found the same effect in the SDSS.  They use different definitions of star-forming and quiescent galaxies, as well as different measures of the SSFR, which suggests the result does not strongly depend on those choices. 
\newline\indent
The most likely, and perhaps only reasonable explanation for a lack of a dependence of the SSFR of star-forming galaxies on environment is that the timescale over which environment quenches star formation in galaxies is very rapid.  If so, environmental quenching will move galaxies out of the star-forming classification and into the quiescent classification before a drop in their SSFRs is measured.  Such a process may have a measurable effect on the D$_{n}$(4000) of quiescent galaxies as a function of environment as well as the number of poststarburst galaxies as function of environment, issues we investigate further in $\S$ 6.3 and $\S$ 7.
\subsection{The D$_{n}$(4000) of Star-Forming and Quiescent Galaxies at Fixed Environment and Fixed Stellar Mass}
In Figure 11 we plot the D$_{n}$(4000) measures from the stacked spectra as a function of stellar mass at fixed environment (left panels) and as a function of environment at fixed stellar mass (right panels).  We can now consider the properties of both the star-forming and quiescent galaxies, which are shown in the bottom and top panels of Figure 11, respectively.  We have also listed these parameters in Tables 4 and 5.  The trends in D$_{n}$(4000) with stellar mass and environment are qualitatively similar to the trends in the SSFR with these parameters.  For both the star-forming and quiescent galaxies the D$_{n}$(4000) measures are strongly correlated with stellar mass, regardless of environment, and the D$_{n}$(4000) appears to be independent of environment, regardless of the stellar mass.  The exception is the lowest-mass quiescent galaxies, which suggest a trend in D$_{n}$(4000) with environment.  This trend is quite interesting if real; however, low-mass quiescent galaxies are rare at $z \sim$ 1, and the statistics in these stacks are still somewhat limited.  Even though the bootstrap error bars are small, those bins contain between 3 - 15 galaxies, and more data are probably required to confirm this interesting potential trend.
\newline\indent
Together, Figures 10 and 11 comprise one of the main conclusions of this paper, namely that it appears that at $z \sim$ 1 {\it stellar mass is the parameter that is primarily responsible for determining the stellar populations of both star-forming and quiescent galaxies, not their environment}.  Instead, the main role of environment is that it determines the f$_{SF}$, and also at some level, the mean stellar mass of galaxies.  Once the covariance of f$_{SF}$ with environment is controlled for, as well as the covariance of stellar mass with environment, it appears that environment does not affect the mean properties of both star-forming and quiescent galaxies, only the relative fraction of each.
\newline\indent
In $\S$ 6.2 we suggested that the lack of a dependence of the SSFR of star-forming galaxies on environment could be explained by a rapid environmental quenching timescale that quickly transforms star-forming galaxies into quiescent galaxies.  If so, we would also expect that the D$_{n}$(4000) of the star-forming galaxies would be independent of environment, as is observed.  However, we might also expect that the D$_{n}$(4000) of the quiescent galaxies would correlate with environment, because a higher fraction of galaxies in high-density regions have been quenched by their environment, hence they may have different D$_{n}$(4000) at fixed stellar mass compared to lower-density environments.  Except for the lowest-mass bin, where the data are fairly sparse, we do not see evidence for such a dependence.  As we will show with some simple quenching models in $\S$ 8, the lack of a correlation between the D$_{n}$(4000) of quiescent galaxies and their environment is still possible under certain circumstances, even if there is a substantial amount of environmental quenching. 
\section{The Poststarburst Population}
\begin{figure*}
\plotone{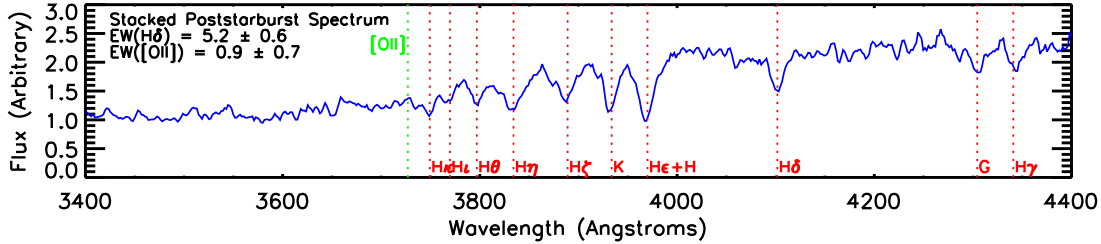}
\caption{\footnotesize Mean stacked spectrum of candidate poststarburst galaxies with 1.0 $<$ D$_{n}$(4000) $<$ 1.45 and no detectable [OII] emission.  The spectrum has been smoothed with a 3\AA~boxcar.  Prominent absorption(emission) features are marked in red(green). The stacked spectrum clearly shows the full Balmer series of absorption lines.  It has an EW(H$\delta$) = 5.2 $\pm$ 0.6, and EW([OII]) = 0.9 $\pm$ 0.7 and is classified as a K+A galaxy.  The stacked spectrum suggests that on average the [OII]- and D$_{n}$(4000)-based poststarburst selection technique selects similar galaxies as the more typical K+A selection.}
\end{figure*}
Our current hypothesis is that the timescale for environmental quenching is rapid enough to transform star-forming galaxies into quiescent galaxies before a decline in their SSFRs can be observed.  One way to test this is to try to identify the population of recently-quenched galaxies and see how it correlates with both environment and stellar mass.  Defining a complete sample of recently-quenched, or "poststarburst" galaxies is difficult because the quantitative description of this class of galaxies has varied within the literature.  Typically, when rest-frame optical spectroscopy is available, many authors have considered galaxies classified as "K+A galaxies" \citep[e.g.,][]{Balogh1999, Dressler1999, Poggianti1999, Leborgne2006, Poggianti2009a, Yan2009} as poststarburst galaxies or regular star-forming galaxies whose star formation was abruptly terminated before the epoch of observation.  The quantitative definition of K+A in itself has taken a range of values in different studies; however, in general it is the subset of galaxies with weak [OII] emission and strong H$\delta$ absorption-line strength.  The lack of [OII] emission is an indicator of a lack of ongoing star formation, and the strong H$\delta$ absorption is indicative of a population of A-stars which have lifetimes of $\sim$ 1 Gyr.  Therefore, K+A galaxies can be considered to have no ongoing star formation but a young stellar population less than 1 Gyr old.  
\newline\indent
The GCLASS spectra are of sufficient quality that the EW([OII]) can be measured in individual spectra as weak as 1-3\AA~(see $\S$ 5); however, H$\delta$ is a weak absorption feature that at $z \sim$ 1 is in a wavelength range contaminated with numerous sky lines, making it difficult to use the common K+A definition with these data.  Fortunately, at $z \sim$ 1 we can take advantage of the relatively young absolute age of most galaxies and the rapid non-linear evolution of D$_{n}$(4000) as a function of galaxy age.  As Figure 4 shows, the D$_{n}$(4000) of a single-burst population evolves from a value of D$_{n}$(4000) = 1.0 at zero-age to D$_{n}$(4000) = 1.5 at one Gyr after the initial burst.  Thereafter the strength of the break begins to evolve much more slowly with time.  
In principle, a weak D$_{n}$(4000) can be used as a proxy for a young stellar population, much in the same way that the K+A definition uses H$\delta$ as a proxy for a young stellar population.  
\newline\indent
\cite{Balogh1999} showed using synthetic spectra tracks that galaxies that have had their star formation abruptly truncated show a correlation between EW(H$\delta$) and D$_{n}$(4000) (their Figure 11).  They showed that for galaxies with EW(H$\delta$) $>$ 5\AA~(the typical definition for a K+A galaxy) the typical range of D$_{n}$(4000) values is 1.0 $<$ D$_{n}$(4000) $<$ 1.45.  Therefore, it might be expected that selecting galaxies with, 1) no detectable [OII] emission, and 2) 1.0 $<$ D$_{n}$(4000) $<$ 1.45, these may be analogous to the standard K+A selection with the caveat that this selection will miss older galaxies that experienced a rejuvenation period that was subsequently truncated.
\newline\indent
\begin{figure*}
\plotone{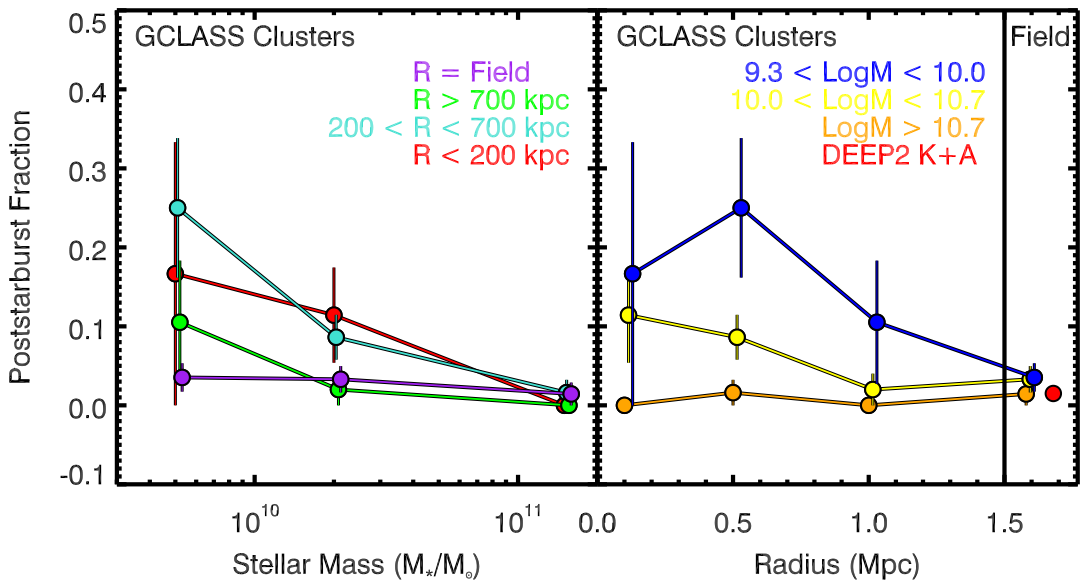}
\caption{\footnotesize Left Panel: Fraction of poststarburst galaxies as a function of stellar mass in different environments.  Right Panel: Fraction of poststarburst galaxies as a function of environment in different stellar mass bins.  The number of poststarbursts increases with decreasing clustercentric radius, except for the most massive galaxies, of which few are poststarbursts in any environment.  Likewise, poststarbursts appear to be rare ($<$ 2\% of the population) at all stellar masses in the field comparison sample.  The DEEP2 K+A selection from \cite{Yan2009} is shown for comparison and agrees well with our result.}
\end{figure*}
We tested what type of stellar population this selection criteria selects by stacking the spectra of all galaxies that met the criteria.  This stacked spectrum is plotted in Figure 12.  The spectrum is dominated by Balmer lines and is clearly similar to the spectra of K+A galaxies.  The stack has an EW(H$\delta$) = 5.2\AA~$\pm$ 0.6\AA, and an EW([OII]) = 0.9\AA~$\pm$ 0.7\AA.  Using the \cite{Balogh1999} definition of K+A galaxies: EW(H$\delta$) $>$ 5\AA, and EW([OII]) $<$ 5\AA, the mean galaxy selected with the D$_{n}$(4000)-based and EW([OII])-based criteria is in fact a K+A galaxy, suggesting this selection is also a good criteria for a poststarburst galaxy.  We note that this criteria almost certainly does not a select the complete sample of poststarbursts.  As has been discussed by \cite{Yan2009}, \cite{Lemaux2010} and \cite{Kocevski2010}, even the more rigorous K+A definition misses poststarburst galaxies with [OII] emission from an AGN, and can also misclassify star-forming galaxies as poststarbursts when the [OII] line is highly extinguished by dust.  In our analysis we are primarily interested in the environmental and stellar-mass dependence of the poststarburst population.  Even if the definition selects only a subset of all poststarbursts, provided it does so consistently the comparison of the poststarburst fraction at different stellar masses and environments should still be meaningful.
\newline\indent
With a clear definition of a poststarburst galaxy, we now plot the fraction of poststarburst galaxies as a function of stellar mass in different environments in the left panel of Figure 13.  Likewise, in the right panel we plot the fraction of poststarburst galaxies as a function of environment in several bins of stellar mass.  As a reference we denote the DEEP2 measurement of the K+A fraction at $z \sim$ 0.8 from \cite{Yan2009} as the red point.  This value is consistent with our field measurement, and further support the claim that our poststarburst selection criteria selects similar galaxies as the K+A criteria.
\newline\indent
Figure 13 shows that the fraction of poststarburst galaxies is simultaneously a function of stellar mass, and environment.  Less massive galaxies are more frequently poststarbursts compared to massive galaxies in all environments, and galaxies in the highest-density environments such as the cluster cores, are more frequently poststarbursts.  The latter statement is not true for the Log$M_{\odot}$ $>$ 10.7 galaxies, which have almost a zero poststarburst fraction in the clusters.
\newline\indent
It is interesting that poststarbursts are more common in the high-density cluster environment at $z \sim$ 1 compared to the field.  Taking an average over galaxies with 9.3 $<$ LogM$_{*}$/M$_{\odot}$ $<$ 10.7 shows that they are 3.1 $\pm$ 1.1 times more common in galaxy clusters compared to the field.   Not only is this excess of poststarburst galaxies in the high-density cluster environment more direct evidence for the importance of environmental quenching of star formation at $z \sim$ 1, it also shows that {\it environmental quenching is sufficiently abrupt to create the poststarburst signature in galaxies}.  
\newline\indent
A similar overdensity of poststarburst galaxies in and around clusters has also been seen by \cite{Poggianti2009a} in the EDisCS cluster sample at 0.4 $< z <$ 0.8.  They find that the K+A fraction is a factor of $\sim$ 2 higher in clusters compared to a control sample of field galaxies.  Likewise, \cite{Balogh2011} have also seen evidence for a large population of "green valley" galaxies in galaxy groups at $z \sim$ 1.  Taken together, these results all consistently point to high-density regions playing an important role in quenching galaxies up to $z \sim$ 1.  
\newline\indent
Indeed, given that self quenching should occur in all environments, if we take Figure 13 at face value it seems to imply that the primary cause of the poststarburst phase is environmental quenching.  Quantitatively the two match as well, as the both the fraction of quiescent galaxies and the poststarburst fraction increase by a factor of $\sim$ 3 as we move from the field to the cluster environment.  We have speculated that a rapid environmental-quenching timescale can explain the lack of a correlation between the SSFRs of star-forming galaxies and their environment, and the increasing fraction of poststarbursts with increasing environmental density lends further support to this hypothesis.  Our results also suggest that the environmental-quenching timescale may be more rapid than the stellar-mass quenching timescale.  The correlation between the SSFR of star-forming galaxies and their stellar mass ($\S$ 6.2) suggests that at some level, galaxies self-regulate their growth by winding down their SSFRs as they grow in stellar mass.  Environment appears to be a stochastic process that can quickly truncate the star formation in a galaxy once it has fallen into a sufficiently large overdensity.  
\newline\indent
Another interesting point from Figure 13 is that although poststarbursts are substantially more common in the cluster environment, they still make up a relatively small fraction of the overall population, even for lower-mass galaxies.  Galaxies that fit our poststarburst definition are 10\% $\pm$ 2\% of all cluster galaxies with 10.7 $<$ LogM$_{*}$/M$_{\odot}$ $<$ 9.3, and 0.5\% $\pm$ 0.5\% of cluster galaxies with LogM$>$ 11.0.  If we compare the number of poststarbursts to the number of quiescent galaxies in the cluster, and assume that environment is the primary source of the poststarburst population, it suggests that 16\% $\pm$ 3\% of quiescent galaxies with 9.3 $<$ LogM$_{*}$/M$_{\odot}$ $<$ 10.7 have had their star formation truncated by their environment within the last $\sim$ 1 Gyr.  It also suggests that very few of the quiescent galaxies with LogM$_{*}$/M$_{\odot}$ $>$ 10.7 have been truncated by their environment within the last Gyr.  This recently-quenched fraction is modest and suggests that even if environmental quenching dominates the growth of the low-mass end of the quiescent population, the quenching rate for such galaxies is still fairly low at $z \sim$ 1, and that the vast majority of quenched galaxies at that epoch were quenched at least $>$ 1 Gyr prior (i.e., $z >$ 1.3).  
\begin{figure*}
\plotone{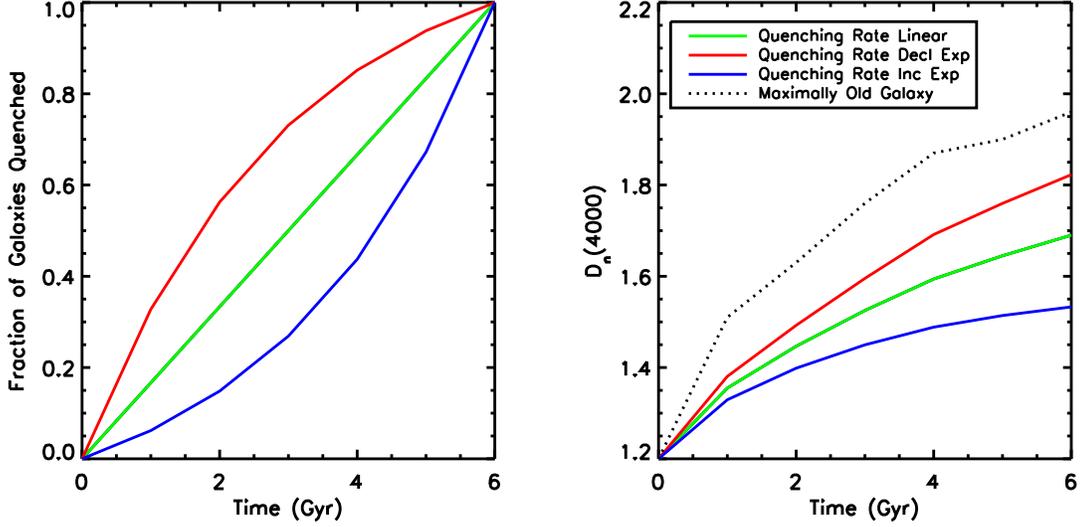}
\caption{\footnotesize Left Panel: Model tracks of the fraction of quenched galaxies as a function of time for simple analytic functions for the quenching rate: linear (green), exponentially declining (red), exponentially increasing (blue).  Right Panel:  Evolution of the average D$_{n}$(4000) for the quiescent galaxy population using the three quenching rate functions.  A galaxy that is quenched at t = 0 and hence is maximally old is shown for reference.  The exponentially declining quenching rate has the strongest average D$_{n}$(4000) because more galaxies quench early and hence have more time to evolve passively.  An exponentially increasing quenching rate has the weakest average D$_{n}$(4000) because the population of quiescent galaxies is dominated by galaxies that were recently quenched.}
\end{figure*}
\begin{figure*}
\plotone{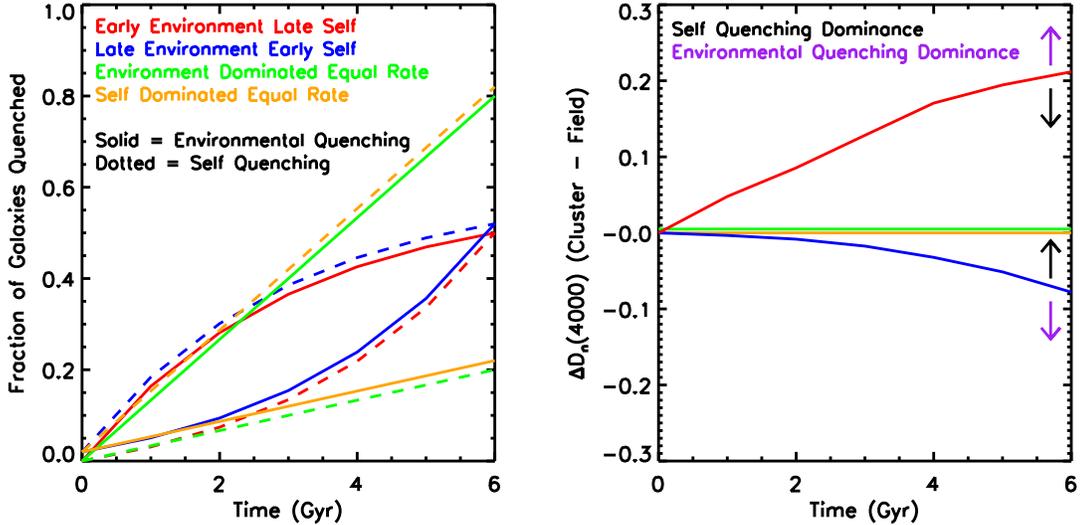}
\caption{\footnotesize Left Panel: Model tracks for the fraction of quenched galaxies as a function of time using the analytic quenching functions in Figure 14.  These models include separate mass-quenching components (dotted lines), and a environmental-quenching components (solid lines).  The red and blue models have the same amplitude for mass and environmental quenching but the quenching rates evolve differently with time.  The orange and green models have different amplitudes for the mass and environmental quenching components, but the rates evolve the identically with time.  Right Panel:  Relative difference in the inferred D$_{n}$(4000) for the model that has both environmental and mass quenching as in the left panel (simulating a cluster) compared to a model that has just mass-quenching (simulating the field).  If the rate of mass and environmental quenching evolves identically, there is no difference in the D$_{n}$(4000) between the cluster and field, even if the amplitudes of each are different.  The model where environmental quenching dominates at early times (red) has a larger D$_{n}$(4000) in the cluster compared to the field.  When environmental quenching dominates at late times (blue), the quiescent galaxies in the cluster have weaker D$_{n}$(4000) than the field, because they are dominated by recently-quenched galaxies.  The arrows show the effect of increasing (purple) or decreasing (black) the importance of environmental quenching relative to mass quenching.}
\end{figure*}
\section{A Simple Quenching Model to Explain the Lack of a D$_{\lowercase{n}}$(4000)-Environment Correlation At Fixed Stellar Mass}
The existence of a poststarburst population that is associated primarily with high-density regions appears to support a rapid environmental-quenching timescale, which in turn provides a good explanation for the lack of a correlation between the SSFRs and D$_{n}$(4000)s of star-forming galaxies and their environment.  However, in $\S$ 6.3 we also showed that the D$_{n}$(4000) of the quiescent galaxies is independent of their environment.  We know from both the strong decline in the f$_{SF}$ with increasing environmental density, as well as the increase in the poststarburst fraction with increasing density that environmental processes must be responsible for quenching a significant fraction of the quiescent population.  It seems then difficult to understand why the D$_{n}$(4000) of quiescent galaxies does not correlate with environment at fixed mass, but instead correlates with stellar mass at fixed environment.  In this section we construct some simple quenching models to investigate how much the D$_{n}$(4000) of quiescent galaxies can differ between environments in different quenching scenarios.
\subsection{A Basic D$_{n}$(4000) Evolution Model}
On purpose, our model is simple and designed to provide a basic illustration of the problem, not a detailed quantitative prediction for the evolution of D$_{n}$(4000) in different environments.  Our approach is to simulate a population of star-forming galaxies, assuming they are all the same stellar mass, and then quench them at different rates to follow the evolution of the average D$_{n}$(4000) of {\it only} the quenched galaxies at fixed stellar mass.   To make this model requires that we have a prediction of how the D$_{n}$(4000) evolves with time for both star-forming and quiescent galaxies.  
\newline\indent
We assume that star-forming galaxies form stars at a constant rate until they are quenched, either by their environment, or by some self-regulating process related to their stellar mass.  As the star-formation histories in Figure 3 show, galaxies with a constant SFR have a slowly evolving D$_{n}$(4000) that remains near a value of $\sim$ 1.2.  Therefore, for simplicity, we assume that the D$_{n}$(4000) of star-forming galaxies is exactly equal to 1.2 at all times while they are star forming.  In terms of the quiescent galaxies, the red track in Figure 3 shows that the D$_{n}$(4000) of a single-burst population evolves rapidly after the initial burst, and thereafter the rate of evolution begins to slow.   We will assume that once star-forming galaxies have been quenched, they follow along the D$_{n}$(4000) track for a passively-evolving single-burst galaxy.  We note this is not a completely fair representation, as the D$_{n}$(4000)'s of galaxies that have been star forming for longer periods will evolve slightly slower than single-bursts; however, given that we are more interested in illustrating the difference in D$_{n}$(4000) between different environments and different quenching scenarios, not predicting its precise value, a more complex treatment of the star formation histories is not warranted.
\newline\indent
As a first step in the model, we define three scenarios by which the quenching rate of galaxies might evolve, regardless of whether environmental quenching or self quenching is responsible.  We consider the evolution of the quenching rate over a period of 6 Gyr, which is the age of the universe at $z \sim$ 1.  This is the maximum amount of time over which galaxies have had to quench, and will illustrate the maximum difference between the properties of galaxies in different environments that can be achieved.   
\newline\indent
The first quenching rate we consider is where the quenching rate of galaxies proceeds linearly, and the rate does not evolve with time.  In the left panel of Figure 14 we plot the fraction of quenched galaxies as a function of time for this model in green.  We also consider a model where most of the quenching occurs at early times, where we parameterize the quenching rate as a decaying exponential (red).  Lastly, we consider a model where most of the quenching occurs at late times, where the quenching rate is parameterized as an increasing exponential (blue).  The models are normalized so that 100\% of the quiescent galaxies are quenched by t = 6 Gyr.  We have chosen an exponential to parameterize the "early" and "late" quenching scenarios; however, we note that any sharply increasing or decreasing function would be sufficient.  
\newline\indent
In the right panel of Figure 14 we plot the predicted evolution of the mean D$_{n}$(4000) of quiescent galaxies as a function of time for the three quenching functions.  For reference we also plot the mean D$_{n}$(4000) of a purely passively evolving galaxy to show the upper limit of D$_{n}$(4000) as a function of time.  As expected, the mean D$_{n}$(4000) of quiescent galaxies depends on the epoch when they are quenched.  The exponentially declining quenching model has the largest mean D$_{n}$(4000) because most of the galaxies have quenched at early times and hence have had more time to age.  The exponentially increasing quenching model has the smallest mean D$_{n}$(4000), because at all times the quiescent population is dominated by galaxies that have only recently been quenched.
\subsection{A Model for the Competing Effects of Environmental and Self Quenching on D$_{n}$(4000)}
We can now add complexity to the model by assuming two independent quenching mechanisms: one related to stellar mass, which we will call self quenching, the other related to environment, which we refer to as environmental quenching.  In particular, we would like to understand the difference in the D$_{n}$(4000) of quiescent galaxies that results from a model that has both self quenching and environmental quenching -- which should represent a high-density region such as a cluster -- as compared to the predicted D$_{n}$(4000) of quiescent galaxies from models that have only self quenching -- which should represent the low-density regions found in the field population.
\newline\indent
There are a large range of quenching models that can be built to make this comparison; however, the most relevant things to test in terms of a difference in the D$_{n}$(4000) of quiescent galaxies with environment is 1) the amplitude of self quenching as compared to environmental quenching, i.e., which process is most important for creating the quiescent population, and 2) how the time evolution of the self-quenching and environmental-quenching rates affects the D$_{n}$(4000)s.  
\newline\indent
In the left panel of Figure 15 we plot the fraction of quenched galaxies as a function of time for models that include both self quenching and environmental quenching.  Each model for the quenching rate is plotted in a different color, and within those colors the fraction of galaxies quenched by environment is shown as the solid curve and the fraction of galaxies that are self quenched is shown as the dotted curve.  
\newline\indent
We test four models to explore the importance of items 1) and 2) above.  Firstly, we examine the case where environmental quenching and self quenching have equal importance, each quenching 50\% of the population, but the rate at which they quench galaxies evolves differently with time.  The red model shows the case where environmental quenching dominates at early times, and self quenching dominates at late times.  The blue model shows the converse, with self quenching dominating at early times and environmental quenching dominating at late times.  
\newline\indent
In the right panel of Figure 15 we plot the {\it difference} in the evolution of the mean D$_{n}$(4000) of the quiescent galaxies for these models that include both self and environmental quenching compared to the mean D$_{n}$(4000) for a model that just has self quenching (with the same self-quenching rate evolution as in the combined model).  The red line shows that when environmental quenching dominates early, quiescent galaxies in high-density environments will have stronger D$_{n}$(4000)s than those in low-density environments.  This occurs because the galaxies in high-density environments quench earlier on average and hence have more time to evolve.  The blue curve shows that when environmental quenching dominates at late times, quiescent galaxies in high-density regions actually have $weaker$ D$_{n}$(4000) than those in low-density regions.  This occurs because high-density regions have a larger population of recently-quenched galaxies that are young and move the mean of D$_{n}$(4000)s to a lower value.  
\newline\indent
The blue curve is interesting and is worth commenting on.  A prediction from the \cite{Peng2010} empirical quenching model is that self quenching should dominate the early evolution of galaxies and environmental quenching should dominate their later evolution as the growth of structure in the universe proceeds.  Our simple model for the  D$_{n}$(4000) evolution of quiescent galaxies shows that if this is the case, we might expected to see something quite counter-intuitive, that at fixed stelar mass, quiescent galaxies in clusters should actually be $younger$ than those in the field.   In general, the converse is seen \citep[e.g.,][]{Thomas2005,Nelan2005,Gallazzi2005,vandokkum2007,Thomas2010}, which would seem to argue against environmental quenching dominating at late times.  Indeed, if anything, it would seem to suggest that environmental quenching should dominate at earlier epochs.   On the other hand, we should also consider that most of those studies have focused on primarily elliptical galaxies, not all quenched galaxies.  In general clusters contain a much higher fraction of S0 galaxies than the field \citep[e.g.,][]{Dressler1980} and it may be that when S0's are included with the elliptical galaxies, the population of quenched galaxies in clusters (E + S0) is in fact younger than the population of quenched galaxies in the field (dominated by E). 
\newline\indent
Although it is unclear how to reconcile this issue, it is important to point out that the absolute difference in the D$_{n}$(4000)s between our simulated cluster and field is quite small, with $\Delta$D$_{n}$(4000) = 0.1 to 0.2.   Our model assumes an equal contribution from self quenching and environmental quenching.  In principle, the $\Delta$D$_{n}$(4000) between cluster and field can be increased or decreased by adjusting the amplitude of self quenching compared to environmental quenching.  Increasing the importance of self quenching over environmental quenching decreases the difference between cluster and field, and likewise, increasing the importance of environmental quenching over self quenching increases the difference between cluster and field.  This shows that a lack of a difference in the D$_{n}$(4000) of quiescent galaxies as a function of environment would be expected if self quenching dominates over environmental quenching at all epochs.  Indeed, if greater than two-thirds of the cluster galaxies are self quenched, this would produce a $\Delta$D$_{n}$(4000) that is $<$ 0.1, and hence is unlikely to be detected with our data.
\newline\indent
While it is the simplest answer, a dominance of self quenching over environmental quenching is not a fully satisfactory explanation for the lack of a D$_{n}$(4000)-environment correlation. We know that the change in the f$_{SF}$ with environment is similar to the change in the f$_{SF}$ with stellar mass, and that there are a substantial number of poststarbursts associated with clusters.  Both of these suggest that environmental quenching is at least as important in the transformation of galaxies as is self quenching.  In order to explore this further, we now consider a second set of models.  This time where the amplitude of self quenching and environmental quenching are quite different, but where the quenching rates evolve in a similar way.  For this comparison we use the linear quenching rate, where in fact the rate does not evolve with time, but we note that the exponential models that do evolve with time provide a similar comparison. 
\newline\indent
The green curves in Figure 15 show the model where environmental quenching dominates, quenching 80\% of the population by t = 6 Gyr, and the orange curve shows the converse, where self quenching dominates and quenches 80\% of the population by t = 6 Gyr.  Interestingly, the $\Delta$D$_{n}$(4000) between cluster and field for these models is precisely zero.  This demonstrates that it is not critical that self quenching dominates over environmental quenching in order to produce similar D$_{n}$(4000) in high-density and low-density regions.  All that is required is that the self quenching rate evolves in the same way as the environmental quenching rate.  If they do, then no matter the amplitude of each, the $\Delta$D$_{n}$(4000) between cluster and field is precisely zero.  Put another way, the D$_{n}$(4000) depends on $when$ galaxies quench in cosmic time, not what is the source of the quenching.  Provided that the environmental-quenching and self-quenching rates evolve so as to quench galaxies similarly at all epochs, the fraction of quenched galaxies will be higher in clusters; however, then the mean D$_{n}$(4000) of the quiescent galaxies will be the same in all environments.
\newline\indent
Putting both sets of models together provides two plausible explanations for why the D$_{n}$(4000) of quiescent galaxies does not depend on environment.  First, if self quenching dominates over environmental quenching, then the D$_{n}$(4000) will be the same in all environments, regardless of how the two quenching rates evolve.  Alternatively, provided that the self-quenching and environmental-quenching rates evolve similarly with time, then regardless of which process dominates no difference in the D$_{n}$(4000) of quiescent galaxies with environment will be seen.  The latter would be a better explanation for our data, which suggests that environmental quenching is roughly equally important as mass quenching.
\newline\indent
\section{Summary}
In this paper we have studied the correlations between the properties of galaxies and their environment and stellar mass.  We first considered the average properties of a mass-limited sample of galaxies with LogM$_{*}$/M$_{\odot}$ $>$ 9.3 ($\S$ 5), and then subdivided the galaxies into different types (star-forming or quiescent) and studied the properties as a function of environment at fixed stellar mass, and vice versa.  We also identified the population of recently-quenched galaxies and studied their abundance as a function of environment and stellar mass.  Lastly, we constructed several simple quenching models to explain some of the (lack of) correlations in the data. The main results of the paper are,
\newline
1. For galaxies with LogM$_{*}$/M$_{\odot}$ $>$ 9.3, the well-known correlations between f$_{SF}$, SSFR, SFR, D$_{n}$(4000), and color with environment seen in the local universe are already clearly in place by $z \sim$ 1.
\newline
2. For cluster galaxies, the well-known correlations between properties such as f$_{SF}$, SSFR, SFR, D$_{n}$(4000), and color with stellar mass seen in the local universe are also in place at $z \sim$ 1.
\newline
3. Similar to the local universe, there is a covariance between the mean stellar mass of galaxies and their environment, with more massive galaxies living in higher-density environments.
\newline
4. The relative decline in the f$_{SF}$ with environment at fixed stellar mass is a factor of $\sim$ 2-4 and the relative decline in the f$_{SF}$ with stellar mass at fixed environment is a factor of $\sim$ 2-3.  Both stellar mass and environment affect the f$_{SF}$, and their relative effects are separable up to $z \sim$ 1.
\newline
5. Controlling for the covariance in stellar mass and f$_{SF}$ with environment we find that the SSFRs of star-forming galaxies are correlated with stellar mass at fixed environment, but is nearly independent of environment at fixed stellar mass.
\newline
6. The D$_{n}$(4000)s of both star-forming and quiescent galaxies are correlated with their stellar mass at fixed environment, but is nearly independent of environment at fixed stellar mass.
\newline
7. The data suggest that stellar mass is the primary parameter that determines the stellar populations of galaxies.  Environment plays a different role, controlling the f$_{SF}$, and to a lesser degree, the mean stellar mass of galaxies.
\newline
8. The population of poststarburst galaxies is correlated with both environment and stellar mass, in that poststarbursts are more common amongst lower-mass galaxies and in higher-density environments.  Poststarburst galaxies with 9.3 $<$ LogM$_{*}$/M$_{\odot}$ $<$ 10.7 are 3.1 $\pm$ 1.1 times more common in clusters at $z \sim$ 1 compared to the field.
\newline
9. The lack of a correlation of the SSFRs and D$_{n}$(4000) of star-forming galaxies and their environment is best-explained by a rapid environmental quenching timescale.  The rapid environmental quenching timescale is further supported by the clear association of poststarbursts with high-density environments.
\newline
10. Despite an overabundance of poststarbursts in clusters compared to the field, poststarbursts still make a relatively small fraction of the total cluster population.  Only 10\% $\pm$ 0.2\% of cluster galaxies at $z \sim$ 1 with 9.3 $<$ LogM$_{*}$/M$_{\odot}$ $<$ 11.0 are poststarbursts.  
\newline
11. Simple quenching models show that the lack of a correlation between the D$_{n}$(4000) of quiescent galaxies and their environment can be explained two ways: Either self quenching dominates over environmental quenching at $z >$ 1, or else the evolution of the self quenching rate mirrors the evolution of the environmental quenching rate.   
\newline\indent
Overall, this study of $z \sim$ 1 clusters shows that the trends in galaxy properties with environment and stellar mass that are seen in the local universe \cite[e.g.,][]{Kauffmann2004,Balogh2004a,Baldry2006,Weinmann2006,Peng2010} are already well-established at $z \sim$ 1.   It appears clear from our study as well as both lower- and higher-redshift studies that while the fraction of star-forming galaxies is a strong function of their environment, \cite[e.g.,][]{Balogh2004a,Kauffmann2004,Peng2010,Patel2011,Quadri2011}, the mean properties of the star-forming and quiescent galaxies at fixed stellar mass are not.  This means that stellar mass is the primary parameter linked to determining the stellar populations of galaxies.
\newline\indent
Our simple modeling shows that the lack of correlation of the properties of star-forming and quiescent galaxies with their environment can be understood if the environmental-quenching timescale is rapid, and that the evolution of the self-quenching and environmental-quenching rates mirrors each other -- regardless of which processes dominates the overall quenching process.  This hypothesis is further supported by the substantial population of poststarburst galaxies that are only found in the high-density cluster environment.
\newline\indent
The next obvious step to better understanding of the process of environmental quenching will be to connect the evolution of the cluster galaxy population over a larger redshift baseline.  Doing so will allow us to differentiate between different quenching models and get a better handle on the physical processes behind environmental quenching.\

\acknowledgements
The GCLASS project is the result of a generous investment in observing time from both the Canadian and U.S. Gemini partners.  The GCLASS data was taken in 8 separate allocations over the period of 2007-2010 and we would like to thank members of both Can-TAC and NOAO-TAC for their continued long-term support of the project.  GCLASS would also not have been possible without the hard work of members of the Gemini Staff, who carefully checked the many mask designs, phase2 skeletons, and overall ensured the delivery of high-quality data products.  We would like to acknowledge all members of the Gemini staff who have been involved in executing the GCLASS observations.  GW gratefully acknowledges support from NSF grant AST-0909198.  RD gratefully acknowledges the support provided by the BASAL Center for Astrophysics and Associated Technologies (CATA), and by FONDECYT grant N. 1100540.

\bibliographystyle{apj}
\bibliography{apj-jour,myrefs}

\begin{thebibliography}{98}
\expandafter\ifx\csname natexlab\endcsname\relax\def\natexlab#1{#1}\fi

\bibitem[{{Abraham} {et~al.}(2004){Abraham}, {Glazebrook}, {McCarthy},
  {Crampton}, {Murowinski}, {J{\o}rgensen}, {Roth}, {Hook}, {Savaglio}, {Chen},
  {Marzke}, \& {Carlberg}}]{Abraham2004}
{Abraham}, R.~G., {et~al.} 2004, \aj, 127, 2455

\bibitem[{{Baldry} {et~al.}(2006){Baldry}, {Balogh}, {Bower}, {Glazebrook},
  {Nichol}, {Bamford}, \& {Budavari}}]{Baldry2006}
{Baldry}, I.~K., {Balogh}, M.~L., {Bower}, R.~G., {Glazebrook}, K., {Nichol},
  R.~C., {Bamford}, S.~P., \& {Budavari}, T. 2006, \mnras, 373, 469

\bibitem[{{Balogh} {et~al.}(2004{\natexlab{a}}){Balogh}, {Eke}, {Miller},
  {Lewis}, {Bower}, {Couch}, {Nichol}, {Bland-Hawthorn}, {Baldry}, {Baugh},
  {Bridges}, {Cannon}, {Cole}, {Colless}, {Collins}, {Cross}, {Dalton}, {de
  Propris}, {Driver}, {Efstathiou}, {Ellis}, {Frenk}, {Glazebrook}, {Gomez},
  {Gray}, {Hawkins}, {Jackson}, {Lahav}, {Lumsden}, {Maddox}, {Madgwick},
  {Norberg}, {Peacock}, {Percival}, {Peterson}, {Sutherland}, \&
  {Taylor}}]{Balogh2004b}
{Balogh}, M., {et~al.} 2004{\natexlab{a}}, \mnras, 348, 1355

\bibitem[{{Balogh} {et~al.}(2004{\natexlab{b}}){Balogh}, {Baldry}, {Nichol},
  {Miller}, {Bower}, \& {Glazebrook}}]{Balogh2004a}
{Balogh}, M.~L., {Baldry}, I.~K., {Nichol}, R., {Miller}, C., {Bower}, R., \&
  {Glazebrook}, K. 2004{\natexlab{b}}, \apjl, 615, L101

\bibitem[{{Balogh} {et~al.}(1999){Balogh}, {Morris}, {Yee}, {Carlberg}, \&
  {Ellingson}}]{Balogh1999}
{Balogh}, M.~L., {Morris}, S.~L., {Yee}, H.~K.~C., {Carlberg}, R.~G., \&
  {Ellingson}, E. 1999, \apj, 527, 54

\bibitem[{{Balogh} {et~al.}(2011){Balogh}, {McGee}, {Wilman}, {Finoguenov},
  {Parker}, {Connelly}, {Mulchaey}, {Bower}, {Tanaka}, \&
  {Giodini}}]{Balogh2011}
{Balogh}, M.~L., {et~al.} 2011, \mnras, 412, 2303

\bibitem[{{Bamford} {et~al.}(2009){Bamford}, {Nichol}, {Baldry}, {Land},
  {Lintott}, {Schawinski}, {Slosar}, {Szalay}, {Thomas}, {Torki}, {Andreescu},
  {Edmondson}, {Miller}, {Murray}, {Raddick}, \& {Vandenberg}}]{Bamford2009}
{Bamford}, S.~P., {et~al.} 2009, \mnras, 393, 1324

\bibitem[{{Bauer} {et~al.}(2011){Bauer}, {Gr{\"u}tzbauch}, {J{\o}rgensen},
  {Varela}, \& {Bergmann}}]{Bauer2011}
{Bauer}, A.~E., {Gr{\"u}tzbauch}, R., {J{\o}rgensen}, I., {Varela}, J., \&
  {Bergmann}, M. 2011, \mnras, 411, 2009

\bibitem[{{Blanton} \& {Moustakas}(2009)}]{blanton09}
{Blanton}, M.~R., \& {Moustakas}, J. 2009, \araa, 47, 159

\bibitem[{{Bolzonella} {et~al.}(2010){Bolzonella}, {Kova{\v c}}, {Pozzetti},
  {Zucca}, {Cucciati}, {Lilly}, {Peng}, {Iovino}, {Zamorani}, {Vergani},
  {Tasca}, {Lamareille}, {Oesch}, {Caputi}, {Kampczyk}, {Bardelli}, {Maier},
  {Abbas}, {Knobel}, {Scodeggio}, {Carollo}, {Contini}, {Kneib}, {Le
  F{\`e}vre}, {Mainieri}, {Renzini}, {Bongiorno}, {Coppa}, {de la Torre}, {de
  Ravel}, {Franzetti}, {Garilli}, {Le Borgne}, {Le Brun}, {Mignoli},
  {Pell{\'o}}, {Perez-Montero}, {Ricciardelli}, {Silverman}, {Tanaka},
  {Tresse}, {Bottini}, {Cappi}, {Cassata}, {Cimatti}, {Guzzo}, {Koekemoer},
  {Leauthaud}, {Maccagni}, {Marinoni}, {McCracken}, {Memeo}, {Meneux},
  {Porciani}, {Scaramella}, {Aussel}, {Capak}, {Halliday}, {Ilbert},
  {Kartaltepe}, {Salvato}, {Sanders}, {Scarlata}, {Scoville}, {Taniguchi}, \&
  {Thompson}}]{Bolzonella2010}
{Bolzonella}, M., {et~al.} 2010, \aap, 524, A76+

\bibitem[{{Bower} {et~al.}(1992){Bower}, {Lucey}, \& {Ellis}}]{Bower1992}
{Bower}, R.~G., {Lucey}, J.~R., \& {Ellis}, R.~S. 1992, \mnras, 254, 601

\bibitem[{{Bruzual} \& {Charlot}(2003)}]{Bruzual2003}
{Bruzual}, G., \& {Charlot}, S. 2003, \mnras, 344, 1000

\bibitem[{{Calzetti} {et~al.}(2000){Calzetti}, {Armus}, {Bohlin}, {Kinney},
  {Koornneef}, \& {Storchi-Bergmann}}]{Calzetti2000}
{Calzetti}, D., {Armus}, L., {Bohlin}, R.~C., {Kinney}, A.~L., {Koornneef}, J.,
  \& {Storchi-Bergmann}, T. 2000, \apj, 533, 682

\bibitem[{{Chabrier}(2003)}]{Chabrier2003}
{Chabrier}, G. 2003, \pasp, 115, 763

\bibitem[{{Conroy} {et~al.}(2009){Conroy}, {Gunn}, \& {White}}]{Conroy2009}
{Conroy}, C., {Gunn}, J.~E., \& {White}, M. 2009, \apj, 699, 486

\bibitem[{{Cooper} {et~al.}(2010{\natexlab{a}}){Cooper}, {Gallazzi}, {Newman},
  \& {Yan}}]{Cooper2010a}
{Cooper}, M.~C., {Gallazzi}, A., {Newman}, J.~A., \& {Yan}, R.
  2010{\natexlab{a}}, \mnras, 402, 1942

\bibitem[{{Cooper} {et~al.}(2005){Cooper}, {Newman}, {Madgwick}, {Gerke},
  {Yan}, \& {Davis}}]{Cooper2005}
{Cooper}, M.~C., {Newman}, J.~A., {Madgwick}, D.~S., {Gerke}, B.~F., {Yan}, R.,
  \& {Davis}, M. 2005, \apj, 634, 833

\bibitem[{{Cooper} {et~al.}(2006){Cooper}, {Newman}, {Croton}, {Weiner},
  {Willmer}, {Gerke}, {Madgwick}, {Faber}, {Davis}, {Coil}, {Finkbeiner},
  {Guhathakurta}, \& {Koo}}]{Cooper2006}
{Cooper}, M.~C., {et~al.} 2006, \mnras, 370, 198

\bibitem[{{Cooper} {et~al.}(2007){Cooper}, {Newman}, {Coil}, {Croton}, {Gerke},
  {Yan}, {Davis}, {Faber}, {Guhathakurta}, {Koo}, {Weiner}, \&
  {Willmer}}]{Cooper2007}
---. 2007, \mnras, 376, 1445

\bibitem[{{Cooper} {et~al.}(2008){Cooper}, {Newman}, {Weiner}, {Yan},
  {Willmer}, {Bundy}, {Coil}, {Conselice}, {Davis}, {Faber}, {Gerke},
  {Guhathakurta}, {Koo}, \& {Noeske}}]{Cooper2008a}
---. 2008, \mnras, 383, 1058

\bibitem[{{Cooper} {et~al.}(2010{\natexlab{b}}){Cooper}, {Coil}, {Gerke},
  {Newman}, {Bundy}, {Conselice}, {Croton}, {Davis}, {Faber}, {Guhathakurta},
  {Koo}, {Lin}, {Weiner}, {Willmer}, \& {Yan}}]{Cooper2010b}
---. 2010{\natexlab{b}}, \mnras, 409, 337

\bibitem[{{Cucciati} {et~al.}(2006){Cucciati}, {Iovino}, {Marinoni}, {Ilbert},
  {Bardelli}, {Franzetti}, {Le F{\`e}vre}, {Pollo}, {Zamorani}, {Cappi},
  {Guzzo}, {McCracken}, {Meneux}, {Scaramella}, {Scodeggio}, {Tresse}, {Zucca},
  {Bottini}, {Garilli}, {Le Brun}, {Maccagni}, {Picat}, {Vettolani},
  {Zanichelli}, {Adami}, {Arnaboldi}, {Arnouts}, {Bolzonella}, {Charlot},
  {Ciliegi}, {Contini}, {Foucaud}, {Gavignaud}, {Marano}, {Mazure}, {Merighi},
  {Paltani}, {Pell{\`o}}, {Pozzetti}, {Radovich}, {Bondi}, {Bongiorno},
  {Busarello}, {de la Torre}, {Gregorini}, {Lamareille}, {Mathez}, {Mellier},
  {Merluzzi}, {Ripepi}, {Rizzo}, {Temporin}, \& {Vergani}}]{Cucciati2006}
{Cucciati}, O., {et~al.} 2006, \aap, 458, 39

\bibitem[{{Cucciati} {et~al.}(2010){Cucciati}, {Iovino}, {Kova{\v c}},
  {Scodeggio}, {Lilly}, {Bolzonella}, {Bardelli}, {Vergani}, {Tasca}, {Zucca},
  {Zamorani}, {Pozzetti}, {Knobel}, {Oesch}, {Lamareille}, {Caputi},
  {Kampczyk}, {Tresse}, {Maier}, {Carollo}, {Contini}, {Kneib}, {Le F{\`e}vre},
  {Mainieri}, {Renzini}, {Bongiorno}, {Coppa}, {de la Torre}, {de Ravel},
  {Franzetti}, {Garilli}, {Le Borgne}, {Le Brun}, {Mignoli}, {Pell{\`o}},
  {Peng}, {Perez-Montero}, {Ricciardelli}, {Silverman}, {Tanaka}, {Koekemoer},
  {Scoville}, {Abbas}, {Bottini}, {Cappi}, {Cassata}, {Cimatti}, {Guzzo},
  {Leauthaud}, {Maccagni}, {Marinoni}, {McCracken}, {Memeo}, {Meneux},
  {Porciani}, \& {Scaramella}}]{Cucciati2010}
---. 2010, \aap, 524, A2+

\bibitem[{{Demarco} {et~al.}(2007){Demarco}, {Rosati}, {Lidman}, {Girardi},
  {Nonino}, {Rettura}, {Strazzullo}, {van der Wel}, {Ford}, {Mainieri},
  {Holden}, {Stanford}, {Blakeslee}, {Gobat}, {Postman}, {Tozzi}, {Overzier},
  {Zirm}, {Ben{\'{\i}}tez}, {Homeier}, {Illingworth}, {Infante}, {Jee}, {Mei},
  {Menanteau}, {Motta}, {Zheng}, {Clampin}, \& {Hartig}}]{Demarco2007}
{Demarco}, R., {et~al.} 2007, \apj, 663, 164

\bibitem[{{Demarco} {et~al.}(2010{\natexlab{a}}){Demarco}, {Wilson}, {Muzzin},
  {Lacy}, {Surace}, {Yee}, {Hoekstra}, {Blindert}, \& {Gilbank}}]{Demarco2010}
---. 2010{\natexlab{a}}, \apj, 711, 1185

\bibitem[{{Demarco} {et~al.}(2010{\natexlab{b}}){Demarco}, {Gobat}, {Rosati},
  {Lidman}, {Rettura}, {Nonino}, {van der Wel}, {Jee}, {Blakeslee}, {Ford}, \&
  {Postman}}]{Demarco2010b}
---. 2010{\natexlab{b}}, \apj, 725, 1252

\bibitem[{{Dressler}(1980)}]{Dressler1980}
{Dressler}, A. 1980, \apj, 236, 351

\bibitem[{{Dressler} {et~al.}(1999){Dressler}, {Smail}, {Poggianti}, {Butcher},
  {Couch}, {Ellis}, \& {Oemler}}]{Dressler1999}
{Dressler}, A., {Smail}, I., {Poggianti}, B.~M., {Butcher}, H., {Couch}, W.~J.,
  {Ellis}, R.~S., \& {Oemler}, Jr., A. 1999, \apjs, 122, 51

\bibitem[{{Elbaz} {et~al.}(2007){Elbaz}, {Daddi}, {Le Borgne}, {Dickinson},
  {Alexander}, {Chary}, {Starck}, {Brandt}, {Kitzbichler}, {MacDonald},
  {Nonino}, {Popesso}, {Stern}, \& {Vanzella}}]{Elbaz2007}
{Elbaz}, D., {et~al.} 2007, \aap, 468, 33

\bibitem[{{F{\"o}rster Schreiber} {et~al.}(2004){F{\"o}rster Schreiber}, {van
  Dokkum}, {Franx}, {Labb{\'e}}, {Rudnick}, {Daddi}, {Illingworth}, {Kriek},
  {Moorwood}, {Rix}, {R{\"o}ttgering}, {Trujillo}, {van der Werf}, {van
  Starkenburg}, \& {Wuyts}}]{ForsterSchreiber2004}
{F{\"o}rster Schreiber}, N.~M., {et~al.} 2004, \apj, 616, 40

\bibitem[{{Gallazzi} {et~al.}(2006){Gallazzi}, {Charlot}, {Brinchmann}, \&
  {White}}]{Gallazzi2006}
{Gallazzi}, A., {Charlot}, S., {Brinchmann}, J., \& {White}, S.~D.~M. 2006,
  \mnras, 370, 1106

\bibitem[{{Gallazzi} {et~al.}(2005){Gallazzi}, {Charlot}, {Brinchmann},
  {White}, \& {Tremonti}}]{Gallazzi2005}
{Gallazzi}, A., {Charlot}, S., {Brinchmann}, J., {White}, S.~D.~M., \&
  {Tremonti}, C.~A. 2005, \mnras, 362, 41

\bibitem[{{Gilbank} {et~al.}(2010){Gilbank}, {Baldry}, {Balogh}, {Glazebrook},
  \& {Bower}}]{Gilbank2010}
{Gilbank}, D.~G., {Baldry}, I.~K., {Balogh}, M.~L., {Glazebrook}, K., \&
  {Bower}, R.~G. 2010, \mnras, 405, 2594

\bibitem[{{Gilbank} {et~al.}(2011){Gilbank}, {Gladders}, {Yee}, \&
  {Hsieh}}]{Gilbank2011}
{Gilbank}, D.~G., {Gladders}, M.~D., {Yee}, H.~K.~C., \& {Hsieh}, B.~C. 2011,
  \aj, 141, 94

\bibitem[{{Gladders} \& {Yee}(2000)}]{Gladders2000}
{Gladders}, M.~D., \& {Yee}, H.~K.~C. 2000, \aj, 120, 2148

\bibitem[{{Gladders} \& {Yee}(2005)}]{Gladders2005}
---. 2005, \apjs, 157, 1

\bibitem[{{Glazebrook} \& {Bland-Hawthorn}(2001)}]{Glazebrook2001}
{Glazebrook}, K., \& {Bland-Hawthorn}, J. 2001, \pasp, 113, 197

\bibitem[{{Gobat} {et~al.}(2008){Gobat}, {Rosati}, {Strazzullo}, {Rettura},
  {Demarco}, \& {Nonino}}]{Gobat2008}
{Gobat}, R., {Rosati}, P., {Strazzullo}, V., {Rettura}, A., {Demarco}, R., \&
  {Nonino}, M. 2008, \aap, 488, 853

\bibitem[{{G{\'o}mez} {et~al.}(2003){G{\'o}mez}, {Nichol}, {Miller}, {Balogh},
  {Goto}, {Zabludoff}, {Romer}, {Bernardi}, {Sheth}, {Hopkins}, {Castander},
  {Connolly}, {Schneider}, {Brinkmann}, {Lamb}, {SubbaRao}, \&
  {York}}]{Gomez2003}
{G{\'o}mez}, P.~L., {et~al.} 2003, \apj, 584, 210

\bibitem[{{Graves} {et~al.}(2009){Graves}, {Faber}, \& {Schiavon}}]{Graves2009}
{Graves}, G.~J., {Faber}, S.~M., \& {Schiavon}, R.~P. 2009, \apj, 693, 486

\bibitem[{{Gr{\"u}tzbauch} {et~al.}(2011{\natexlab{a}}){Gr{\"u}tzbauch},
  {Chuter}, {Conselice}, {Bauer}, {Bluck}, {Buitrago}, \&
  {Mortlock}}]{Grutzbauch2011b}
{Gr{\"u}tzbauch}, R., {Chuter}, R.~W., {Conselice}, C.~J., {Bauer}, A.~E.,
  {Bluck}, A.~F.~L., {Buitrago}, F., \& {Mortlock}, A. 2011{\natexlab{a}},
  \mnras, 412, 2361

\bibitem[{{Gr{\"u}tzbauch} {et~al.}(2011{\natexlab{b}}){Gr{\"u}tzbauch},
  {Conselice}, {Varela}, {Bundy}, {Cooper}, {Skibba}, \&
  {Willmer}}]{Gruztbauch2011a}
{Gr{\"u}tzbauch}, R., {Conselice}, C.~J., {Varela}, J., {Bundy}, K., {Cooper},
  M.~C., {Skibba}, R., \& {Willmer}, C.~N.~A. 2011{\natexlab{b}}, \mnras, 411,
  929

\bibitem[{{Hatch} {et~al.}(2011){Hatch}, {Kurk}, {Pentericci}, {Venemans},
  {Kuiper}, {Miley}, \& {R{\"o}ttgering}}]{Hatch2011}
{Hatch}, N.~A., {Kurk}, J.~D., {Pentericci}, L., {Venemans}, B.~P., {Kuiper},
  E., {Miley}, G.~K., \& {R{\"o}ttgering}, H.~J.~A. 2011, ArXiv e-prints

\bibitem[{{Hayashi} {et~al.}(2010){Hayashi}, {Kodama}, {Koyama}, {Tanaka},
  {Shimasaku}, \& {Okamura}}]{Hayashi2010}
{Hayashi}, M., {Kodama}, T., {Koyama}, Y., {Tanaka}, I., {Shimasaku}, K., \&
  {Okamura}, S. 2010, \mnras, 402, 1980

\bibitem[{{Hilton} {et~al.}(2010){Hilton}, {Lloyd-Davies}, {Stanford}, {Stott},
  {Collins}, {Romer}, {Hosmer}, {Hoyle}, {Kay}, {Liddle}, {Mehrtens}, {Miller},
  {Sahl{\'e}n}, \& {Viana}}]{Hilton2010}
{Hilton}, M., {et~al.} 2010, \apj, 718, 133

\bibitem[{{Iovino} {et~al.}(2010){Iovino}, {Cucciati}, {Scodeggio}, {Knobel},
  {Kova{\v c}}, {Lilly}, {Bolzonella}, {Tasca}, {Zamorani}, {Zucca}, {Caputi},
  {Pozzetti}, {Oesch}, {Lamareille}, {Halliday}, {Bardelli}, {Finoguenov},
  {Guzzo}, {Kampczyk}, {Maier}, {Tanaka}, {Vergani}, {Carollo}, {Contini},
  {Kneib}, {Le F{\`e}vre}, {Mainieri}, {Renzini}, {Bongiorno}, {Coppa}, {de la
  Torre}, {de Ravel}, {Franzetti}, {Garilli}, {Le Borgne}, {Le Brun},
  {Mignoli}, {Pell{\`o}}, {Peng}, {Perez-Montero}, {Ricciardelli}, {Silverman},
  {Tresse}, {Abbas}, {Bottini}, {Cappi}, {Cassata}, {Cimatti}, {Koekemoer},
  {Leauthaud}, {Maccagni}, {Marinoni}, {McCracken}, {Memeo}, {Meneux},
  {Porciani}, {Scaramella}, {Schiminovich}, \& {Scoville}}]{Iovino2010}
{Iovino}, A., {et~al.} 2010, \aap, 509, A40+

\bibitem[{{Kauffmann} {et~al.}(2004){Kauffmann}, {White}, {Heckman},
  {M{\'e}nard}, {Brinchmann}, {Charlot}, {Tremonti}, \&
  {Brinkmann}}]{Kauffmann2004}
{Kauffmann}, G., {White}, S.~D.~M., {Heckman}, T.~M., {M{\'e}nard}, B.,
  {Brinchmann}, J., {Charlot}, S., {Tremonti}, C., \& {Brinkmann}, J. 2004,
  \mnras, 353, 713

\bibitem[{{Kauffmann} {et~al.}(2003){Kauffmann}, {Heckman}, {White}, {Charlot},
  {Tremonti}, {Peng}, {Seibert}, {Brinkmann}, {Nichol}, {SubbaRao}, \&
  {York}}]{Kauffmann2003}
{Kauffmann}, G., {et~al.} 2003, \mnras, 341, 54

\bibitem[{{Kocevski} {et~al.}(2010){Kocevski}, {Lemaux}, {Lubin}, {Gal},
  {McGrath}, {Fassnacht}, {Squires}, {Surace}, \& {Lacy}}]{Kocevski2010}
{Kocevski}, D.~D., {et~al.} 2010, ArXiv e-prints

\bibitem[{{Koyama} {et~al.}(2010){Koyama}, {Kodama}, {Shimasaku}, {Hayashi},
  {Okamura}, {Tanaka}, \& {Tokoku}}]{Koyama2010}
{Koyama}, Y., {Kodama}, T., {Shimasaku}, K., {Hayashi}, M., {Okamura}, S.,
  {Tanaka}, I., \& {Tokoku}, C. 2010, \mnras, 403, 1611

\bibitem[{{Koyama} {et~al.}(2008){Koyama}, {Kodama}, {Shimasaku}, {Okamura},
  {Tanaka}, {Lee}, {Im}, {Matsuhara}, {Takagi}, {Wada}, \&
  {Oyabu}}]{Koyama2008}
{Koyama}, Y., {et~al.} 2008, \mnras, 391, 1758

\bibitem[{{Kriek} {et~al.}(2010){Kriek}, {Labb{\'e}}, {Conroy}, {Whitaker},
  {van Dokkum}, {Brammer}, {Franx}, {Illingworth}, {Marchesini}, {Muzzin},
  {Quadri}, \& {Rudnick}}]{Kriek2010}
{Kriek}, M., {et~al.} 2010, \apjl, 722, L64

\bibitem[{{Kuiper} {et~al.}(2010){Kuiper}, {Hatch}, {R{\"o}ttgering}, {Miley},
  {Overzier}, {Venemans}, {De Breuck}, {Croft}, {Kajisawa}, {Kodama}, {Kurk},
  {Pentericci}, {Stanford}, {Tanaka}, \& {Zirm}}]{Kuiper2010}
{Kuiper}, E., {et~al.} 2010, \mnras, 405, 969

\bibitem[{{Le Borgne} {et~al.}(2006){Le Borgne}, {Abraham}, {Daniel},
  {McCarthy}, {Glazebrook}, {Savaglio}, {Crampton}, {Juneau}, {Carlberg},
  {Chen}, {Marzke}, {Roth}, {J{\o}rgensen}, \& {Murowinski}}]{Leborgne2006}
{Le Borgne}, D., {et~al.} 2006, \apj, 642, 48

\bibitem[{{Lemaux} {et~al.}(2010){Lemaux}, {Lubin}, {Shapley}, {Kocevski},
  {Gal}, \& {Squires}}]{Lemaux2010}
{Lemaux}, B.~C., {Lubin}, L.~M., {Shapley}, A., {Kocevski}, D., {Gal}, R.~R.,
  \& {Squires}, G.~K. 2010, \apj, 716, 970

\bibitem[{{Li} {et~al.}(2011){Li}, {Glazebrook}, {Gilbank}, {Balogh}, {Bower},
  {Baldry}, {Davies}, {Hau}, \& {McCarthy}}]{Li2011}
{Li}, I.~H., {et~al.} 2011, \mnras, 411, 1869

\bibitem[{{Lonsdale} {et~al.}(2003){Lonsdale}, {Smith}, {Rowan-Robinson},
  {Surace}, {Shupe}, {Xu}, {Oliver}, {Padgett}, {Fang}, {Conrow},
  {Franceschini}, {Gautier}, {Griffin}, {Hacking}, {Masci}, {Morrison},
  {O'Linger}, {Owen}, {P{\'e}rez-Fournon}, {Pierre}, {Puetter}, {Stacey},
  {Castro}, {Polletta}, {Farrah}, {Jarrett}, {Frayer}, {Siana}, {Babbedge},
  {Dye}, {Fox}, {Gonzalez-Solares}, {Salaman}, {Berta}, {Condon}, {Dole}, \&
  {Serjeant}}]{Lonsdale2003}
{Lonsdale}, C.~J., {et~al.} 2003, \pasp, 115, 897

\bibitem[{{Maraston}(2005)}]{Maraston2005}
{Maraston}, C. 2005, \mnras, 362, 799

\bibitem[{{Maraston} {et~al.}(2006){Maraston}, {Daddi}, {Renzini}, {Cimatti},
  {Dickinson}, {Papovich}, {Pasquali}, \& {Pirzkal}}]{Maraston2006}
{Maraston}, C., {Daddi}, E., {Renzini}, A., {Cimatti}, A., {Dickinson}, M.,
  {Papovich}, C., {Pasquali}, A., \& {Pirzkal}, N. 2006, \apj, 652, 85

\bibitem[{{Marchesini} {et~al.}(2009){Marchesini}, {van Dokkum}, {F{\"o}rster
  Schreiber}, {Franx}, {Labb{\'e}}, \& {Wuyts}}]{Marchesini2009}
{Marchesini}, D., {van Dokkum}, P.~G., {F{\"o}rster Schreiber}, N.~M., {Franx},
  M., {Labb{\'e}}, I., \& {Wuyts}, S. 2009, \apj, 701, 1765

\bibitem[{{Muzzin} {et~al.}(2009{\natexlab{a}}){Muzzin}, {Marchesini}, {van
  Dokkum}, {Labb{\'e}}, {Kriek}, \& {Franx}}]{Muzzin2009b}
{Muzzin}, A., {Marchesini}, D., {van Dokkum}, P.~G., {Labb{\'e}}, I., {Kriek},
  M., \& {Franx}, M. 2009{\natexlab{a}}, \apj, 701, 1839

\bibitem[{{Muzzin} {et~al.}(2008){Muzzin}, {Wilson}, {Lacy}, {Yee}, \&
  {Stanford}}]{Muzzin2008}
{Muzzin}, A., {Wilson}, G., {Lacy}, M., {Yee}, H.~K.~C., \& {Stanford}, S.~A.
  2008, \apj, 686, 966

\bibitem[{{Muzzin} {et~al.}(2009{\natexlab{b}}){Muzzin}, {Wilson}, {Yee},
  {Hoekstra}, {Gilbank}, {Surace}, {Lacy}, {Blindert}, {Majumdar}, {Demarco},
  {Gardner}, {Gladders}, \& {Lonsdale}}]{Muzzin2009a}
{Muzzin}, A., {et~al.} 2009{\natexlab{b}}, \apj, 698, 1934

\bibitem[{{Nair} \& {Abraham}(2010)}]{Nair2010}
{Nair}, P.~B., \& {Abraham}, R.~G. 2010, \apjs, 186, 427

\bibitem[{{Nelan} {et~al.}(2005){Nelan}, {Smith}, {Hudson}, {Wegner}, {Lucey},
  {Moore}, {Quinney}, \& {Suntzeff}}]{Nelan2005}
{Nelan}, J.~E., {Smith}, R.~J., {Hudson}, M.~J., {Wegner}, G.~A., {Lucey},
  J.~R., {Moore}, S.~A.~W., {Quinney}, S.~J., \& {Suntzeff}, N.~B. 2005, \apj,
  632, 137

\bibitem[{{Patel} {et~al.}(2009){Patel}, {Holden}, {Kelson}, {Illingworth}, \&
  {Franx}}]{Patel2009}
{Patel}, S.~G., {Holden}, B.~P., {Kelson}, D.~D., {Illingworth}, G.~D., \&
  {Franx}, M. 2009, \apjl, 705, L67

\bibitem[{{Patel} {et~al.}(2011){Patel}, {Kelson}, {Holden}, {Franx}, \&
  {Illingworth}}]{Patel2011}
{Patel}, S.~G., {Kelson}, D.~D., {Holden}, B.~P., {Franx}, M., \&
  {Illingworth}, G.~D. 2011, ArXiv e-prints

\bibitem[{{Peng} {et~al.}(2010){Peng}, {Lilly}, {Kova{\v c}}, {Bolzonella},
  {Pozzetti}, {Renzini}, {Zamorani}, {Ilbert}, {Knobel}, {Iovino}, {Maier},
  {Cucciati}, {Tasca}, {Carollo}, {Silverman}, {Kampczyk}, {de Ravel},
  {Sanders}, {Scoville}, {Contini}, {Mainieri}, {Scodeggio}, {Kneib}, {Le
  F{\`e}vre}, {Bardelli}, {Bongiorno}, {Caputi}, {Coppa}, {de la Torre},
  {Franzetti}, {Garilli}, {Lamareille}, {Le Borgne}, {Le Brun}, {Mignoli},
  {Perez Montero}, {Pello}, {Ricciardelli}, {Tanaka}, {Tresse}, {Vergani},
  {Welikala}, {Zucca}, {Oesch}, {Abbas}, {Barnes}, {Bordoloi}, {Bottini},
  {Cappi}, {Cassata}, {Cimatti}, {Fumana}, {Hasinger}, {Koekemoer},
  {Leauthaud}, {Maccagni}, {Marinoni}, {McCracken}, {Memeo}, {Meneux}, {Nair},
  {Porciani}, {Presotto}, \& {Scaramella}}]{Peng2010}
{Peng}, Y., {et~al.} 2010, \apj, 721, 193

\bibitem[{{Poggianti} {et~al.}(1999){Poggianti}, {Smail}, {Dressler}, {Couch},
  {Barger}, {Butcher}, {Ellis}, \& {Oemler}}]{Poggianti1999}
{Poggianti}, B.~M., {Smail}, I., {Dressler}, A., {Couch}, W.~J., {Barger},
  A.~J., {Butcher}, H., {Ellis}, R.~S., \& {Oemler}, Jr., A. 1999, \apj, 518,
  576

\bibitem[{{Poggianti} {et~al.}(2009){Poggianti}, {Arag{\'o}n-Salamanca},
  {Zaritsky}, {De Lucia}, {Milvang-Jensen}, {Desai}, {Jablonka}, {Halliday},
  {Rudnick}, {Varela}, {Bamford}, {Best}, {Clowe}, {Noll}, {Saglia},
  {Pell{\'o}}, {Simard}, {von der Linden}, \& {White}}]{Poggianti2009a}
{Poggianti}, B.~M., {et~al.} 2009, \apj, 693, 112

\bibitem[{{Popesso} {et~al.}(2011){Popesso}, {Rodighiero}, {Saintonge},
  {Santini}, {Grazian}, {Lutz}, {Brusa}, \& {PEP Consortium}}]{Popesso2011}
{Popesso}, P., {Rodighiero}, G., {Saintonge}, A., {Santini}, P., {Grazian}, A.,
  {Lutz}, D., {Brusa}, M., \& {PEP Consortium}, f.~t. 2011, ArXiv e-prints

\bibitem[{{Postman} \& {Geller}(1984)}]{Postman1984}
{Postman}, M., \& {Geller}, M.~J. 1984, \apj, 281, 95

\bibitem[{{Quadri} {et~al.}(2007){Quadri}, {van Dokkum}, {Gawiser}, {Franx},
  {Marchesini}, {Lira}, {Rudnick}, {Herrera}, {Maza}, {Kriek}, {Labb{\'e}}, \&
  {Francke}}]{Quadri2007}
{Quadri}, R., {et~al.} 2007, \apj, 654, 138

\bibitem[{{Quadri} {et~al.}(2011){Quadri}, {Williams}, {Franx}, \&
  {Hildebrandt}}]{Quadri2011}
{Quadri}, R.~F., {Williams}, R.~J., {Franx}, M., \& {Hildebrandt}, H. 2011,
  ArXiv e-prints

\bibitem[{{Rettura} {et~al.}(2010){Rettura}, {Rosati}, {Nonino}, {Fosbury},
  {Gobat}, {Menci}, {Strazzullo}, {Mei}, {Demarco}, \& {Ford}}]{Rettura2010}
{Rettura}, A., {et~al.} 2010, \apj, 709, 512

\bibitem[{{Rettura} {et~al.}(2011){Rettura}, {Mei}, {Stanford}, {Raichoor},
  {Moran}, {Holden}, {Rosati}, {Ellis}, {Nakata}, {Nonino}, {Treu},
  {Blakeslee}, {Demarco}, {Eisenhardt}, {Ford}, {Fosbury}, {Illingworth},
  {Huertas-Company}, {Jee}, {Kodama}, {Postman}, {Tanaka}, \&
  {White}}]{Rettura2011}
---. 2011, \apj, 732, 94

\bibitem[{{Rosati} {et~al.}(2009){Rosati}, {Tozzi}, {Gobat}, {Santos},
  {Nonino}, {Demarco}, {Lidman}, {Mullis}, {Strazzullo}, {B{\"o}hringer},
  {Fassbender}, {Dawson}, {Tanaka}, {Jee}, {Ford}, {Lamer}, \&
  {Schwope}}]{Rosati2009}
{Rosati}, P., {et~al.} 2009, \aap, 508, 583

\bibitem[{{Scodeggio} {et~al.}(2009){Scodeggio}, {Vergani}, {Cucciati},
  {Iovino}, {Franzetti}, {Garilli}, {Lamareille}, {Bolzonella}, {Pozzetti},
  {Abbas}, {Marinoni}, {Contini}, {Bottini}, {Le Brun}, {Le F{\`e}vre},
  {Maccagni}, {Scaramella}, {Tresse}, {Vettolani}, {Zanichelli}, {Adami},
  {Arnouts}, {Bardelli}, {Cappi}, {Charlot}, {Ciliegi}, {Foucaud}, {Gavignaud},
  {Guzzo}, {Ilbert}, {McCracken}, {Marano}, {Mazure}, {Meneux}, {Merighi},
  {Paltani}, {Pell{\`o}}, {Pollo}, {Radovich}, {Zamorani}, {Zucca}, {Bondi},
  {Bongiorno}, {Brinchmann}, {de La Torre}, {de Ravel}, {Gregorini}, {Memeo},
  {Perez-Montero}, {Mellier}, {Temporin}, \& {Walcher}}]{Scodeggio2009}
{Scodeggio}, M., {et~al.} 2009, \aap, 501, 21

\bibitem[{{Skibba} {et~al.}(2009){Skibba}, {Bamford}, {Nichol}, {Lintott},
  {Andreescu}, {Edmondson}, {Murray}, {Raddick}, {Schawinski}, {Slosar},
  {Szalay}, {Thomas}, \& {Vandenberg}}]{Skibba2009}
{Skibba}, R.~A., {et~al.} 2009, \mnras, 399, 966

\bibitem[{{Smith} {et~al.}(2006){Smith}, {Hudson}, {Lucey}, {Nelan}, \&
  {Wegner}}]{Smith2006}
{Smith}, R.~J., {Hudson}, M.~J., {Lucey}, J.~R., {Nelan}, J.~E., \& {Wegner},
  G.~A. 2006, \mnras, 369, 1419

\bibitem[{{Sobral} {et~al.}(2011){Sobral}, {Best}, {Smail}, {Geach},
  {Cirasuolo}, {Garn}, \& {Dalton}}]{Sobral2011}
{Sobral}, D., {Best}, P.~N., {Smail}, I., {Geach}, J.~E., {Cirasuolo}, M.,
  {Garn}, T., \& {Dalton}, G.~B. 2011, \mnras, 411, 675

\bibitem[{{Strazzullo} {et~al.}(2010){Strazzullo}, {Rosati}, {Pannella},
  {Gobat}, {Santos}, {Nonino}, {Demarco}, {Lidman}, {Tanaka}, {Mullis},
  {Nu{\~n}ez}, {Rettura}, {Jee}, {B{\"o}hringer}, {Bender}, {Bouwens},
  {Dawson}, {Fassbender}, {Franx}, {Perlmutter}, \& {Postman}}]{Strazzullo2010}
{Strazzullo}, V., {et~al.} 2010, \aap, 524, A17+

\bibitem[{{Tanaka} {et~al.}(2010){Tanaka}, {De Breuck}, {Venemans}, \&
  {Kurk}}]{Tanaka2010}
{Tanaka}, M., {De Breuck}, C., {Venemans}, B., \& {Kurk}, J. 2010, \aap, 518,
  A18+

\bibitem[{{Thomas} {et~al.}(2005){Thomas}, {Maraston}, {Bender}, \& {Mendes de
  Oliveira}}]{Thomas2005}
{Thomas}, D., {Maraston}, C., {Bender}, R., \& {Mendes de Oliveira}, C. 2005,
  \apj, 621, 673

\bibitem[{{Thomas} {et~al.}(2010){Thomas}, {Maraston}, {Schawinski}, {Sarzi},
  \& {Silk}}]{Thomas2010}
{Thomas}, D., {Maraston}, C., {Schawinski}, K., {Sarzi}, M., \& {Silk}, J.
  2010, \mnras, 404, 1775

\bibitem[{{Tran} {et~al.}(2010){Tran}, {Papovich}, {Saintonge}, {Brodwin},
  {Dunlop}, {Farrah}, {Finkelstein}, {Finkelstein}, {Lotz}, {McLure},
  {Momcheva}, \& {Willmer}}]{Tran2010}
{Tran}, K., {et~al.} 2010, \apjl, 719, L126

\bibitem[{{Tran} {et~al.}(2007){Tran}, {Franx}, {Illingworth}, {van Dokkum},
  {Kelson}, {Blakeslee}, \& {Postman}}]{Tran2007}
{Tran}, K.-V.~H., {Franx}, M., {Illingworth}, G.~D., {van Dokkum}, P.,
  {Kelson}, D.~D., {Blakeslee}, J.~P., \& {Postman}, M. 2007, \apj, 661, 750

\bibitem[{{Tremonti} {et~al.}(2004){Tremonti}, {Heckman}, {Kauffmann},
  {Brinchmann}, {Charlot}, {White}, {Seibert}, {Peng}, {Schlegel}, {Uomoto},
  {Fukugita}, \& {Brinkmann}}]{Tremonti2004}
{Tremonti}, C.~A., {et~al.} 2004, \apj, 613, 898

\bibitem[{{van den Bosch} {et~al.}(2008){van den Bosch}, {Aquino}, {Yang},
  {Mo}, {Pasquali}, {McIntosh}, {Weinmann}, \& {Kang}}]{vandenBosch2008}
{van den Bosch}, F.~C., {Aquino}, D., {Yang}, X., {Mo}, H.~J., {Pasquali}, A.,
  {McIntosh}, D.~H., {Weinmann}, S.~M., \& {Kang}, X. 2008, \mnras, 387, 79

\bibitem[{{van Dokkum} \& {van der Marel}(2007)}]{vandokkum2007}
{van Dokkum}, P.~G., \& {van der Marel}, R.~P. 2007, \apj, 655, 30

\bibitem[{{von der Linden} {et~al.}(2010){von der Linden}, {Wild}, {Kauffmann},
  {White}, \& {Weinmann}}]{vonderlinden2010}
{von der Linden}, A., {Wild}, V., {Kauffmann}, G., {White}, S.~D.~M., \&
  {Weinmann}, S. 2010, \mnras, 404, 1231

\bibitem[{{Vulcani} {et~al.}(2010){Vulcani}, {Poggianti}, {Finn}, {Rudnick},
  {Desai}, \& {Bamford}}]{Vulcani2010}
{Vulcani}, B., {Poggianti}, B.~M., {Finn}, R.~A., {Rudnick}, G., {Desai}, V.,
  \& {Bamford}, S. 2010, \apjl, 710, L1

\bibitem[{{Weinmann} {et~al.}(2010){Weinmann}, {Kauffmann}, {von der Linden},
  \& {De Lucia}}]{Weinmann2010}
{Weinmann}, S.~M., {Kauffmann}, G., {von der Linden}, A., \& {De Lucia}, G.
  2010, \mnras, 406, 2249

\bibitem[{{Weinmann} {et~al.}(2006){Weinmann}, {van den Bosch}, {Yang}, \&
  {Mo}}]{Weinmann2006}
{Weinmann}, S.~M., {van den Bosch}, F.~C., {Yang}, X., \& {Mo}, H.~J. 2006,
  \mnras, 366, 2

\bibitem[{{Wilson} {et~al.}(2009){Wilson}, {Muzzin}, {Yee}, {Lacy}, {Surace},
  {Gilbank}, {Blindert}, {Hoekstra}, {Majumdar}, {Demarco}, {Gardner},
  {Gladders}, \& {Lonsdale}}]{Wilson2009}
{Wilson}, G., {et~al.} 2009, \apj, 698, 1943

\bibitem[{{Wuyts} {et~al.}(2007){Wuyts}, {Labb{\'e}}, {Franx}, {Rudnick}, {van
  Dokkum}, {Fazio}, {F{\"o}rster Schreiber}, {Huang}, {Moorwood}, {Rix},
  {R{\"o}ttgering}, \& {van der Werf}}]{Wuyts2007}
{Wuyts}, S., {et~al.} 2007, \apj, 655, 51

\bibitem[{{Yan} {et~al.}(2009){Yan}, {Newman}, {Faber}, {Coil}, {Cooper},
  {Davis}, {Weiner}, {Gerke}, \& {Koo}}]{Yan2009}
{Yan}, R., {et~al.} 2009, \mnras, 398, 735

\bibitem[{{Yee} {et~al.}(2007){Yee}, {Gladders}, {Gilbank}, {Majumdar},
  {Hoekstra}, \& {Ellingson}}]{Yee2007}
{Yee}, H.~K.~C., {Gladders}, M.~D., {Gilbank}, D.~G., {Majumdar}, S.,
  {Hoekstra}, H., \& {Ellingson}, E. 2007, in Astronomical Society of the
  Pacific Conference Series, Vol. 379, Cosmic Frontiers, ed. {N.~Metcalfe \&
  T.~Shanks}, 103--+

\end{thebibliography}




\end{document}